\documentclass{JHEP3}
\usepackage{amsmath}
\usepackage{cite}
\usepackage{epsfig}
\usepackage{xspace}
\usepackage{graphics}
\usepackage[latin1]{inputenc}

\newcommand{\rivet}{R{\smaller IVET}\xspace}
\newcommand{\thepeg}{T{\smaller HE}PEG\xspace}
\newcommand{\herwig}{H{\smaller ERWIG++}\xspace}

\newcommand{\eqRef}[1]{eq.~\eqref{#1}\xspace}

\usepackage{cite}
\usepackage{xspace}
\usepackage{graphics}
\usepackage[latin1]{inputenc}
\usepackage{relsize}
\usepackage{axodraw}
\usepackage{epsfig}
\usepackage{amssymb}
\usepackage{mathrsfs}
\usepackage{mathcomp}

\newcommand{\ariadne}{A{\smaller RIADNE}\xspace}
\newcommand{\dipsy}{{\smaller DIPSY}\xspace}
\newcommand{\pythia}{P{\smaller YTHIA}\,8\xspace}

\newcommand{\si}[0]{\ensuremath{\sigma}}

\newcommand{\ga}[0]{\ensuremath{\gamma}}

\newcommand{\La}[0]{\ensuremath{\Lambda}}

\newcommand{\al}[0]{\ensuremath{\alpha}}

\newcommand{\rh}[0]{\ensuremath{\rho}}

\newcommand{\rmax}[0]{\ensuremath{R_{\max}}}

\def\pmb#1{{\mbox{\boldmath$#1$}}}
\def\ie{\emph{i.e.}}
\def\eg{\emph{e.g.}}

\keywords{Small-$x$ physics, Saturation, Diffraction, Dipole Model, DIS}
\preprint{arXiv:1103.4321 [hep-ph]\\
LU-TP 11-13\\
CERN-PH-TH-2011-058\\
MCnet-11-08\\
\today}

\title{Inclusive and Exclusive observables from
    dipoles in high energy collisions\footnote{Work supported in
    parts by the EU Marie Curie RTN MCnet (MRTN-CT-2006-035606), and
    the Swedish research council (contracts 621-2008-4252 and 621-2009-4076).}}

\author{Christoffer Flensburg$^1$, Gösta Gustafson$^1$,
  and Leif Lönnblad$^{1,2}$\\
  $^1$Dept.~of Astronomy and Theoretical Physics, Lund University, \\
  Sölvegatan 14A, Lund, Sweden\\
  $^2$CERN Theory Unit,\\
  1211 Geneva 23, Switzerland \vspace{2mm}\\
  E-mail: christoffer.flensburg@thep.lu.se, gosta@thep.lu.se, leif.lonnblad@thep.lu.se}

\abstract{We present a new model for simulating exclusive final states
  in minimum-bias collisions between hadrons.

  In a series of papers we have developed a Monte Carlo model based on
  Mueller's dipole picture of BFKL-evolution, supplemented with
  non-leading corrections, which has shown to be very successful in
  describing inclusive and semi-inclusive observables in hadron
  collisions. In this paper we present a further extension of this
  model to also describe exclusive final states. This is a highly
  non-trivial extension, and we have encountered many details that
  influence the description, and for which no guidance from
  perturbative QCD could be found. Hence we have had to make many
  choices based on semi-classical and phenomenological arguments.

  The end result is a new event generator called \dipsy which can be
  used to simulate complete minimum-bias non-diffractive hadronic
  collision events. Although the description of data from the Tevatron
  and LHC is not quite as good as for \pythia, the most advanced of
  the general purpose event generator programs for these processes,
  our results are clearly competitive, and can be expected to improve
  with careful tuning. In addition, as our model is very different
  from conventional multiple scattering scenaria, the \dipsy program
  can be used to gain deeper insight in the soft and semi-hard
  processes involved both in hadronic and heavy ion collisions.

}

\begin{document}

\sloppy

\section{Introduction}
\label{sec:int}

In high energy collisions the high density of gluons at small $x$
imply that typical events contain several hard partonic subcollisions,
and that saturation effects become important.  A dynamical model based
on BFKL evolution \cite{Kuraev:1977fs,Balitsky:1978ic} for small $x$
and including saturation effects, has been presented in a series of
papers
\cite{Avsar:2005iz,Avsar:2006jy,Avsar:2007xg,Flensburg:2008ag,Flensburg:2010kq}.
It is based on Mueller's dipole cascade
model~\cite{Mueller:1993rr,Mueller:1994jq,Mueller:1994gb}, which is a
formulation of the leading logarithmic (LL) BFKL evolution
approximation in impact-parameter space.  In our model we also include the most
essential beyond LL corrections to BFKL. In a MC simulation these
effects are resummed to all orders (see the discussion in
ref.~\cite{Salam:1998tj}). The transverse coordinate
space is particularly suitable for treating unitarity constraints,
multiple collisions, and saturation effects.  Mueller's model includes
multiple subcollisions in the frame used for the analysis, with
similarities to saturation effects in the Color Glass Condensate
\cite{Iancu:2002xk,Iancu:2003xm,Gelis:2010nm, Iancu:2000hn,JalilianMarian:1997jx} or the BK equation \cite{Balitsky:1997mk, Kovchegov:1999ua}. In our cascade model we also
include saturation effects corresponding to pomeron loops within the
evolution, in a way which is almost independent of the Lorentz frame used. 
Also effects of confinement
are included and important for the result.

In the dipole picture the colour charge in a gluon is screened by a 
corresponding anticharge in a neighbouring gluon. Gluon radiation implies that
one dipole is split in two dipoles, and the result of the cascade is
a chain of colour connected gluons. When two cascades collide, the can interact
via gluon exchange. This implies an exchange of colour, producing two colour 
chains forming a BFKL ladder between projectile and target. Multiple
interaction and saturation represent multiple pomeron exchange and pomeron 
loops. Note here that what is multiple interaction in one Lorentz frame is 
related to saturation within the cascade in another frame. 

The BFKL formalism is directly applicable to inclusive observables,
and the cited articles include results for total, (quasi)elastic, and
diffractive cross sections in $pp$ scattering and in DIS, in very good
agreement with experimental data.  The numerical results are obtained
with a Monte Carlo (MC) event generator called \dipsy\footnote{Acronym
  for Dipoles in Impact-Parameter Space and rapidity (Y)}.

In the present paper we want to generalize the model to describe
exclusive final states. Our aim is here not to generate the most precise predictions, and compete
with traditional generators like \pythia and \herwig.
Instead we want to test our understanding of small $x$ evolution and
saturation effects. In the traditional programs the evolution to small $x$
is determined by structure functions, for which the behaviour at a cutoff
scale $Q_0^2$ is tuned to experimental data, and saturation effects are
tuned by the value of a cutoff $p_{\perp 0}$ for hard subcollisions,
determined individually for every energy. In our model these effects are
instead generated dynamically, based on a perturbative BFKL pomeron and
screening effects from a dipole picture, including a model for pomeron
loops. In addition, in our approach we can also study nontrivial effects
of fluctuations and correlations, which are neglected in most analyses. In
the present paper we study $pp$ collisions, but the model can also be
applied to interactions with nuclei, where it gives a unique possibility
to study finite size effects and effects of fluctuations. We want to
return to these problems in future publications.

In order to generate exclusive states it is necessary to take into account
colour coherence and angular ordering as well as soft radiation. The
latter includes contributions from the $z=1$ singularity in the gluon
splitting function. These effects are taken into account in the CCFM
formalism\cite{Catani:1989sg, Ciafaloni:1987ur}, which also reproduces
the BFKL result for the inclusive cross section. In
ref.~\cite{Andersson:1995ju}, describing the Linked Dipole Chain
model, it was demonstrated that the inclusive cross section is fully
determined by a subset of the gluons in the CCFM approach, denoted
``$k_\perp$-changing'' gluons. As illustrated in fig.~\ref{fig:defqk},
we denote the real emitted gluons in a gluon ladder $q_{\perp i}$, and
the virtual links $k_{\perp i}$. Momentum conservation here implies
that $\mathbf{k}_{\perp i-1}=\mathbf{k}_{\perp i}+\mathbf{q}_{\perp
  i}$.  In a $k_\perp$-changing emission the difference is large
between two adjacent virtual links, $k_{\perp i-1}$ and $k_{\perp i}$.
This means that $q_{\perp i}\approx \max (k_{\perp i-1}, k_{\perp
  i})$.  Softer emissions with smaller $q_{\perp i}$ have $k_{\perp
  i-1}\approx k_{\perp i}$, and are ``$k_\perp$-conserving''.  In CCFM
gluon emission is associated with ``non-Sudakov'' form factors, and in
ref.~\cite{Andersson:1995ju} it was demonstrated that these gluons can
be summed up in such a way, that the total cross section is fully
specified by the $k_\perp$-changing emissions with no associated
non-Sudakov form factors. The remaining $k_\perp$-changing gluons
$q_{\perp i}^{\mathrm{prim}}$ are called ``primary gluons'' in
ref.~\cite{Andersson:1995ju} and ``backbone gluons'' in
ref.~\cite{Salam:1999ft}. They are ordered in both lightcone variables
$q_+$ and $q_-$, and therefore also in rapidity or angle, in
accordance with QCD coherence. (In the LDC model
\cite{Andersson:1995ju} emissions with $k_{\perp i}\approx k_{\perp
  i-1}$, on the boundary between $k_\perp$-conserving and
$k_\perp$-changing emissions, were treated slightly different from the
BFKL prescription, which also implied a slightly different value for
the exponent $\lambda$ in the power-like increase $\propto
1/x^\lambda$. In the present model the LL BFKL result is exactly
reproduced.)

\FIGURE[t]{
\scalebox{1.0}{\mbox{
\begin{picture}(240,150)(-10,0)
  \Text(-10,15)[]{\large $P$}
  \Line(0,15)(40,15)
  \Line(40,20)(70,20)
  \Line(40,15)(70,15)
  \Line(40,10)(70,10)
  \GOval(40,15)(10,7)(0){1}
  \Line(50,20)(60,40)\Text(47,34)[]{\large $k_{0}$}
    \Line(60,40)(90,40)\Text(100,40)[]{\large $q_{1}$}
  \Line(60,40)(70,70)\Text(55,55)[]{\large $k_{1}$}
    \Line(70,70)(105,70)\Text(115,70)[]{\large $q_{2}$}
  \Line(70,70)(75,100)\Text(65,85)[]{\large $k_{2}$}
    \Line(75,100)(110,100)\Text(120,100)[]{\large $q_{3}$}
  \DashLine(75,100)(75,130){2}

\end{picture}}}
\caption{\label{fig:defqk} A chain of initial state gluon
  emissions. We denote the real emitted gluons in a gluon ladder
  $q_{\perp i}$, and the virtual links $k_{\perp i}$. }
}

The chains of primary gluons also determine the structure of the final state,
but as discussed above, to get the full exclusive states softer emissions 
must be added as final state radiation. In summary the 
generation of exclusive final states contains the following steps:
\begin{enumerate}
\item The generation of two dipole cascades from the projectile and the
  target, in accordance with BFKL dynamics and saturation.
\item Calculating which pairs of one parton in the projectile and one parton in
  the target do interact. In BFKL the emission of gluons is a Poissonian 
  type process, and the interaction probability is calculated in the eikonal
  approximation.
\item Extracting the primary, $k_\perp$-changing, gluons, and checking that
  they are given the correct weight. These gluons form colour connected chains
  between the projectile and the target, including branchings and
  loops. This step includes the removal of branches in the
  cascade which do not interact, and restoring energy-momentum conservation.
\item Adding final state radiation in relevant parts of phase space. The
  result consists of chains of colour connected gluons.
\item Hadronization. Here we use Lund string hadronization, where colour 
  strings are stretched between the colour-connected gluons.
\end{enumerate}

The outline of the paper is as follows: In section~\ref{sec:inclusive}
we describe our dipole cascade model for inclusive observables.  The
problems encountered when generating exclusive final states are
discussed in section~\ref{sec:exclusive}.  In
section~\ref{sec:generation} we describe the key procedures used in
the MC for solving the problems presented in
section~\ref{sec:exclusive}, while some of the more technical details
are left for the appendices.  Some results and predictions are given
in section~\ref{sec:results}, together with a discussion of tunable
parameters in the program. We conclude with a summary and outlook in
section~\ref{sec:conclusions}.

\section{The Lund dipole cascade model for inclusive cross sections}
\label{sec:inclusive}
\subsection{Mueller's cascade model and the eikonal formalism}

Mueller's dipole cascade model 
\cite{Mueller:1993rr,Mueller:1994jq,Mueller:1994gb} is a formulation
of LL BFKL evolution in transverse coordinate space. 
Gluon radiation from the colour charge in a parent quark or gluon is screened 
by the accompanying anticharge 
in the colour dipole. This suppresses emissions at large transverse separation,
which corresponds to the suppression of small $k_\perp$ in BFKL.
For a dipole $(\pmb{x},\pmb{y})$ the probability per unit rapidity ($Y$) for
emission of a gluon at transverse position $\pmb{z}$ is given by
\begin{eqnarray}
\frac{d\mathcal{P}}{dY}=\frac{\bar{\alpha}}{2\pi}d^2\pmb{z}
\frac{(\pmb{x}-\pmb{y})^2}{(\pmb{x}-\pmb{z})^2 (\pmb{z}-\pmb{y})^2},
\,\,\,\,\,\,\, \mathrm{with}\,\,\, \bar{\alpha} = \frac{3\alpha_s}{\pi}.
\label{eq:dipkernel1}
\end{eqnarray}
This emission implies that the dipole is split into two dipoles, which
(in the large $N_c$ limit) emit new gluons independently, as illustrated in
fig.~\ref{fig:dipev}. The result reproduces the BFKL evolution, with the 
number of dipoles growing exponentially with $Y$.

\FIGURE[t]{
\scalebox{1.0}{\mbox{

\begin{picture}(250,80)(0,5)
\Vertex(10,80){2}
\Vertex(10,0){2}
\Text(5,80)[]{$\pmb{x}$}
\Text(5,0)[]{$\pmb{y}$}
\Line(10,80)(10,0)
\LongArrow(30,40)(60,40)
\Vertex(100,80){2}
\Vertex(100,0){2}
\Vertex(120,50){2}
\Text(95,80)[]{$\pmb{x}$}
\Text(95,0)[]{$\pmb{y}$}
\Text(128,50)[]{$\pmb{z}$}
\Line(100,80)(120,50)
\Line(120,50)(100,0)
\LongArrow(140,40)(170,40)
\Vertex(205,80){2}
\Vertex(225,50){2}
\Vertex(233,30){2}
\Vertex(205,0){2}
\Text(200,80)[]{$\pmb{x}$}
\Text(200,0)[]{$\pmb{y}$}
\Text(233,50)[]{$\pmb{z}$}
\Text(241,30)[]{$\pmb{w}$}
\Line(205,80)(225,50)
\Line(225,50)(233,30)
\Line(233,30)(205,0)
\end{picture}
}}
\caption{\label{fig:dipev} The evolution of the dipole cascade in transverse
  coordinate space. In each step, a dipole can split into two new dipoles 
  with decay probability given by \eqRef{eq:dipkernel1}. }
}

\FIGURE[t]{
\scalebox{1.3}{\mbox{

\begin{picture}(280,80)(-50,0)

\Text(27,35)[r]{{\footnotesize $i$}}
\Text(53,37)[l]{{\footnotesize $j$}}
\Vertex(150,32){1}
\Vertex(130,30){1}
\Vertex(130,40){1}
\Vertex(151,43){1}
\Vertex(50,32){1}
\Vertex(30,30){1}
\Vertex(30,40){1}
\Vertex(51,43){1}
\Text(2,40)[l]{{\footnotesize $proj.$}}
\Text(60,40)[l]{{\footnotesize $targ.$}}
\LongArrowArcn(100,20)(20,120,60)

\Line(10,20)(30,30)

\Line(30,40)(20,50)
\Line(20,50)(28,57)
\Line(28,57)(18,70)

\Vertex(10,20){1}
\Vertex(20,50){1}
\Vertex(28,57){1}
\Vertex(18,70){1}

\Vertex(70,20){1}
\Vertex(60,10){1}
\Vertex(60,52){1}
\Vertex(60,65){1}

\Vertex(110,20){1}
\Vertex(120,50){1}
\Vertex(128,57){1}
\Vertex(118,70){1}
\Vertex(170,20){1}
\Vertex(160,10){1}
\Vertex(160,52){1}
\Vertex(160,65){1}

\Line(70,20)(60,10)
\Line(60,10)(50,32)
\Line(51,43)(60,52)
\Line(60,52)(60,65)
\ArrowLine(130,30)(110,20)
\ArrowLine(120,50)(130,40)
\Line(120,50)(128,57)
\Line(128,57)(118,70)

\Line(170,20)(160,10)
\ArrowLine(160,10)(150,32)
\ArrowLine(151,43)(160,52)
\Line(160,52)(160,65)

\ArrowLine(30,40)(30,30)
\ArrowLine(130,40)(151,43)
\ArrowLine(50,32)(51,43)
\ArrowLine(150,32)(130,30)
\end{picture}
}}
\caption{\label{fig:dipint} A dipole--dipole interaction implies
  exchange of colour and reconnection of the dipole chains in the
  colliding cascades. The arrows indicate the direction of the dipole
  going from colour charge to anticharge.}
}

In a high energy collision, the dipole cascades in the projectile and
the target are evolved from their rest frames to the rapidities they
will have in the specific Lorentz frame chosen for the analysis.  Two
colliding dipoles interact via gluon exchange, which implies a colour
connection between the projectile and target remnants, as indicated in
fig.~\ref{fig:dipint}.  In the Born approximation the interaction
probability between one dipole with coordinates
$(\pmb{x}_i,\pmb{y}_i)$ in the projectile, and one with coordinates
$(\pmb{x}_j,\pmb{y}_j)$ in the target, is given by (the factor 2 in
the definition of $f$ is a convention)
\begin{equation}
  2f_{ij} = 2f(\pmb{x}_i,\pmb{y}_i|\pmb{x}_j,\pmb{y}_j) =
  \frac{\alpha_s^2}{4}\biggl[\log\biggl(\frac{(\pmb{x}_i-\pmb{y}_j)^2
    (\pmb{y}_i-\pmb{x}_j)^2}
  {(\pmb{x}_i-\pmb{x}_j)^2(\pmb{y}_i-\pmb{y}_j)^2}\biggr)\biggr]^2.
\label{eq:dipamp}
\end{equation}

At high energies the strong increase in the number of dipoles
gives a large probability for multiple dipole--dipole subcollisions,
and the transverse coordinate space is
particularly suitable for the treatment of multiple
interactions and unitarity constraints. Assuming that the subcollisions are
uncorrelated, multiple collisions are taken into account in the eikonal 
approximation, where the probability for an inelastic interaction is given by
\begin{equation}
\mathrm{Int.\,\, prob.} = 1-e^{-2F}, \,\,\,\mathrm{with}\,\, F=\sum f_{ij}. \label{eq:nondiff}
\end{equation}
The multiple interactions produce loops of dipole chains, corresponding to 
the pomeron loops in the reggeon formalism.

Assuming also that the elastic scattering amplitude, $T$, is driven by
absorption into inelastic states, we find via the optical theorem
\begin{equation}
T=1-e^{-F},
\label{tf-relationmueller}
\end{equation}
and thus the probability for an elastic interaction given by $T^2$. (We use a
definition such that $T$ in this case is purely real.)

To get the final result for the proton--proton cross sections we have to take
averages over the projectile and target cascades, and integrate over impact
parameter $\pmb{b}$. For the total non-diffractive cross section
we get
\begin{equation}
\sigma_{\mathrm{inel}}=\int d^2\pmb{b} \langle 1-e^{-2F(\pmb{b})}\rangle=
\int d^2\pmb{b} \langle 1-(1-T(\pmb{b}))^2 \rangle.
\end{equation}
When the projectile has an internal structure, which can be excited, the
purely elastic cross section is obtained by taking the average of the amplitude 
$T$, before taking the square:
\begin{equation}
\si_{\text{el}} = \int d^2\pmb{b} \langle T(\pmb{b}) \rangle^2.
\end{equation}
Taking first the square gives the total diffractive scattering (see 
ref.~\cite{Good:1960ba}):
\begin{equation}
\si_{\text{diff}} = \int d^2\pmb{b} \langle T(\pmb{b})^2 \rangle.
\end{equation}
The cross section for diffractive excitation is consequently given by the 
difference
\begin{equation}
\si_{\text{diff ex}} = \si_{\text{diff}} - \si_{\text{el}} = \int d^2\pmb{b} 
\left(\langle T(\pmb{b})^2 \rangle - \langle T(\pmb{b}) \rangle^2\right).
\end{equation}
Thus diffractive excitation is determined by the fluctuations in the
scattering amplitude. 
It is also possible to calculate \emph{e.g.} the cross section for single
diffractive excitation of the (right-moving) projectile by taking the average
over the target cascade before, but over the projectile cascade after, 
squaring the amplitude. We
also here have to subtract the elastic scattering, and thus get
\begin{equation}
\si_{\text{SD}} = \int d^2\pmb{b} \left(\langle \langle T(\pmb{b}) \rangle_L^2
\rangle_R-\langle T(\pmb{b}) \rangle^2\right),
\label{eq:SD}
\end{equation}
where $\langle \cdot \rangle_{L(R)}$ refers to the average over the left
(right) cascade only.

\subsection{The Lund dipole cascade model}

The Lund dipole cascade introduces a number of corrections to Mueller's
original formulation,
which are all beyond the leading logarithmic approximation. The
corrections are described in greater detail in previous articles \cite{Avsar:2005iz,Avsar:2006jy,Avsar:2007xg},
and a short summary is presented here.

\subsubsection{Beyond LL BFKL evolution}

The NLL corrections to BFKL evolution have three major sources
\cite{Salam:1999cn}:
\vspace{2mm}

\emph {Non-singular terms in the splitting function:}

These terms suppress large $z$-values in the individual parton branchings,
and correspond approximately to a veto for $\ln 1/z < 11/12$. It prevents
the child from being faster than its recoiling parent. Most of this effect
is taken care of
by including energy-momentum conservation in the evolution, and ordering
in the lightcone momentum $p_+$. To first approximation this corresponds
to a veto for $z>0.5$
\cite{Andersson:1995ju}. This cut is effectively taken
into account by associating a dipole with transverse size $r$ with a
transverse momentum
$k_\perp = 1/r$, and demanding conservation of $p_+$ in every step in the
evolution. In addition this gives an effective cutoff for small dipoles,
which eliminates the numerical
problems encountered in the MC implementation by Mueller and Salam \cite{Mueller:1996te}.
\vspace{2mm}

\emph {Projectile-target symmetry:}

A parton chain should look the same if generated from the target end
as from the projectile end. The corresponding corrections are also
called energy scale terms, and is essentially equivalent to the so
called consistency constraint\cite{Kwiecinski:1996td}. This effect is
taken into account by conservation of both positive and negative
lightcone momentum components, $p_+$ and $p_-$.  \vspace{2mm}

\emph {The running coupling:}

This is relatively easily included in a MC simulation process. The scale in
the running coupling is chosen as the largest transverse momentum in the
vertex, in accordance with the results in ref.~\cite{Balitsky:2008zzb}.
\vspace{2mm}

It is well known that the NLL corrections are not sufficient to give a
realistic result for small $x$ evolution. The characteristic function
$\chi(\gamma)$ has spurious singularities at $\gamma=0$ and 1, the point
$\gamma=1/2$ does not correspond to a saddle point, and the  power
$\lambda_{\text{eff}}$, determining the growth for small $x$, is negative for
$\alpha_s = 0.2$.
As discussed in refs.~\cite{Salam:1998tj, Ciafaloni:1998iv,
Ciafaloni:1999au}, these problems can be cured by a resummation of
subleading logs to all orders, which is executed automatically in our MC
simulation.

As shown in ref.~\cite{Andersson:1995ju} the projectile-target symmetry
and ordering in $p_-$ shifts the spurious singularities in $\chi(\gamma)$
to $-\omega/2$ and $1+\omega/2$ (when the symmetric scale $k_1 k_2$ is
used in the Mellin transform). Forshaw \emph{et al.} have demonstrated
that if this shift is taken into account, the sensitivity to the exact
size of the veto in $z$ is very small \cite{Forshaw:1999xm}. Salam has
made a detailed study of different resummation schemes, and finds that the
result is very stable, estimating the uncertainty to about
15\% \cite{Salam:1998tj, Salam:1999cn}.

\subsubsection{Non-linear effects and saturation}

As mentioned above, multiple interactions produce loops of dipole chains
corresponding to pomeron loops. Mueller's model includes all loops cut in 
the particular Lorentz frame used for the analysis, but not loops contained
within the evolution of the individual cascades.  As for dipole scattering
the probability for such loops is given by $\alpha_s$, and therefore
formally colour suppressed compared to dipole splitting, which is
proportional to $\bar{\alpha}=N_c \alpha_s/\pi$. These loops are therefore
related to the probability that two dipoles have the same colour. Two dipoles
with the same colour form a quadrupole field. Such a field may be better
approximated by two dipoles formed by the closest colour--anticolour
charges. This corresponds to a recoupling of the colour dipole chains. We call
this process a dipole ``swing''. 

Thus double dipole scattering in one Lorentz frame corresponds to a swing in
another frame. This is illustrated in fig.~\ref{fig:loops2}, which shows the
collision of two dipole cascades in two different Lorentz frames. The
horizontal axis represents rapidity, and the vertical axis shows symbolically
the two transverse coordinates. The left column illustrates the collision 
in a central Lorentz frame,
where the rest frame is given by the dashed vertical line, while the right
column shows the same process in an asymmetric frame. Row ($a$) shows the
starting configuration with two dipoles. In the left column row ($b$) shows 
the evolved dipole cascades just before the collision, row ($c$) the
situation when dipoles (23) and (46) have interacted, and row ($d$) after a
simultanous interaction between (02) and (57). In the right column row ($b$)
shows how the left cascade has a longer development, while in this example
there is no emission in the right dipole. It is now assumed that dipoles (34)
and (25) have the same colour, and row ($c$) shows the cascade after a swing 
between these dipoles. Finally in row ($d$) dipoles (01) and (35) have
interacted. We see that the final result is the same. A double interaction in
one frame is equivalent to a swing and a single interaction in another frame. 
We note that in this formulation the number of dipoles is not reduced. The 
saturation effect is obtained because the smaller dipoles have smaller cross 
sections. However, when as in this case, the dipoles in the loop (247) in the 
right figure on row ($c$) do not interact, the process can also be 
interpreted as a $2 \rightarrow 1$ process, which corresponds to the emission of
gluon 2 from the right cascade as seen in the central Lorentz frame.  

\FIGURE[t]{
  \begin{minipage}{0.4\linewidth}
    \begin{center}
      \includegraphics[scale=0.7]{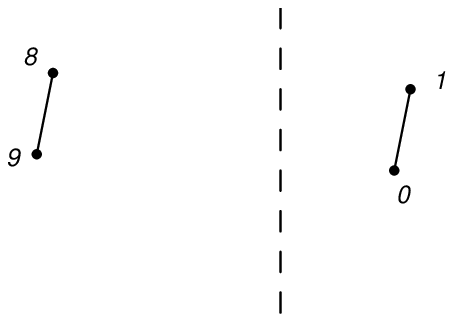}

      \includegraphics[scale=0.7]{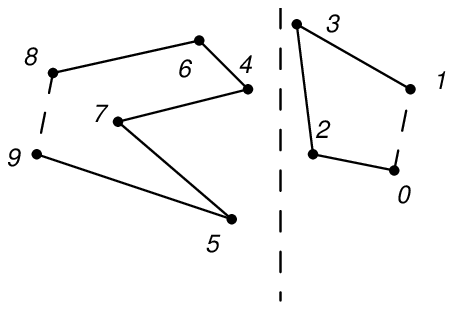}

      \includegraphics[scale=0.7]{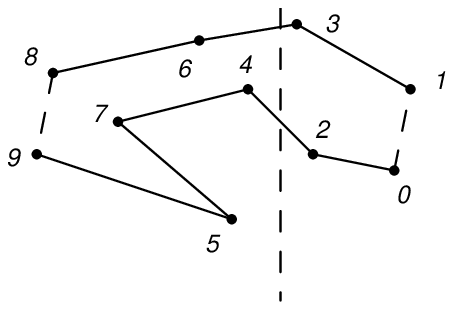}

      \includegraphics[scale=0.7]{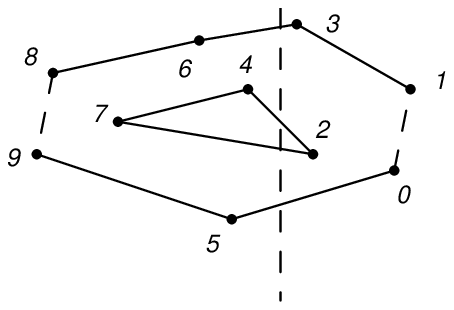}
    \end{center}
  \end{minipage}
  \begin{minipage}{0.08\linewidth}
    \begin{center}
      \vspace{9mm}

      $(a)$

      \vspace{15mm}

      $(b)$

      \vspace{15mm}

      $(c)$

      \vspace{15mm}

      $(d)$

      \vspace{9mm}
    \end{center}
  \end{minipage}
\hspace{10mm}
  \begin{minipage}{0.4\linewidth}
      \includegraphics[scale=0.7]{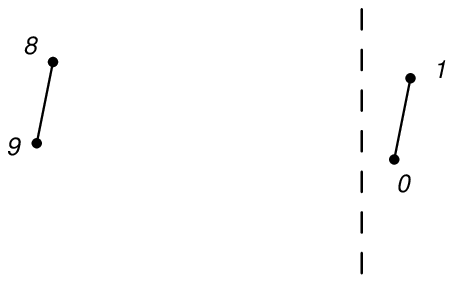}

      \includegraphics[scale=0.7]{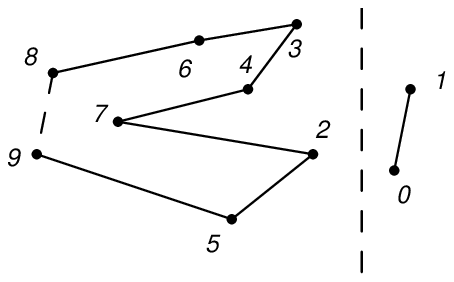}

      \includegraphics[scale=0.7]{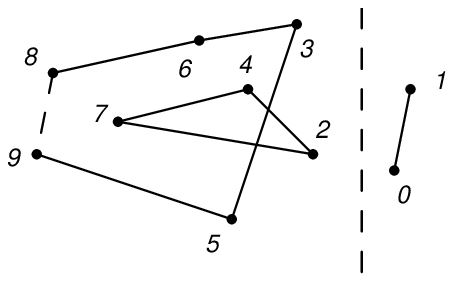}

      \includegraphics[scale=0.7]{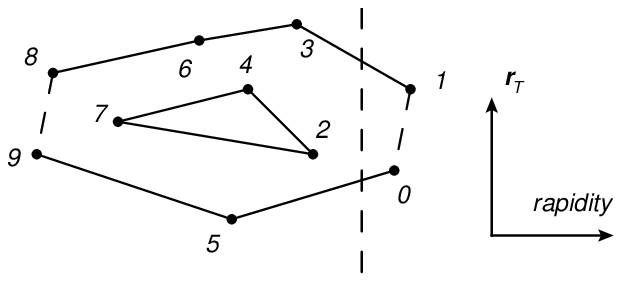}
  \end{minipage}

  \caption{ \label{fig:loops2} Creation of the same loop diagram in different
    interaction frames, shown as the vertical dashed lines. The horizontal
    axis represents rapidity, and the vertical axis the two transverse
    coordinates. The left side shows a central Lorentz frame where multiple interactions (dipole (23) interacting with (46) and dipole (02) interacting with (57)) forms the loop. On the right side, the same loop is formed in an asymmetric frame with a single interaction, but with a swing between (34) and (25).}
}

In the MC, dipoles with the same colour are
allowed to swing with a probability proportional to 
$(r_1^2 r_2^2)/(r_3^2 r_4^2)$,
where $r_1$ and $r_2$ are the sizes of the original dipoles and $r_3$ 
and $r_4$ are the sizes of the recoupled dipoles. Dipoles with the same 
colour are allowed to swing back and forth, which
results in an equilibrium, where the smaller dipoles have a larger weight. 
The given weights favour the formation of smaller 
dipoles, and although this scheme does not give an exactly frame independent
result, the MC simulations show that it is a fairly good approximation.

\subsubsection{Confinement effects}

Confinement effects are included via an effective gluon mass, which
gives an exponential suppression for very large dipoles \cite{Avsar:2007xg}. 
This prevents the proton from growing too fast in transverse size, and is also
essential to satisfy Froisart's bound at high energies \cite{Avsar:2008dn}.

\subsection{Initial dipole configurations}

In DIS an initial photon is split into a $q\bar{q}$ pair, and
for larger $Q^2$ the wavefunction for a virtual photon can be determined 
perturbatively. The internal structure of the proton is, however,
governed by soft QCD, and is not 
possible to calculate perturbatively. In our model it is represented by
an equilateral triangle formed by three dipoles, and with a radius of
$3$ GeV$^{-1} \approx 0.6$ fm. The model should be used at low $x$,
and when the system is evolved over a large 
rapidity range the observable results depend only weakly on the exact 
configuration of the initial dipoles, or whether the charges are treated as 
(anti)quarks or gluons.

\section{From inclusive to exclusive observables}
\label{sec:exclusive}

In this section we discuss the necessary steps when going from
inclusive observables to exclusive final states. The technical details
of the MC implementation are left to sec.~\ref{sec:generation} and the
appendices.

\subsection{The chain of $k_\perp$-changing gluons}

In the introduction we discussed briefly how the structure of an
exclusive final state is determined by a backbone of gluons. The first
step is therefore to extract this backbone chain of $k_\perp$-changing
or primary gluons, with their proper weights. As discussed above, the
different subcollisions are
assumed to be uncorrelated, and the interaction probability for two dipoles,
$i$ and $j$, is given by $1-\exp(-2f_{ij})$, with $f_{ij}$ given by 
\eqRef{eq:dipamp}. When the interacting dipoles are determined, the
backbone chains can be found by tracing their parents and previous 
ancestors backwards in rapidity. A resulting chain is shown in
fig.~\ref{fig:backbone}a, and we here use the notation $q_i$ for the real
emitted gluons, and $k_i$ for the virtual links in the evolution. 
The same chain is presented in fig.~\ref{fig:backbone}b in a
($y, \ln{q_\perp^2}$) diagram, with $y$ being the rapidity. In this
plot $\ln q_+$ and $\ln q_-$ ($=y\pm \ln k_\perp$) increase towards
the upper right and upper left corners.  Due to the ``consistency
constraint''\cite{Kwiecinski:1996td}, which is part of the NLL
corrections to BFKL, the primary gluons are ordered not only in $q_+$,
but also in $q_-$.  This also implies an ordering in rapidity or
angle, in accordance with QCD coherence. The real emissions satisfy the relation
$q_{i+}q_{i-}=q_{i\perp}^2$, and are represented by points in
fig.~\ref{fig:backbone}b. The space-like momenta for the virtual links $k_i$ are
not constrained by such a relation, and are represented by horizontal lines 
indicating their transverse momentum.

The weight for a particular chain with (real) gluons $q_i$ is given by
\cite{Andersson:1995ju,Salam:1999ft} (assuming $k_{\perp 0}=k_{\perp n}=0$)
\begin{equation}
\prod_i \frac{\bar{\alpha}}{\pi} \frac{d^2 q_{\perp i}}{q_{\perp i}^2}dy_i
\delta(\Sigma \mathbf{q}_{\perp i}),\,\,\,
\mathrm{with}\,\,\,\bar{\alpha} \equiv \frac{N_C\alpha_s}{\pi}.
\end{equation}
We note in particular that this expression is symmetric under exchange of
projectile and target. 
The result can also be expressed in terms of the 
virtual links $k_i$. We then have
\begin{equation}
\prod_{i=1}^n d^2 q_{\perp i}\delta(\Sigma \mathbf{q}_{\perp i}) =\prod_{i =1}^{n-1} d^2 k_{\perp i}, 
\end{equation}
and transverse momentum conservation implies that 
we have two different cases:

\begin{enumerate}\itemsep 0mm
\item Step up: $k_{\perp i}\approx q_{\perp i} \gg k_{\perp i-1}$,
  with weight $d^2 q_{\perp i}/q_{\perp i}^2 \approx d^2 k_{\perp
    i}/k_{\perp i}^2$.
\item Step down: $k_{\perp i} \ll q_{\perp i} \approx k_{\perp i-1}$,
  with weight $d^2 q_{\perp i}/q_{\perp i}^2 \approx d^2 k_{\perp
    i}/k_{\perp i-1}^2= d^2 k_{\perp i}/k_{\perp i}^2\times k_{\perp
    i}^2/k_{\perp i-1}^2$.
\end{enumerate}
Thus steps down in $k_\perp$ are suppressed by a factor $k_{\perp
  i}^2/k_{\perp i-1}^2$. This implies that links with a local maximum
$k_\perp$, as $k_3$ in fig.~\ref{fig:backbone}, is given the weight $
d^2 k_{\perp i}/k_{\perp i}^4$, and can be interpreted as a hard
scattering between $k_2$ and $k_4$, producing the two high-$p_\perp$
gluons $q_3$ and $q_4$. For a link with $k_\perp$ lower than the adjacent
real gluons, like $k_4$ in fig.~\ref{fig:backbone}, the transverse momentum of
the real gluons will be determined by the neighbouring links $k_3$
and $k_5$. Thus in such cases the associated weight will be
 $ d^2 k_{\perp i}$, which is
non-singular for small $k_\perp$. Note that no particular rest frame
is specified for the collision between the two cascades. The result is
the same if the ``hard subcollision'' is part of the projectile or
target evolution, and thus independent of the Lorentz frame used for
the analysis.

\FIGURE[t]{
\scalebox{1.0}{\mbox{
\begin{picture}(140,230)(-20,-50)
  \Text(-10,15)[]{\large $P_a$}
  \Line(0,15)(40,15)
  \Line(40,20)(70,20)
  \Line(40,15)(70,15)
  \Line(40,10)(70,10)
  \GOval(40,15)(10,7)(0){1}
  \Line(50,20)(60,40)\Text(47,34)[]{\large $k_{0}$}
    \Line(60,40)(90,40)\Text(100,40)[]{\large $q_{1}$}
  \Line(60,40)(70,70)\Text(55,55)[]{\large $k_{1}$}
    \Line(70,70)(105,70)\Text(115,70)[]{\large $q_{2}$}
  \Line(70,70)(75,100)\Text(65,85)[]{\large $k_{2}$}
    \Line(75,100)(110,100)\Text(120,100)[]{\large $q_{3}$}
  \DashLine(75,100)(75,130){2}
  \DashLine(75,130)(70,160){2}
    \Line(70,160)(105,160)\Text(115,160)[]{\large $q_{n-1}$}
  \Line(70,160)(60,190)\Text(55,173)[]{\large $k_{n-1}$}
    \Line(60,190)(90,190)\Text(100,190)[]{\large $q_{n}$}
  \Line(60,190)(50,210)\Text(47,196)[]{\large $k_{n}$}

  \Text(-10,215)[]{\large $P_b$}
  \Line(0,215)(40,215)
  \Line(40,220)(70,220)
  \Line(40,215)(70,215)
  \Line(40,210)(70,210)
  \GOval(40,215)(10,7)(0){1}
  \Text(-20,-30)[]{\large (a)}
  
\end{picture}}}
\scalebox{0.9}{\mbox{
\begin{picture}(300,280)(15,-20)
  \Line(40,20)(300,20)
  \Line(40,20)(170,280)
  \Line(170,280)(300,20)
  \Line(100,80)(70,20)
  \Text(100,90)[]{$q_{6}$}
  \Vertex(100,80){2}
  \Text(110,70)[]{$k_{5}$}
  \Vertex(125,80){2}
  \Text(125,90)[]{$q_{5}$}
  \Line(100,80)(125,80)
  \Line(125,80)(130,70)
  \Line(130,70)(145,70)
  \Text(140,60)[]{$k_{4}$}
  \Line(145,70)(160,100)
  \Line(160,100)(190,100)
  \Vertex(160,100){2}
  \Text(160,110)[]{$q_{4}$}
  \Text(175,90)[]{$k_{3}$}
  \Vertex(190,100){2}
  \Text(190,110)[]{$q_{3}$}
  \Line(190,100)(200,80)
  \Line(200,80)(230,80)
  \Text(215,70)[]{$k_{2}$}
  \Line(230,80)(245,50)
  \Vertex(230,80){2}
  \Text(230,90)[]{$q_{2}$}
  \Line(245,50)(257.5,50)
  \Text(250,43)[]{$k_{1}$}
  \Vertex(257.5,50){2}
  \Text(270,48)[]{$q_{1}$}
  \Line(257.5,50)(272.5,20)
  \LongArrow(250,180)(250,230)
  \LongArrow(250,180)(300,180)
  \Text(250,240)[]{$\ln q_\perp^2$}
  \Text(310,180)[]{$y$}
  \DashLine(257.5,50)(257.5,20){2}
  \Line(257.5,50)(263,15)
  \Line(257.5,20)(263,15)
  \DashLine(100,80)(100,20){2}
  \Line(100,80)(110,10)
  \Line(100,20)(110,10)
  \DashLine(125,80)(125,20){2}
  \Line(125,80)(135,10)
  \Line(125,20)(135,10)
  \DashLine(160,100)(160,20){2}
  \Line(160,100)(180,0)
  \Line(180,0)(160,20)
  \DashLine(190,100)(190,20){2}
  \Line(190,100)(210,0)
  \Line(210,0)(190,20)
  \DashLine(230,80)(230,20){2}
  \Line(230,80)(240,10)
  \Line(240,10)(230,20)
%
  \LongArrow(200,-15)(300,-15)
  \LongArrow(140,-15)(40,-15)
  \Text(170,-15)[]{$\ln s$}



%
  \LongArrow(220,130)(260,150)
  \Text(280,150)[]{$\ln q_+$}
  \LongArrow(120,130)(80,150)
  \Text(60,150)[]{$\ln q_-$}
  \Text(30,3)[]{\large (b)}
\end{picture}}}

\caption{\label{fig:backbone} (a) A parton chain stretched between projectile and
  target. (b) A backbone of $k_\perp$-changing gluons in a 
  $(y, \ln q_\perp^2)$ plane. The transverse momentum of the virtual links 
  $k_i$ are represented by horizontal lines.}
}

\subsection{Reabsorption of virtual emissions}

A dipole cascade generates a chain of dipoles linked together by gluons.
A dipole--dipole interaction via gluon exchange implies exchange of colour,
and the chains are recoupled as shown in figs.~\ref{fig:dipint} and 
\ref{fig:loops2}. Multiple interactions produce dipole loops, as seen in 
fig.~\ref{fig:loops2}. In fig.~\ref{fig:loops} we show a more complicated  
example, in which two cascades are evolved in
rapidity up to the dashed line in the center. Here three dipole pairs interact,
forming two dipole loops. In a projectile cascade with a large lightcone 
momentum 
$p_+$, and initially small $p_-$, the partons need a contribution of $p_-$
momentum from the target, in order to come on shell. Therefore the branches
denoted $B$ and $C$ in fig.~\ref{fig:loops} have to be treated as virtual, and
to be reabsorbed. The remaining gluons are coming on shell as real gluons, and
are colour-connected along the solid lines in fig.~\ref{fig:loops}. A colour
loop formed by a swing, as the one denoted $A$, can come on shell also if it
does not contain any interacting dipole. Such a process produces a closed
colour loop not traversing the collision line at $y=0$. A more detailed example of this effect can be found in appendix~\ref{sec:satswing}.
The gluons in the virtual branches reabsorbed, the momenta for the real 
backbone gluons must be recalculated. This process is
discussed in more detail in sec.~\ref{sec:generation}.

\FIGURE[t]{
  \includegraphics[width=.8\linewidth]{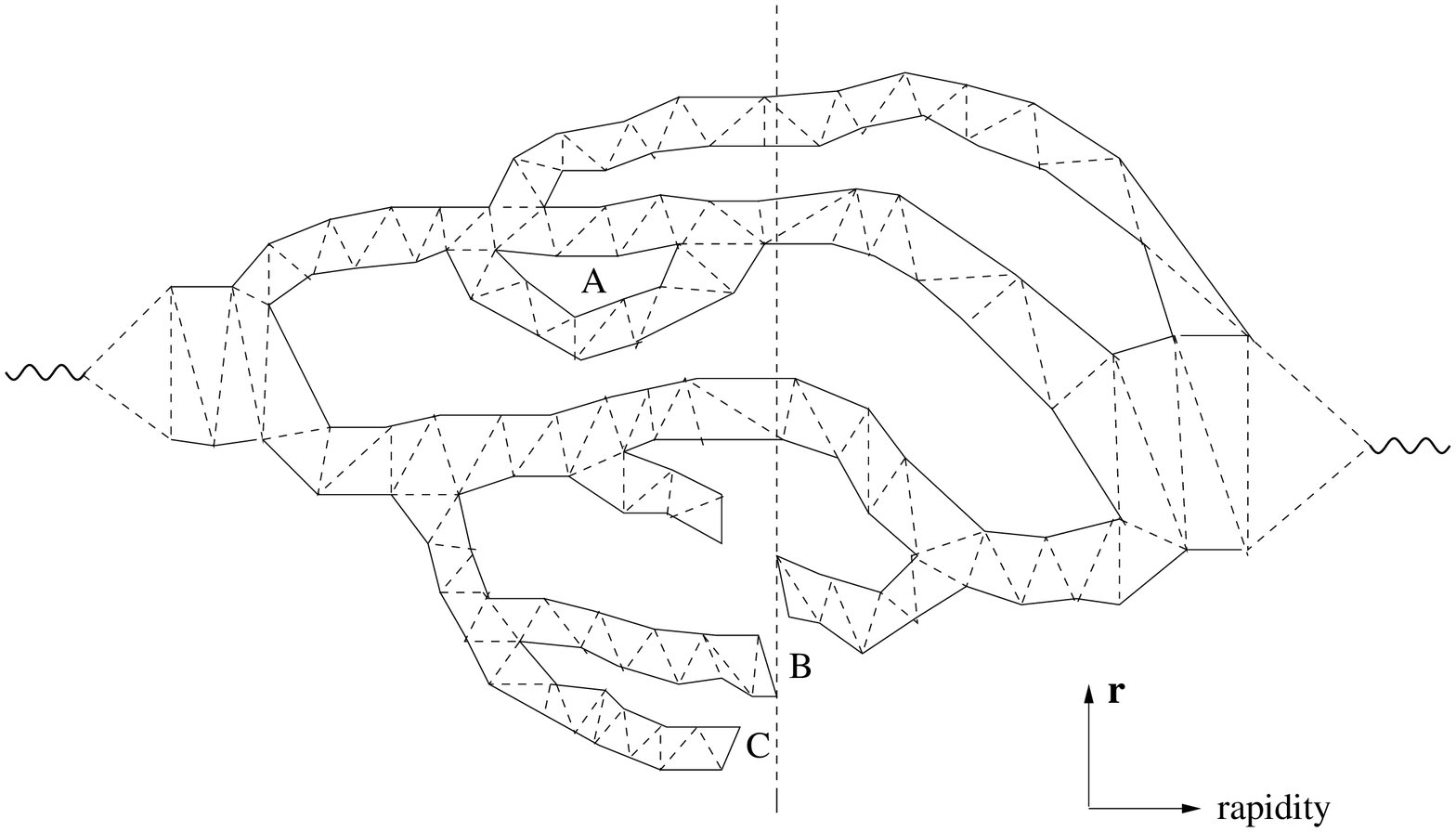}
  \caption{ \label{fig:loops} Collision of two dipole cascades
    in $\pmb{r}$-rapidity space. The dashed vertical line symbolizes
    the Lorentz frame in which the collision is viewed. The
    dipole splitting vertex can result in the formation of different dipole
    branches, and loops are formed
    due to multiple sub-collisions. The loop denoted by $A$ is an effect of
    saturation within the cascade evolution, which can be
    formed via a dipole swing. Branches which do not interact, like those
    denoted $B$ and $C$ are to be treated as virtual, and to be absorbed.}
}

\subsection{Giving proper weights to the emissions}
\label{sec:Rutherford}

\FIGURE[t] {
\scalebox{1.0}{\mbox{
\begin{picture}(400,83)(0,13)
\Vertex(20,50){2.5}
\Vertex(250,70){2.5}
\Vertex(240,30){1}
\Vertex(260,35){1}
\Vertex(260,25){1}
\Vertex(290,25){1}
\Vertex(335,70){1}
\DashLine(20,50)(250,70){6}
\Line(20,50)(240,30)
\DashLine(250,70)(240,30){2}
\Line(250,70)(260,35)
\DashLine(240,30)(260,35){2}
\DashLine(260,35)(260,25){2}
\Line(240,30)(260,25)
\Line(260,25)(290,25)
\DashLine(260,35)(290,25){2}
\Line(290,25)(335,70)
\Line(260,35)(335,70)
\Text(135,70)[]{$a$}
\Text(135,30)[]{$\approx a$}
\Text(10,50)[]{$0$}
\Text(235,50)[]{$\approx b$}
\Text(263,56)[]{$b$}
\Text(250,78)[]{$1$}
\Text(236,24)[]{$2$}
\Text(264,41)[]{$3$}
\Text(260,18)[]{$4$}
\Text(293,19)[]{$5$}
\Text(335,78)[]{$6$}
\Text(250,36)[]{$c$}
\Text(250,22)[]{$c$}
\Text(265,30)[]{$d$}
\Text(279,35)[]{$e$}
\Text(276,20)[]{$e$}
\Text(316,44)[]{$f$}
\end{picture}
}}
\caption{\label{fig:dglap}A dipole cascade in
  $\mathbf{r}_\perp$-space, in which a chain of smaller and
  smaller dipoles is followed by a set of dipoles with increasing
  sizes. The initial dipole between points 0 and 1 is marked by long
  dashes. Those dipoles which have split into two new dipoles, and thus
  disappeared from the chain, are marked by short dashes. The shortest dipole
  (34) corresponds to the  maximum $k_\perp$, and represents a hard
  sub-collision.}}

In the cascade the gluons at the ends of a dipole are given opposite
transverse momenta $k_\perp = 1/r$ (see further sec.~\ref{sec:correctpt}).
In \eqRef{eq:dipkernel1} an emitted dipole of length $r$ is given a weight 
containing the factor $d^2r/r^2$. However, if this dipole emits further 
dipoles, the new
weight  is proportional to $r^2$. Thus the associated weight is just $d^2r$ 
for dipoles which have split and been replaced by new dipoles. In the cascades
shown in figs.~\ref{fig:dglap} and \ref{fig:minkt} these dipoles are marked by
dashed lines. In the following they will be referred to as ``inner
dipoles''. The remaining dipoles, formed by colour-connected gluons, 
are marked by heavy lines, and they all get a weight proportional to $1/r^2$.
They will be referred to as ``outer dipoles''.

\FIGURE{
  \includegraphics[scale=1]{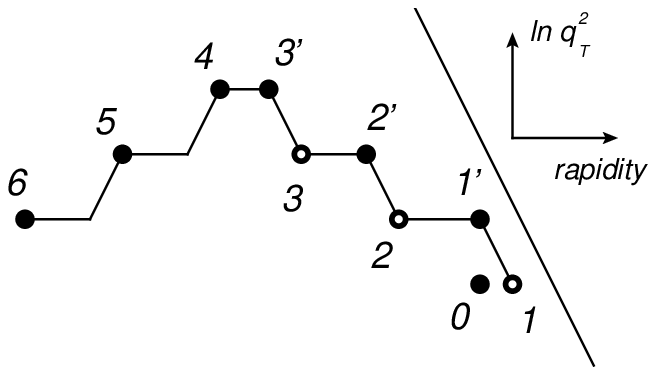}
  \caption{\label{fig:k-dglap} The same cascade as in fig.~\ref{fig:k-dglap}. The $x$ direction shows rapidity, and $y$ direction shows $\ln(k_\perp^2)$. The diagonal line is the constant $p_+$ of the incoming particle.}
}

Let us study the cascade shown in fig.~\ref{fig:dglap}, obtained after the
absorption of the virtual gluons as described in the previous subsection. 
The cascade starts from the dipole (01),
followed by emission of gluons 2, 3, etc. Here
the dipoles are first smaller and smaller, $a \gg b \gg c \gg d$.  The
corresponding $k_\perp$-values therefore become larger and larger in
each step. After the minimum dipole, with size $d$, the subsequent
emissions, 5, and 6, give again larger dipoles with correspondingly
lower $k_\perp$ values. The weight containing factors $1/r^2$ for all
``remaining dipoles'' is proportional to
\begin{equation}
\frac{d^2 \textbf{r}_2}{b^2} \cdot 
\frac{d^2 \textbf{r}_3}{c^2} 
\cdot \frac{d^2 \textbf{r}_4}{d^0} \cdot 
\frac{d^2 \textbf{r}_5}{e^2} \cdot
\frac{d^2 \textbf{r}_6}{f^2}\cdot\frac{1}{f^2}.
\label{eq:dglap}
\end{equation}
In this event gluons 3 and 4 recoil 
against each other with large transverse momenta $k_{\mathrm{max}}=1/d$. 
As all factors of $d$ have canceled in \eqRef{eq:dglap}, this gives
the proper weight $d^2 \textbf{r}_4 \propto
d^2\textbf{k}_{\mathrm{max}}/k_{\mathrm{max}}^4$. This reproduces exactly the 
proper weight for a hard gluon--gluon scattering.

The same cascade is shown in fig.~\ref{fig:k-dglap} in the ($y,\ln(k_\perp^2)$)
diagram of fig.~\ref{fig:backbone}b. In this figure new emissions are
first denoted by open circles, while filled circles mark their position after 
recoil from a later emission. Thus gluon 1 starts at the lower right
corner. After the emission of gluon 2, it is recoiling to the position 
marked $1'$. Similarly gluon 2 is shifted to position $2'$, when it emits gluon
3. When gluon 4 emits the softer gluon 5, its recoil is small and it keeps its
position in the diagram.

Figure \ref{fig:minkt} shows instead a chain with increasing dipole sizes
up to a maximum value, $r_{\mathrm{max}}$, which thus corresponds to a
minimum transverse momentum, $k_{\mathrm{min}}$. Here we also get the
correct weight $d^2 \textbf{r}_{\mathrm{max}}/r_{\mathrm{max}}^4 \propto d^2
\textbf{k}_{\mathrm{min}}$, which implies that there is no singularity for small
$k_\perp$-values. Note that all gluons connected to the long dipoles are also
connected to shorter dipoles, which determine most of their transverse
momentum. Thus no gluon has $q_\perp$ represented by $k_{\mathrm{min}}$.

\FIGURE{
\begin{picture}(300,80)(0,0)
\DashLine(20,30)(20,50){4}
\Line(20,50)(80,60)
\DashLine(20,30)(80,60){2}
\Line(80,60)(220,70)
\DashLine(20,30)(220,70){2}
\DashLine(220,70)(260,52){2}
\Line(220,70)(268,60)
\Line(20,30)(260,52)
\Line(260,52)(268,60)
\Vertex(20,30){2}
\Vertex(20,50){2}
\Vertex(80,60){1}
\Vertex(260,52){1}
\Vertex(220,70){1}
\Vertex(268,60){1}
\Text(10,30)[]{$0$}
\Text(10,50)[]{$1$}
\Text(80,67)[]{$2$}
\Text(262,45)[]{$4$}
\Text(273,62)[]{$5$}
\Text(220,77)[]{$3$}
\Text(50,60)[]{$a$}
\Text(60,45)[]{$a$}
\Text(150,70)[]{$b$}
\Text(130,57)[]{$b$}
\Text(150,35)[]{$b$}
\Text(250,68)[]{$c$}
\Text(240,56)[]{$c$}
\Text(268,55)[]{$d$}
\end{picture}
\caption{\label{fig:minkt}A cascade where the dipole sizes increase to a
  maximum, and then decrease.  Here the size of the largest dipole,
  denoted $b$, corresponds to the minimum $k_\perp$ in the chain.}}

\subsection{Going from transverse coordinate space to momentum space}
\label{sec:correctpt}

\emph{Cascade evolution}

The dipole picture is formulated in transverse coordinate space, but
experimental data are given in momentum space. We must therefore make
a translation of our results from transverse coordinate to transverse
momentum space. Note that most of the inclusive observables studied
previously have been dominated by large dipoles, while the final-state
observables are typically dominated by the large $q_\perp$,
originating from the small dipoles. Thus effects such as proton size
and confinement will be seen to affect observables very little, while
new phenomena important for small dipoles will come into play. The one
important exception to this is the deep inelastic cross section for
large $Q^2$, which gauges small dipoles with an inclusive observable.

The weight $d^2r/r^2$ for gluon emission in the dipole picture
corresponds to $d^2k_\perp/k_\perp^2$ in BFKL
\cite{Kuraev:1977fs,Balitsky:1978ic} or DGLAP
\cite{Gribov:1972ri,Dokshitzer:1977sg,Altarelli:1977zs} evolution. We
also found that a hard scattering, which in momentum space is
proportional to $d^2k_\perp/k_\perp^4$, in the dipole picture
corresponds to $d^2r$. These relations are consistent with the way we
associate a dipole $\mathbf{r}$ with a transverse momentum
$\mathbf{k}=\mathbf{r}/r^2$ (which also implies that
$\mathbf{r}=\mathbf{k}/k^2$). Thus, although the fixed relation
between $r$ and $k_\perp$ is inconsistent with Heisenberg's
uncertainty relation, it still reproduces the correct evolution also
in momentum space. This result would also be obtained by the relation
$k_\perp=c/r$, for any constant $c$. (Any result for a cross section
would get the same number of extra factors of $c$ in the numerator as
in the denominator.) We make the choice of $c=1$ because it gives the
correct $\langle k_\perp^2\rangle$ for a Gaussian distribution in $r$:
A two-dimensional distribution $\propto \exp(-\mathbf{r}^2/R^2)$ has
the average $\langle \mathbf{r}^2\rangle=R^2$. The Fourier transform
of the amplitude $\propto \exp(-\mathbf{r}^2/2R^2)$ is $\propto
\exp(-\mathbf{k}^2R^2/2)$, and thus the density is $\propto
\exp(-\mathbf{k}^2R^2)$ and $\langle \mathbf{k}^2\rangle=1/R^2$.

\emph{Dipole-dipole interaction}

It is, however, not possible to make the same identification in the 
dipole--dipole interaction cross section from gluon exchange. The logarithmic
expression in \eqRef{eq:dipamp} originates from a two-dimensional Fourier
transform of the propagator $1/k_\perp^2$, including interference from
interactions between the equal and unequal charges in the two dipoles. If we
replace $r$ by $1/k_\perp$ in \eqRef{eq:dipamp}, we would obtain a cross 
section which grows like $(d k_\perp^2/k_\perp^4)\cdot \ln^2k_\perp$ 
for large $k_\perp$. Thus, taking the Fourier transform of the scattering
amplitude, and then go back to momentum space via the relation $k_\perp=1/r$
does not give back the original cross section in momentum space.
It is then more correct to use the original
scattering cross section directly in momentum space.

The amplitude for scattering of an elementary charge against a target charge
at position $\mathbf{r}_{\mathrm{target}}$ corresponds to the (two-dimensional)
Fourier transform of the logarithmic Coulomb potential in two dimensions,
denoted $V(\mathbf{r}-\mathbf{r}_{\mathrm{target}})$:
\begin{equation}
A(\mathbf{k})=\frac{1}{2\pi}\int d^2r \,e^{i\mathbf{k}\mathbf{r}} 
V(\mathbf{r}-\mathbf{r}_{\mathrm{target}})\propto 
\frac{\alpha_s\,e^{i\mathbf{k}\mathbf{r}_{\mathrm{target}}}}{k^2}.
\end{equation}
The phase is here inessential for the cross section, which is determined by
$|A|^2$. The phase is, however, important when the target is a dipole. For
scattering against a colour charge at $\mathbf{r}_3$ and an anticharge at 
$\mathbf{r}_4$, we get 
\begin{equation}
A(\mathbf{k})\propto \frac{\alpha_s}{k^2}(e^{i\mathbf{k}\mathbf{r}_3}-
e^{i\mathbf{k}\mathbf{r}_4})=\frac{\alpha_s}{k^2}
2\sin(\mathbf{k}\frac{\mathbf{r}_3-\mathbf{r}_4}{2})e^{i\mathbf{k}(\mathbf{r}_3 + \mathbf{r}_4)/2}.
\end{equation}
For a dipole with charge and anticharge at respectively $\mathbf{r}_1$ and
$\mathbf{r}_2$, scattering against a dipole with coordinates $\mathbf{r}_3$ and
$\mathbf{r}_4$, we get
\begin{equation}
A\propto \frac{\alpha_s}{k^2} 4\sin(\pmb{\rh}_1 \cdot \pmb{k})
\sin(\pmb{\rh}_2\cdot \pmb{k}), 
\end{equation}
with $\pmb{\rh}_1 =(\mathbf{r}_1-\mathbf{r}_2)/2$ and $\pmb{\rh}_2
=(\mathbf{r}_3-\mathbf{r}_4)/2$. Averaging and summing over colours this gives 
the scattering cross section
\begin{equation}
\frac{d\sigma}{d^2k} = 8 \frac{\al_s^2}{k^4} \sin^2(\pmb{\rh}_1 \pmb{k})\sin^2(\pmb{\rh}_2 \pmb{k}).
\end{equation}
To generate $\mathbf{k}$-values according to this distribution we simply
select random uniformly distributed impact-parameters $\pmb{r}_\text{int}$ and make the
identification $\pmb{k}=\pmb{r}_\text{int}/r_\text{int}^2$. With 
$d^2r_\text{int}=d^2k/k^4$ this corresponds to the interaction probability
\begin{equation}
f_{ij} = 8 \al_s^2 \sin^2(\frac{\pmb{\rh}_i \cdot \pmb{r}_\text{int}}
{r^2_\text{int}})\sin^2(\frac{\pmb{\rh}_j \cdot \pmb{r}_\text{int}}
{r^2_\text{int}}). 
\label{eq:sinint}
\end{equation}

 Note that for small $r_\text{int}$, $f_{ij}$ goes as a constant times the 
rapidly oscillating sine functions, with average $(1/2)^2$. This corresponds to 
the expected $d^2k_\perp/k^4_\perp$ for large $k_\perp$. In the limit where the 
interaction distance is large compared to the dipole sizes, $f_{ij}$ falls off 
as $r_\text{int}^4$, which gives the infrared stable result $\propto d^2 k_\perp$
for small $k_\perp$.

\subsection{Final state radiation and hadronization}

As discussed above, the backbone chain of $k_\perp$-changing gluons determines 
the inclusive inelastic cross section. To get the exclusive final states the
soft $k_\perp$-conserving emissions have to be added as final state radiation
\cite{Catani:1989sg, Andersson:1995ju, Salam:1999ft}. As the 
$k_\perp$-conserving emissions have $q_{\perp i}<k_{\perp i} \approx 
k_{\perp i-1}$, they fill the area below the horizontal lines in 
fig.~\ref{fig:backbone}.
Final state radiation is also emitted in the jets produced by the backbone
gluons, here represented by the folds sticking out of the $(y,\ln q_\perp^2)$
plane. The separation of e.g. the colour charge in gluon  $q_3$ and the
corresponding anticharge in gluon $q_2$ forms a colour dipole, and gives 
a gluon cascade similar to the
cascade in an $e^+e^-$-annihilation event. The only difference is that
emissions with $q_\perp > k_{\perp 2}$ are not allowed. 

The final-state radiation is thus added in much the same way as in the
so-called LDC Monte Carlo\cite{Kharraziha:1998dn,Kharraziha:ldcmc}.
In the simulations we use the $p_\perp$ ordered dipole cascade model 
\cite{Gustafson:1986db,Gustafson:1987rq} as implemented in the \ariadne program 
\cite{Lonnblad:1992tz}\footnote{We will here use a
  preliminary reimplementation in \thepeg of the old \ariadne Fortran
  code.}. In this formalism a gluon emission
splits a dipole into two, both of which may continue radiating
independently. Thus, also after final state radiation, the parton state
consists of dipole chains formed by colour connected gluons.

In the final step the partons hadronize into final state particles.
This is treated by the Lund string fragmentation model
\cite{Andersson:1983jt,Andersson:1983ia} as implemented in the \pythia
program \cite{Sjostrand:2007gs,Sjostrand:2006za}. Note that the parameters 
in \ariadne have been
thoroughly tuned together with the string fragmentation parameters to
agree with LEP data, and we do not change any of these parameters
here.


\section{Generating the exclusive final states}
\label{sec:generation}

The previous section outlined some of the conceptual problems encountered when describing exclusive final states in the dipole model, and how they can be solved. This chapter will go more into detail on how the Monte Carlo handles different situations, and how the pieces of the model are fitted together.

The first step is to select the interacting dipoles from the virtual cascade, which makes it possible to identify the on-shell and virtual gluons. The on-shell gluons will form the backbone gluons in the previous section, while the virtual gluons are removed. It will turn out that some of the backbone gluons have to be reweighted and some of the $q_\perp$ maxima will be removed, as will be described in sec.~\ref{sec:RutherfordAlgorithm}. Then a further check of ordering will be made to match phase space with final state radiation, and the colour flow between the remaining backbone gluons will be set based on the virtual cascade. These gluons will then undergo final state radiation, and hadronize to produce the final state.

Some of the more involved details, such as how the allowed phase space
is chosen for emissions and interactions, and how saturation effects
in the cascade and interaction complicate the procedure, will be
discussed further in the appendices.

\subsection{Selecting the interactions}

As in our original Monte Carlo \cite{Avsar:2005iz}, the two incoming
particles are first identified with a dipole state at their initial
rapidity. These starting partons are called the valence partons, and
are then evolved through rapidity to the interaction frame. Here, a
relative transverse distance between the two incoming particles, the
impact-parameter, is selected, and all the dipole--dipole scattering
probabilities $f_{ij}$ can be calculated. Since the individual dipole
interactions are assumed to be independent, \eqRef{eq:sinint} can be
used to find the probability for non-diffractive interaction between
dipole $i$ and $j$:
\begin{equation}
1 - e^{-2f_{ij}} = 1 - \exp \left(-16 \al_s^2 \sin^2 \left( \frac{\pmb{\rh}_i \cdot \pmb{r}_\text{int}}{r^2_\text{int}} \right) \sin^2 \left( \frac{\pmb{\rh}_j \cdot \pmb{r}_\text{int}}{r^2_\text{int}} \right) \right).
\end{equation}
The interaction distance $r_\text{int}$ is set to the distance between two randomly selected partons, one from each dipole.

Just like in the cascade, the kinematics of the interaction is
calculated. In this case, there will be two partons coming in from the
left bringing with them $p_+$, but with a deficit in $p_-$, and
correspondingly two partons from the right bringing $p_-$, but with a
deficit in $p_+$. An interaction where the partons do not bring enough
momentum to set all partons on shell is vetoed, that is $f_{ij}$ is
set to 0. The allowed phase space will be described more in detail in
appendix~\ref{sec:phasespace}.

It should be noted here that although the collision of two dipoles formally is a 4-to-4 reaction, with two incoming partons from each side, one of the partons in each dipole will have emitted the other parton. Thus a dipole interaction only connects the end of one backbone gluon chain, but the amplitude depends on the last two partons in the dipole picture.

\subsection{Identifying the backbone gluons}
\label{sec:realchain}
Once the interaction dipoles are selected, one can trace the parents of the
interacting partons back, and identify the backbone gluons, as illustrated
in figure \ref{fig:loops}. The rest of the partons will not get the necessary $p_-$ from the colliding particle to come on shell, and will be reabsorbed as virtual emissions.

In a dipole picture, each emission is a coherent sum of emissions from the two
partons at the end of the emitting dipole, and a
unique parent cannot be determined. In the conventional parton cascade the
radiation pattern is separated in two components representing independent
emissions from the two different charges. The coherence effect is then
approximated by angular ordering, and the recoil is taken by a single
parent. For time-like cascades as in $e^+e^-$-annihilation or final state
radiation this is not a dramatic effect. For the space-like cascades discussed
here it is, however more important if both parents can get a recoil and be
put on shell, or if the recoil effect for one of them may be so weak that this
gluon has to be regarded as virtual and be reabsorbed. Naturally either
alternative can only be approximate, and the question which of them gives the
best description cannot be answered by perturbative QCD. We have therefore
implemented both schemes in the event generator, and when comparing the
results with experimental data we find best agreement when the parents share
the recoil. For the results presented below, we have assumed that the relative
shares are proportional to $1/r_i^2$, where $r_1$ and $r_2$ are the distances
to the two parents.

\subsection{Reweighting outer $q_\perp$ maxima}
\label{sec:RutherfordAlgorithm}
\FIGURE{
\includegraphics[angle=0, scale=1]{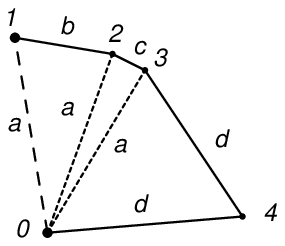}
\hspace{3cm}
\includegraphics[angle=0, scale=1]{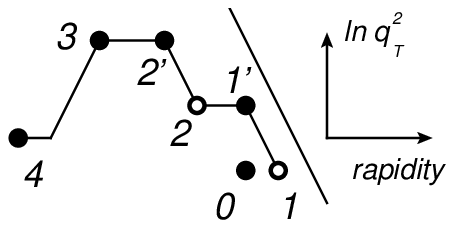}
\caption{\label{fig:outsidemaxkt}A cascade with a maximum in $q_\perp$ where the small dipole corresponding to the large $q_\perp$, does not split. The outer $q_\perp$ maximum is shown in impact-parameter space (left) and $y$-$q_\perp$ space (right). }
}

Recalling sec.~\ref{sec:Rutherford}, two examples showed that giving the
outer dipoles a weight of $d^2r/r^2$, and the inner dipoles a flat weight
$d^2r$ reproduced the correct weights for maxima and minima in $q_\perp$. In
these examples the minimum or maximum dipole were both inner dipoles, that is
dipoles that emitted further dipoles. If an outer dipole corresponds to a
local minimum as in fig.~\ref{fig:outsidemaxkt}, it is given a weight
proportional to $d^2r/r^2$. As the two connected gluons both are given large
transverse momenta $q_\perp=1/r$, we get a distribution proportional to $d^2
q_\perp/q_\perp^2$, giving a too strong tail out to large $q_\perp$-values.

To get the correct weights, these small outer dipoles have to be suppressed
by an extra factor $r^2_</r_>^2$, where $r_<$ is the size of the small dipole
giving the $q_\perp$ maximum, and $r_>$ is the corresponding maximum in dipole
size later in rapidity.  We note here in particular, that this extra 
reweighting was not needed in the calculation of inclusive observables 
presented in earlier articles. These observables depend only on interacting
dipoles, which are never reweighted or absorbed in this way.

In the  \dipsy event generator the reweighting is implemented by finding and 
reabsorbing some of the outer maxima, so that the correct weight is
restored. The inner and outer dipoles are only known after the backbone gluons 
are identified, and therefore
the reweighting can be performed only after selecting the interactions and the
identification of the backbone gluons. This procedure is described in more detail in appendix~\ref{sec:reweighting}.

\subsection{FSR matching and ordering}
\label{sec:PSmatching}
The kinematics for the reweighted backbone gluons can be significantly
different from the kinematics in the virtual cascade. To exactly match
the phase space covered by \ariadne, the $q_+$ and $q_-$ ordering is
checked again for the backbone gluons before being passed on from
\dipsy.

In \ariadne it is then checked that each gluon in the final state
cascade is un-ordered w.r.t.\ the backbone gluons, \ie, that the
positive (and negative) light-cone momentum of an emitted gluon is
less than that of (one of) the backbone gluons from which it is
emitted. As the backbone gluons are required to be ordered, all of phase space is covered exactly once by the backbone gluons with FSR.

\subsection{Colour flow}
\label{sec:colourflow}
The backbone gluons are chosen and corrected looking only at the parent
structure, that is, each gluon only remembers which two partons emitted it,
independently of any subsequent dipole swings, as these are not changing the
momenta of the backbone gluons. For final state radiation and hadronization it
is, however, important to keep track of the colour flow between the backbone
gluons, when the virtual emissions are reabsorbed.

This is done by going back to the colour flow of the virtual cascade, and
absorb one virtual gluon at a time, until only the on-shell gluons are left. When
a virtual gluon is removed, the two neighboring dipoles are combined to a
single dipole. For events with no swing, this will
always return the colour flow as if the virtual emissions never happened. 
For events with dipole swings the colour flow is more complicated,
and will be discussed in appendix~\ref{sec:finalsat}.

\paragraph{The proton remnant:}
The starting configuration of a proton is in our model represented by a
triangle of three dipoles, roughly representing the positions of three valence
quarks. However, three connected dipoles corresponds to the colour flow of
three gluons rather than three quarks, and the extra charge may be regarded as
representing the
contribution from the gluons present in the proton wavefunction already at 
low virtuality.

This extra charge was introduced for inclusive observables, but for exclusive
final states it has other effects that have to be handled. The three valence
partons will in general continue down the beam-pipe, and if all three dipoles
have interacted, then each of the valence partons will have two colour
connections to the colliding projectile. This is an overestimate of the colour
flow to the proton remnant, as the extra colour charges representing the
gluonic component in general do not carry a large fraction of the proton energy. This is corrected for in \dipsy by reconnecting on average half of the colour flow to the valence partons by a dipole swing, which will move the gluonic content of the proton one ``rung'' down the gluon ladder.

\subsection{Higher order corrections}
The above outlines the main points of how the backbone gluons are generated in \dipsy. There are, however, further details which, despite their non-leading nature, have to be accounted for. Below follows short summaries of these corrections, while the detailed description of the algorithms implemented in \dipsy are left for the appendices.

\subsubsection{Coherence as relaxed ordering}
Previously in this section it was described how the outer $q_\perp$ maxima
were reweighted, and absorbed with a certain probability. A $q_\perp$ maximum
will through $q_-$ ordering veto emissions at small $q_\perp$ close in
rapidity, but if the maximum is reabsorbed, the ordered phase space will be
significantly larger. This is solved by overestimating the ordered phase space
in the initial generation, making it possible to emit gluons with lower
$q_\perp$ in case the high-$q_\perp$ gluon is absorbed.
This extra phase space can be regarded as an effect of coherence, in the sense
that several close-by partons can coherently emit a gluon with a longer
wavelength, and the gluon can use the combined energy in all the close-by
partons. This process is described in detail in appendix~\ref{sec:phasespace},
where it is also shown how the ordering and energy conservation in the
interaction is implemented, based on how it is done in the cascade.

\subsubsection{Saturation effects}
Previous considerations were all based on a single chain of dipoles, but at a
7~TeV collision there are on average more than three subcollisions, making
saturation effects very important for the final state. In
appendix~\ref{sec:finalsat} it is shown how the procedures described here has been developed to work also in events with both
multiple interaction and swings. Multiple interactions give rise to splitting chains of backbone gluons while the swing can cause both splitting and merging chains, and
together they are able to build any diagram of gluon chains.

In this section it is demonstrated how the reweighting of outer $q_\perp$ maxima give the correct weight $d^2q_\perp \al_s^2(q_\perp)/q_\perp^4$ in every situation, with the only exception being a special configuration in two merging chains, where one of the running couplings can get an incorrect scale. This is estimated to be a negligible effect.

The colour flow in a saturated cascade is more complicated as the dipole swing will reconnect the colour flow, simulating soft gluon exchanges between different parts of the gluon chains.

\section{Self-consistency and tuning}
\label{sec:results}

There are several details in \dipsy that are not fixed by perturbative
calculations, but nonetheless have a significant effect on the results. These
details will have to be decided by other means, or tuned to experimental
data. Many of them cannot be directly related to a tunable parameter, but are
rather choices between different approximations or models. Other effects were
 tested, only to turn out to not give improved results. Thus, rather than
trying to list which parameters and choices were tuned, we will in this
section describe the constraints and data, which were used to fix the details
of the model. After this we will compare the model to the remaining observables.

One of the most important self-consistency constraint is the
frame-independence, that is the property that all observables should
be the same no matter what collision frame $y_0$ is used. This is a
symmetry that is necessarily present in an all-order calculation, but
can not be expected to be exactly manifested in a fix-order treatment
as the non-leading effects enter differently in the interaction and
the cascade. Thus, unknown non-leading effect will, if possible, be
tuned to fit a known all-order property.

This section will first describe how the frame independence of
different inclusive and exclusive observables fixes many of the
details mention in sec.~\ref{sec:generation}, and then how tuning to
experimental data fixes the last uncertainties. The section is ended
with a comparison of \dipsy with experimental data and other event
generators for a selection of observables.

\subsection{Achieving frame independence}
\label{sec:findep}
Here the frame dependence of some observables are studied, and many
non-leading subtleties can be fixed by requiring an
approximately flat frame dependence. Notice that at this stage no
comparison is made to experimental data.

\FIGURE{
  \begin{minipage}{0.45\linewidth}
    \begin{center}
      \includegraphics[angle=-90,scale=0.4]{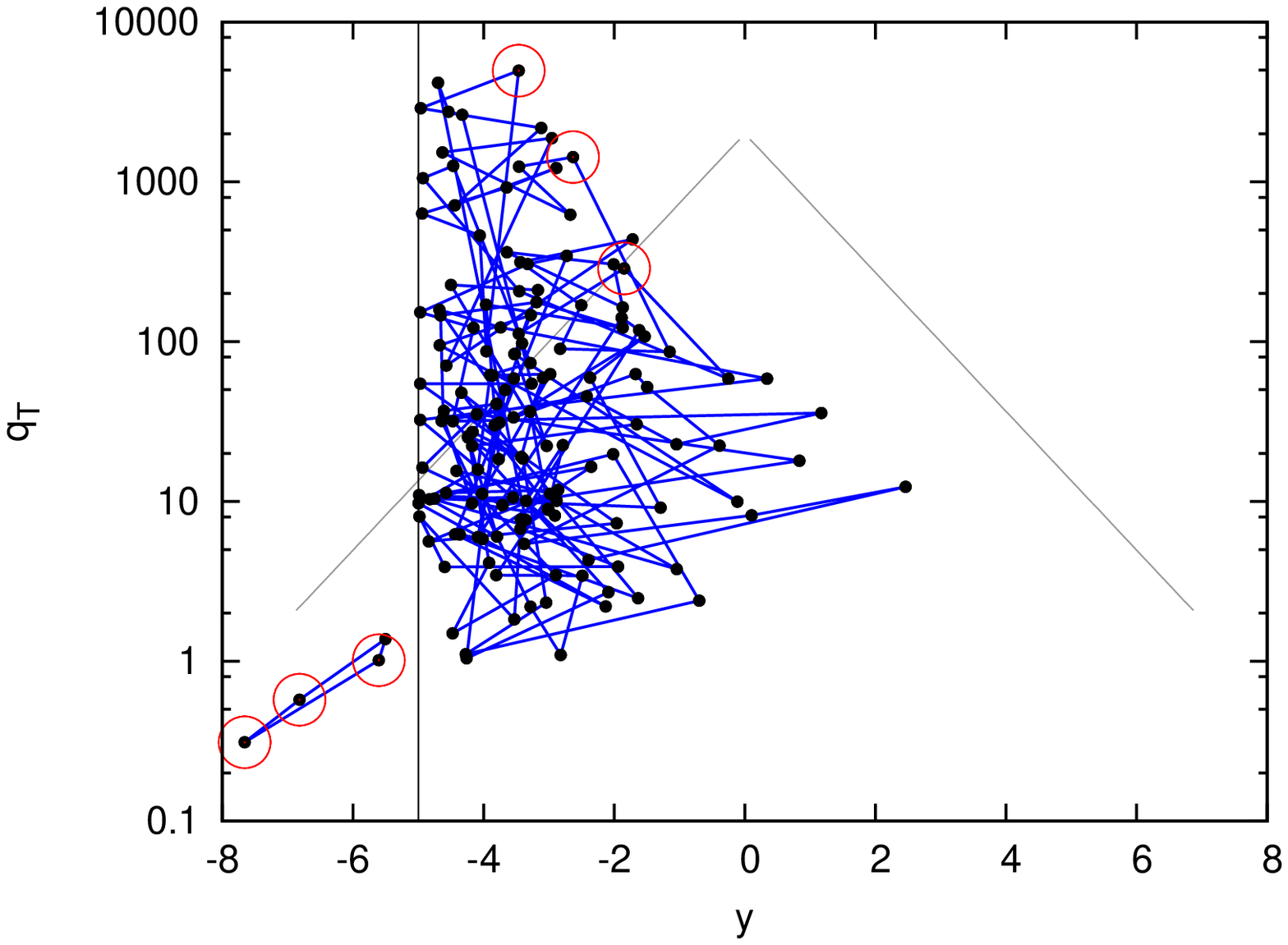}

      $(a)$
    \end{center}
  \end{minipage}
  \begin{minipage}{0.45\linewidth}
    \begin{center}
      \includegraphics[angle=-90,scale=0.4]{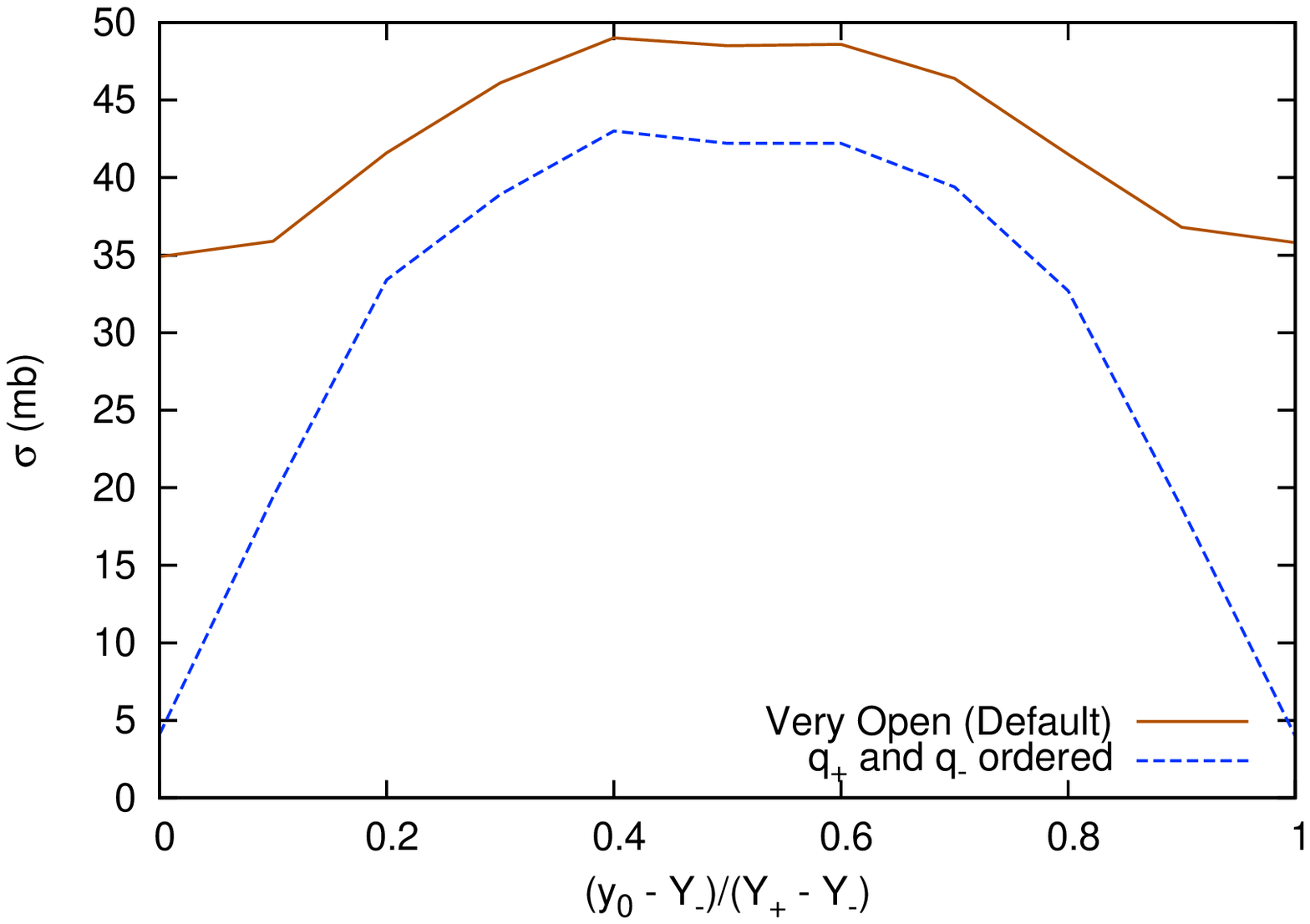}

      $(b)$
    \end{center}
  \end{minipage}
\caption{\label{fig:sampleppoff} (a) Two colliding virtual cascades in the $y$-$q_\perp$ representation, generated by the Monte Carlo in an off-central frame. The lines shows the colour flow between the partons, the dashed vertical line is the collision frame $y_0$ and the circled partons are the starting valence partons. The outside triangle show the incoming $p_+$ and $p_-$ of the protons. (b) The frame dependence of the total $pp$ cross section at $\sqrt{s} = 200$~GeV. The dashed curve has full $q_\pm$ ordering in the interaction, the full one only requires enough energy to set the interaction on shell. }
}
\paragraph{\boldmath$\si_{pp}$ frame independence:}
We find that the restrictions in the interaction have to be very
generous to maintain a constant total $pp$ cross section for $y_0$
close to the valence rapidities. A proton evolved over the full range
to the other protons rest frame, at the Tevatron this is 15 units of
rapidity, will have large transverse momenta, while the valence
partons of the unevolved proton will have very small transverse
momenta. Requiring full ordering in lightcone momentum in this case
will disallow almost all dipole pairs, and the interaction probability
will be strongly suppressed in off-central frames. This is illustrated
in fig.~\ref{fig:sampleppoff}a with a sample virtual cascade at the
Tevatron.  To maintain the cross section for all frames, a very open
ordering in $f_{ij}$ is needed, allowing interaction between partons
even when they are not ordered from the virtual cascades. The minimum
amount of ordering from appendix~\ref{sec:intorder} gives a maximum
deviation of about 20\% at the endpoints as can be seen in
fig.~\ref{fig:sampleppoff}b. Anything stricter results in stronger
frame dependence. The required ordering is thus set to this minimum
\begin{equation}
q_+q_- > 16 q_{\perp \text{int}} \label{eq:intorder}
\end{equation}
\FIGURE[b]{
  \includegraphics[angle=-90,scale=0.4]{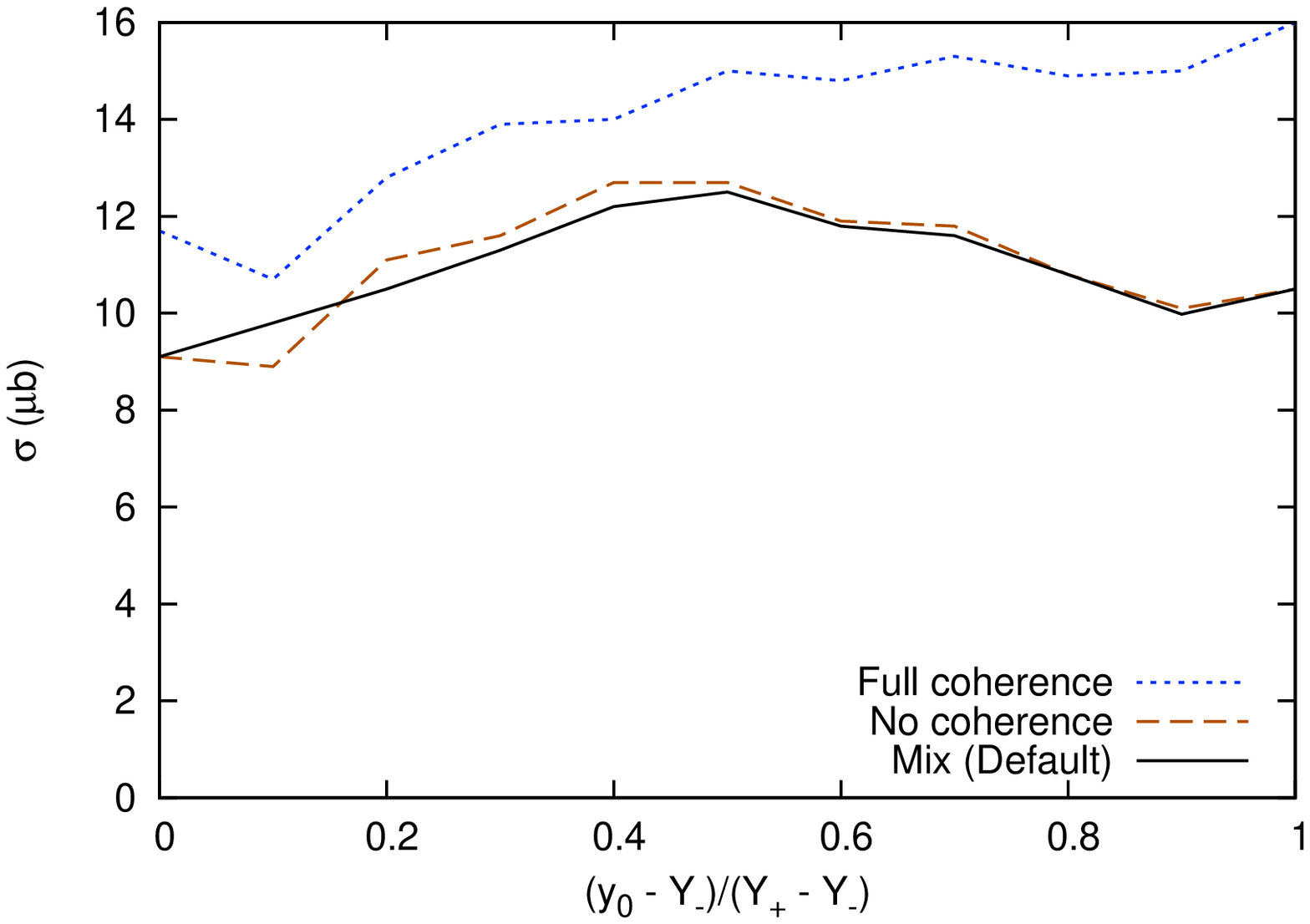}
  \caption{\label{fig:gpframe} (a) Frame dependence of $\ga^*p$ at $Q^2 = 14$ and $W = 220$,
    showing three different orderings in the virtual cascade. For low
    $y_0$ the proton is evolved a larger part of the rapidity
    interval, and vice versa.}
}
where $q_+$ and $q_-$ are the lightcone momenta the incoming partons
bring, and $q_{\perp \text{int}}$ is the interaction momentum used in
$f_{ij}$ in \eqRef{eq:sinint}. This requirement only enforces that the incoming particles have sufficient energy to set the interacting partons on shell, without any additional ordering. This is significantly worse than
previous versions of this model, where deviations between different
frames were within a few percent. This is is a sign that either the
interaction ordering is too strict, blocking the interaction towards
the end frames, or that the evolution does not grow fast enough, and a
cascade evolved over a long rapidity range is not as strong as two
cascades, each evolved only to the middle.

The first option is unlikely, as the ordering is already as generous
as reasonable can be done, and in fact tests with no interaction
ordering at all give no significant improvement. Likely, this problem
is a result on the ordering in the virtual cascade, and how the
coherence is used. During tuning it has shown that this frame
dependence is quite sensitive to these choices, and further tuning
could possibly solve this problem. For this tune however, the result
is the one shown in fig.~\ref{fig:sampleppoff}b.

\paragraph{\boldmath $\si_{\ga^*p}$ frame independences:}
The proton starts at a transverse momentum of about 0.3~GeV, while the two quarks from the photon start with a much higher transverse momentum. Here the key observation is that it is the $q_+$ ordering that limits emissions of gluons with larger $q_\perp$, while the $q_-$ ordering limits the emission of smaller $q_\perp$. Thus, with the interaction frame close to the photon, the cross section will be set by the protons ability to climb up to the higher $q_\perp$, dominated by the $q_+$ ordering, and conversely, with an interaction frame close to the proton, the cross section will be determined by the $q_-$ ordering in the photons cascade.

\FIGURE{
  \includegraphics[angle=-90,scale=0.4]{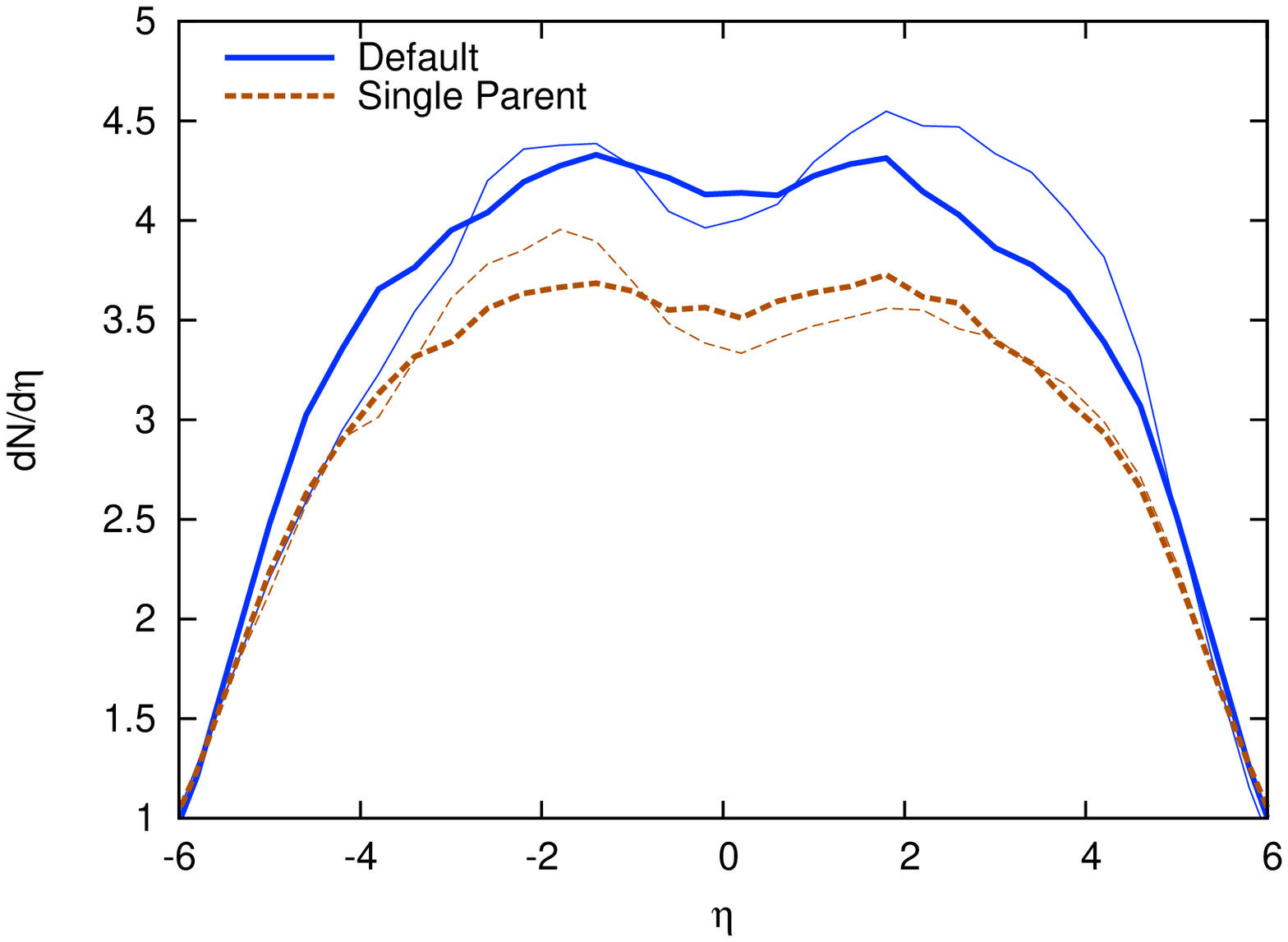}
\caption{\label{fig:dndetaframe} $dN/d\eta$ for $pp$ at 900~GeV with default \dipsy, and \dipsy with the ``single parent'' setting. The thick line is collided in midframe $y_0 = 0$, the thin line in $y_0 = -2.7$.}
}

This provides an opportunity to use self consistency to determine which the ordered phase space in the cascade should be. In agreement with the arguments in appendix~\ref{sec:coherence}, it turns out that a fully coherent ordering will allow all emissions that may be ordered in the backbone chain. This is however done by overestimating emissions of large dipoles in the virtual cascade. This is seen in fig.~\ref{fig:gpframe} where the coherent ordering is tilted with a larger cross section when the dipole is allowed to evolve. Here the $\ga^*$ is incoming from the negative $z$-axis with a large $p_+$, and the proton in incoming form the positive $z$-axis with a large $p_-$. Ignoring coherence provides a symmetric frame dependence (although with the same bump as in $pp$), but will cut away many emissions that may be ordered in the backbone chain.

A combination is used, where the $q_\perp$ of the effective parton is used, but the rapidity of the single parent. This maintains a symmetric frame dependence while still covering as much as possible of the ordered phase space for the backbone gluons. The $q_-$ ordering is also very important in deciding the energy dependence for inclusive observables, such as the $pp$ total cross section. As will be seen later, it turns out that this combined ordering provides an energy dependence that fits very well with experimental data.

\paragraph{Frame independence of \boldmath$dN/d\eta$:}
For this observable to be frame independent it is needed that any
given rapidity produces the same density of charged particles
independently of where the interaction frame is. This means that the
$q_\perp$ must behave the same coming from the cascade or from the
interaction, and it should not depend on whether the cascade goes on
for long, or if it interacts immediately. All this depends heavily on
the ordered phase space in the virtual cascade and interaction, and in
how the backbone gluons are handled. In practice, tuning for frame
independence of $dN/d\eta$ is equivalent to balancing the $q_\perp$ in
the interaction and in the cascade. If the interaction has too much
transverse activity, then the charge particles will tend to clump up
around the interaction frame when it is moved. Conversely, if the
cascade provides too much $q_\perp$, then the side with the longer
cascade will have more charged particles, and the interaction on the
other side in rapidity will not be enough to balance this.

With the virtual cascade and $f_{ij}$ fixed from the inclusive
observables according to the above arguments, it is only the handling
of the backbone gluons left to tune. The problems connected to the
reweighting of the outer $q_\perp$ maxima mainly entered through the
ordered phase space in the virtual cascade. Once the virtual cascade
and interactions are determined, the ordering of the backbone gluons is
fixed by matching to FSR, so few choices are left at this point.

One choice, though, is whether both or just one parent should come on
shell, as was described in sec.~\ref{sec:realchain}. Keeping both
parents will give more activity in the cascade as more particles will
be kept, and more recoils will be done. Thus, the single parent choice
will have the multiplicity more towards the interaction side, and the
double parent choice more towards the cascade side, as is seen in
fig.~\ref{fig:dndetaframe}.

Neither of them deviate more than 10\% at any pseudo-rapidity which is encouraging for the model, but neither of the options is much better than the other, so this observable cannot fix this choice. It will instead be tuned to the experimental value of this observable, which is close to line for double parent.

\subsection{Tuning to experimental data}
\label{sec:PStuning}

\paragraph{Inclusive \boldmath$pp$ cross sections:}
In the last published tune of the Monte Carlo \cite{Flensburg:2008ag}, there were 4 parameters, set to:
\begin{equation}
\La_\text{QCD} = 0.2\text{~GeV} \qquad \rmax = 2.9\text{~GeV}^{-1} \qquad
R_p = 3.0\text{~GeV}^{-1} \qquad W_p = 0\text{~GeV}^{-1},
\end{equation}
being the QCD scale, the confinement range, the proton size, and the fluctuation in the proton size. These parameters were tuned to the total and elastic $pp$ cross section as function of $\sqrt{s}$, and the elastic differential cross section as function of $t$.

In principle the same can be done now, but one of the four parameters
now hold a more important role than in previous papers. Recall from
the discussion in section \ref{sec:correctpt} that the large dipoles
will dominate the inclusive $pp$ cross section, while the small
dipoles will dominate the exclusive observables. One should note that $\rmax$, $R_p$ and
$W_p$ mainly affect large dipoles, while
$\La_\text{QCD}$ also affects small dipoles. Thus the inclusive $pp$ cross section can be tuned using $\rmax$, $R_p$ and
$W_p$ without affecting the exclusive observables much. A comparison of the
default tune with data can be found in fig.~\ref{fig:totxsec}.

\FIGURE{
  \includegraphics[angle=-90,width = 0.45\linewidth]{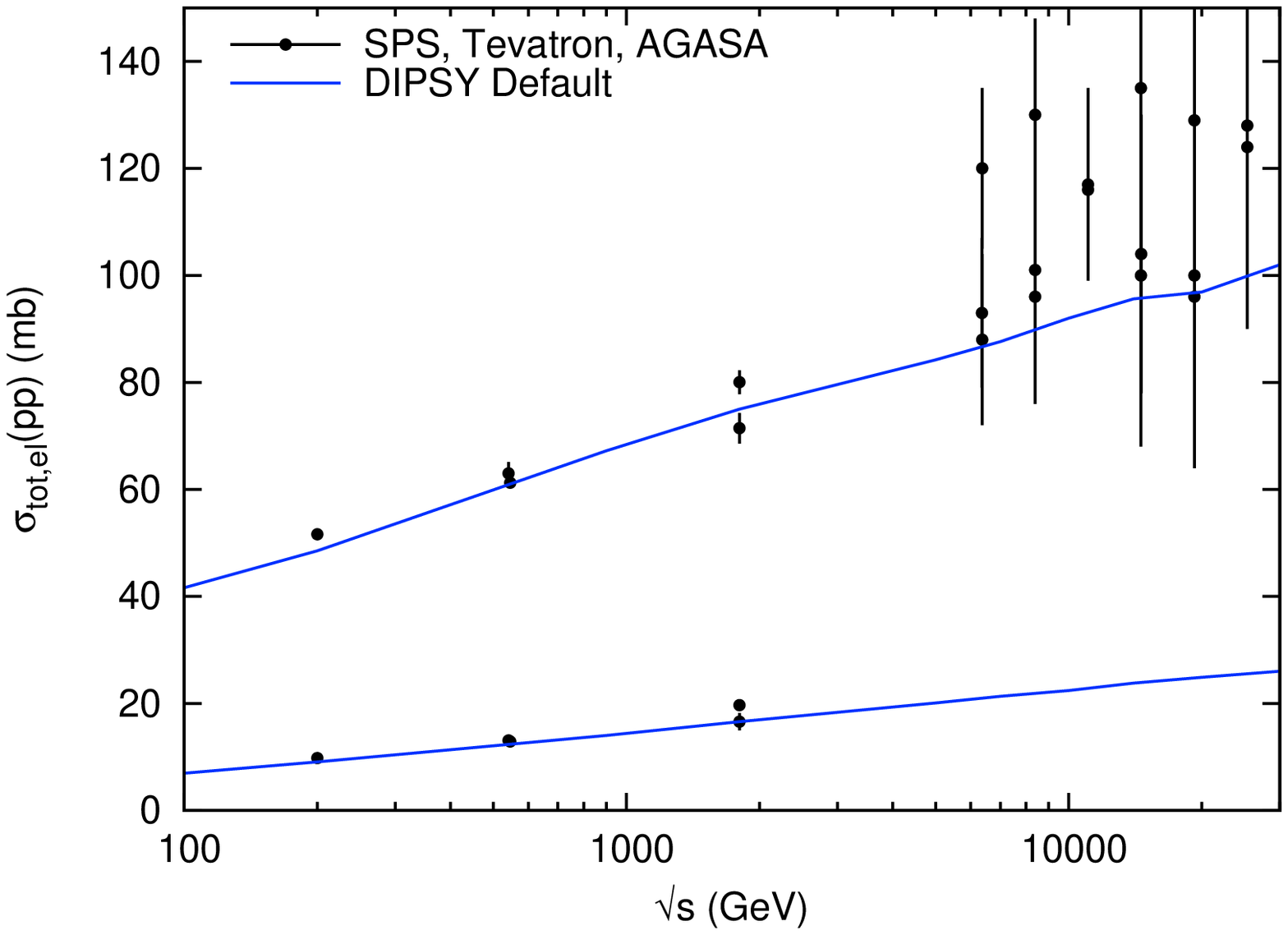}
  \includegraphics[angle=-90,width = 0.45\linewidth]{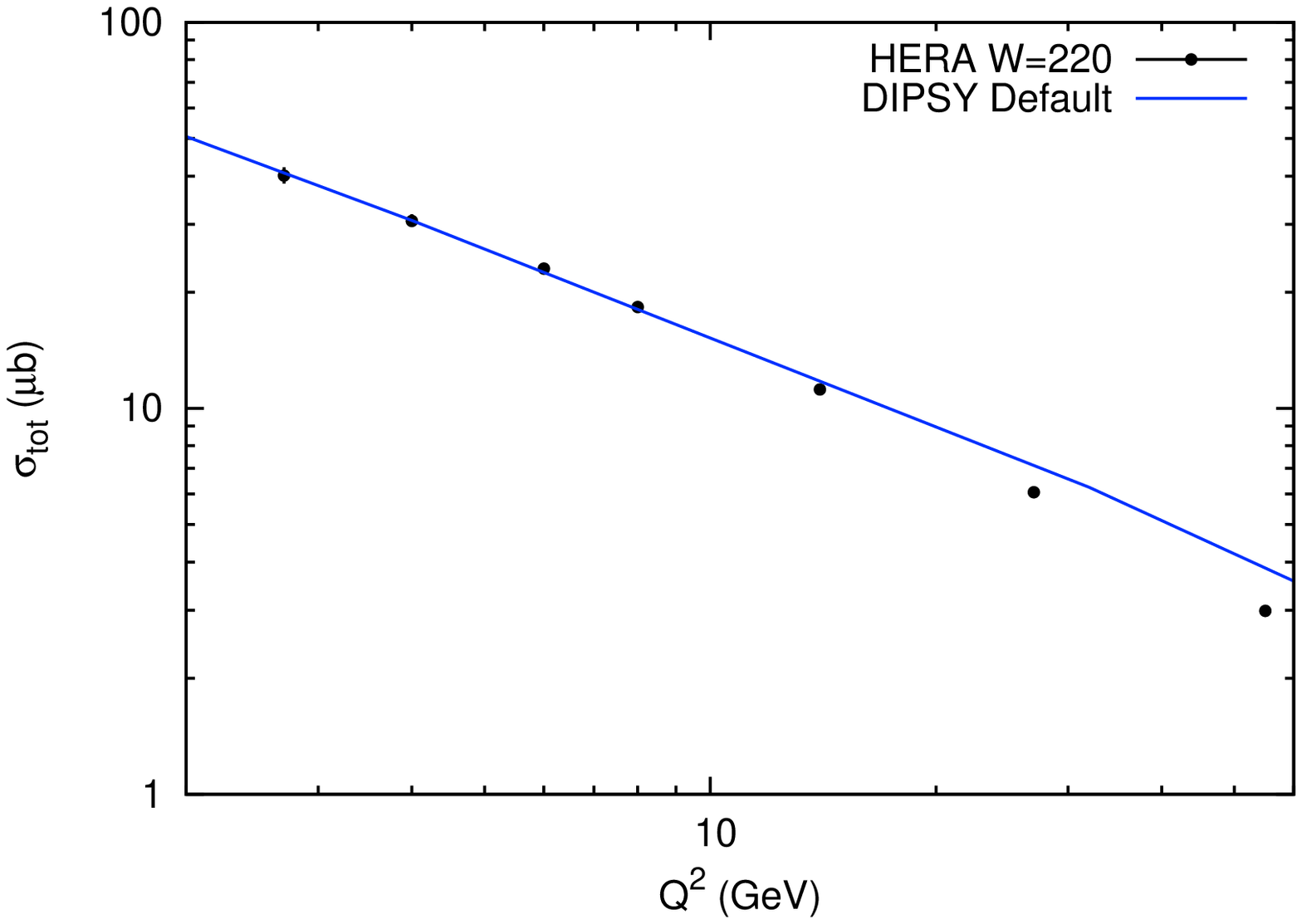}
\caption{\label{fig:totxsec} The inclusive data that was tuned to. (Left) The total and elastic cross section as function of $\sqrt{s}$ for $pp$. Experimental data from~\cite{Bernard:1986ye,Adamus:1987sx,Abe:1993xx,Abe:1993xy,Amos:1990jh,Amos:1990fw,Avila:2002bp,Augier:1994jn,Block:2000pg}. (Right) The total $\ga^\star p$ cross section as function of $Q^2$ for $W = 220$~GeV. Experimental data from~\cite{Chekanov:2005vv,Aktas:2005ty,:2007cz}. }
}

It should be noted that further fluctuations have been included in the
proton wavefunction other than the fluctuation in size. First, the
partons are allowed to fluctuate in shape, making the normally
equilateral triangle of valence partons distorted. This has
essentially the same effect as the fluctuation is size as a flatter
triangle will have a smaller cross section. Second, the valence
partons have a Gaussian smear in rapidity. While they previously were
all placed at the same rapidity, that of the proton, they are now a
bit spread out. This has a small effect also for the exclusive
observables, as the peak in $dN/d\eta$ and $dp_\perp/d\eta$ at the
valence rapidity becomes more smeared out. As in previous
publications, tuning shows that all these fluctuations have to be small to maintain the ratio between the elastic and total cross section. Although the fluctuations are not needed for any observable, the fluctuations in size, angle and rapidity are set to a Gaussian distribution of widths 0.1~GeV, 0.1 and 0.1 respectively.

\paragraph{Inclusive \boldmath$\ga^\star p$ total cross sections:}
As mentioned several times above, this is a key observable for
tuning. Changes that affects the $\ga^*p$ cross section will also
affect exclusive observables such as $dN/d\eta$, and changes to the
virtual cascade that affects the $dN/d\eta$ will also affect the
$\ga^*p$ cross section. However, changes to how the backbone gluons are
handled after the virtual cascade is made, will affect only the final
state, not the inclusive observables. $\si_{\ga^*p}$ is allowing us to
tune the virtual cascade for small dipoles separately, and after that
is fixed, details in how the backbone gluons are treated can be tuned
separately.

This observable is affected, apart from the ordering that was fixed
above, by $\La_\text{QCD}$ and $c$ from sec.~\ref{sec:correctpt}, that
is the parameter that sets $q_\perp=c/r$. With $c=1$ as was motivated
previously, the model can be made to agree with HERA data with a very
reasonable\footnote{For comparison, $\La_\text{QCD}=0.22$~GeV in the
  final state shower in \ariadne, after a tune to LEP data. }
$\La_\text{QCD}$ of 0.23~GeV, see fig.~\ref{fig:totxsec}.

\paragraph{\boldmath$dN/d\eta$:}
Now the virtual cascade is completely fixed, and all that is left is to fix the remaining uncertainties in how the backbone gluons are selected and handled. The remaining details essentially decide how many gluons will come on shell, and how many will be reabsorbed.

The most important choice left to be fixed is the problem in sec.~\ref{sec:realchain}, whether one or both of a partons parents should come on shell. Not keeping both parents produce about 20\% less charged particles than keeping both, and comparing experimental data, this is a bit too low. Thus we will in this tune chose to keep both parents in each emission.

This fixes all the details of the tune in this paper, and are ready to compare the tuned Monte Carlo to experimental data.

\subsection{Comparison with experiments}

We will now compare the results from our program with experimental
data on exclusive final-state observables, both from the LHC and the
Tevatron.

The \dipsy program is written in C++ using \thepeg
\cite{Lonnblad:2006pt}, which is a toolkit and framework for
implementing event generation models. This framework is also used by
\herwig \cite{Bahr:2008pv} and, more importantly here, a new version
of \ariadne \cite{Lonnblad:1992tz}, a pre-release of which we have
been using for the results presented in the following. The Lund string
fragmentation model of \pythia has been interfaced to \thepeg and is
also used with \dipsy. In this way we produce exclusive hadronic
final-state, which can be directly compared to data.

We will use the \rivet framework \cite{Buckley:2010ar} (version 1.5.0)
for validating event generators, which is also available from within
\thepeg. We have selected some representative observables from
underlying event, and minimum bias studies by the ATLAS experiment at
center of mass energies of $900$~GeV and $7$~TeV
\cite{Aad:2010fh,atlas:2010ir}. We also show some comparisons with
results from the CDF experiment at $1.8$~TeV \cite{Affolder:2001xt}.

It should be noted that these observables are very difficult to
describe also for the state-of-the-art general purpose event
generators such as \pythia \cite{Sjostrand:2007gs} and \herwig
\cite{Bahr:2008pv}, which can be plainly seen from a recent review of
event generators \cite{Buckley:2011ms} and on the web site
\href{http://mcplots.cern.ch/}{\tt http://mcplots.cern.ch/} where
up-to-date comparisons between event generators and data are
presented.

The most advanced of the general purpose event generators, \pythia, is
the only one that gives a fair description of most minimum-bias and
underlying-event data, and in the following we will compare our
results, not only to data but also to a recent tuning of \pythia
(called 4C in \cite{Corke:2010yf}). In fact we will show two versions
of this tuning, one which only includes non-diffractive events, and
one which also includes elastic and diffractive scattering. Several
experimental measurements include a correction to what is referred to
non-single-diffractive events, which in principle should be directly
comparable to our results. But this correction is very model-dependent
(see \eg\ \cite{Buckley:2011ms} for a discussion on this) and we
therefore give an indication of the size of this correction by
including both non-diffractive and diffractive results from
\pythia. Note that our results should be compared to the
non-diffractive results from \pythia.

Also for \dipsy we show two sets of curves. As discussed above, a
serious constraint for our model is that it should be independent of
the Lorentz frame in which we perform the collisions. To quantify this
independence we have performed simulations not only at central
rapidity, where most measurements are made, but also in an asymmetric
frame\footnote{The rapidity of the collision frame is taken to be 2.7
  for 900~GeV, 3.0 for 1800~GeV, and 3.5 for 7~TeV.}. In the central
rapidity frame, the transverse momenta of final-state gluons that are
relevant for the observables are predominantly given by the
dipole--dipole interactions, while in the asymmetric case the
transverse momenta mainly comes from recoils in the evolution.

In the following we will only show small fraction of the plots we have
produced with \rivet. Further plots can be inspected on
\href{http://home.thep.lu.se/~leif/DIPSY.html}{\texttt{http://home.thep.lu.se/{$\sim$}leif/DIPSY.html}}.

\paragraph{Minimum-bias observables:}

\FIGURE{
  \epsfig{file=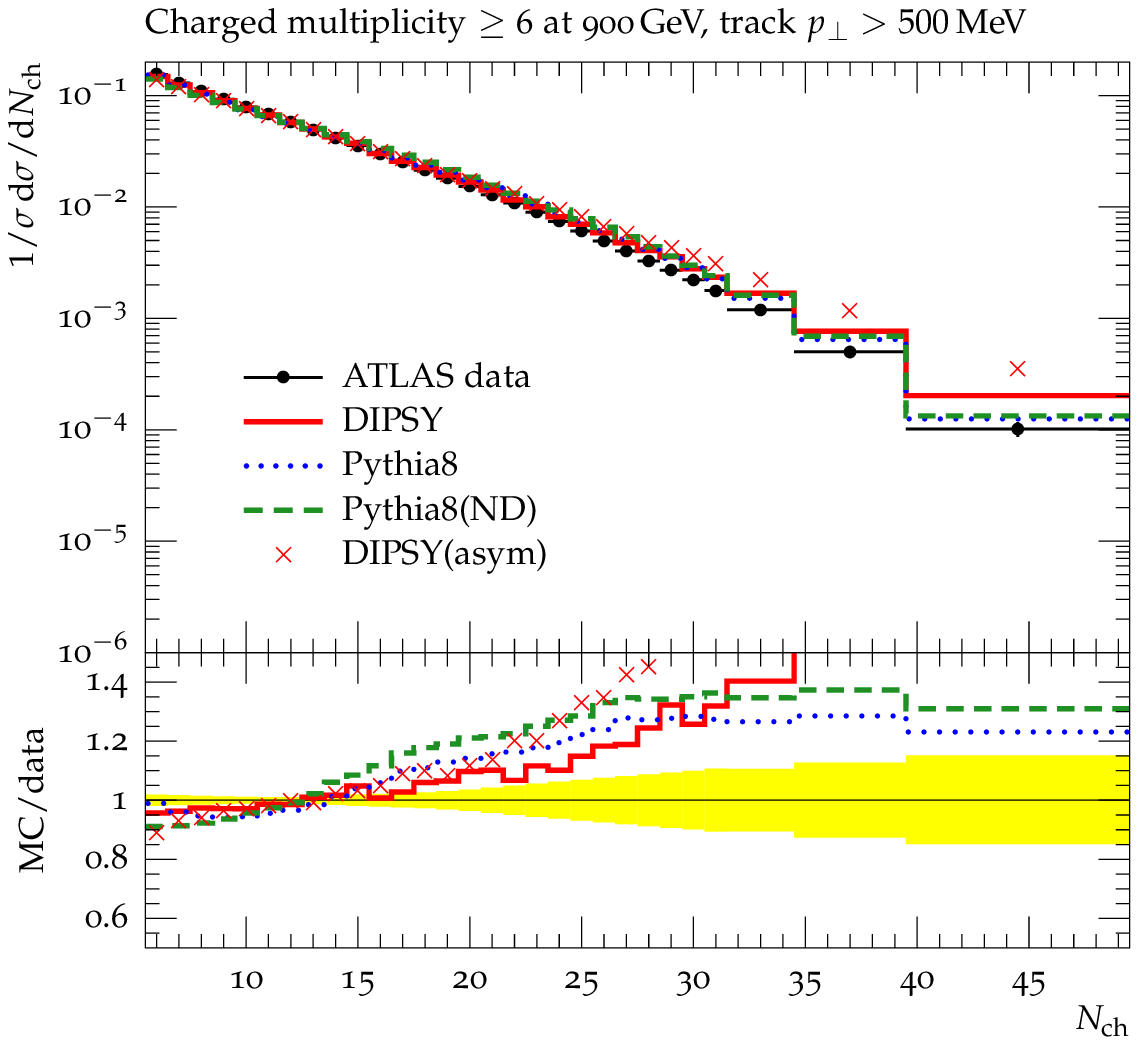,width=0.5\linewidth}%
  \epsfig{file=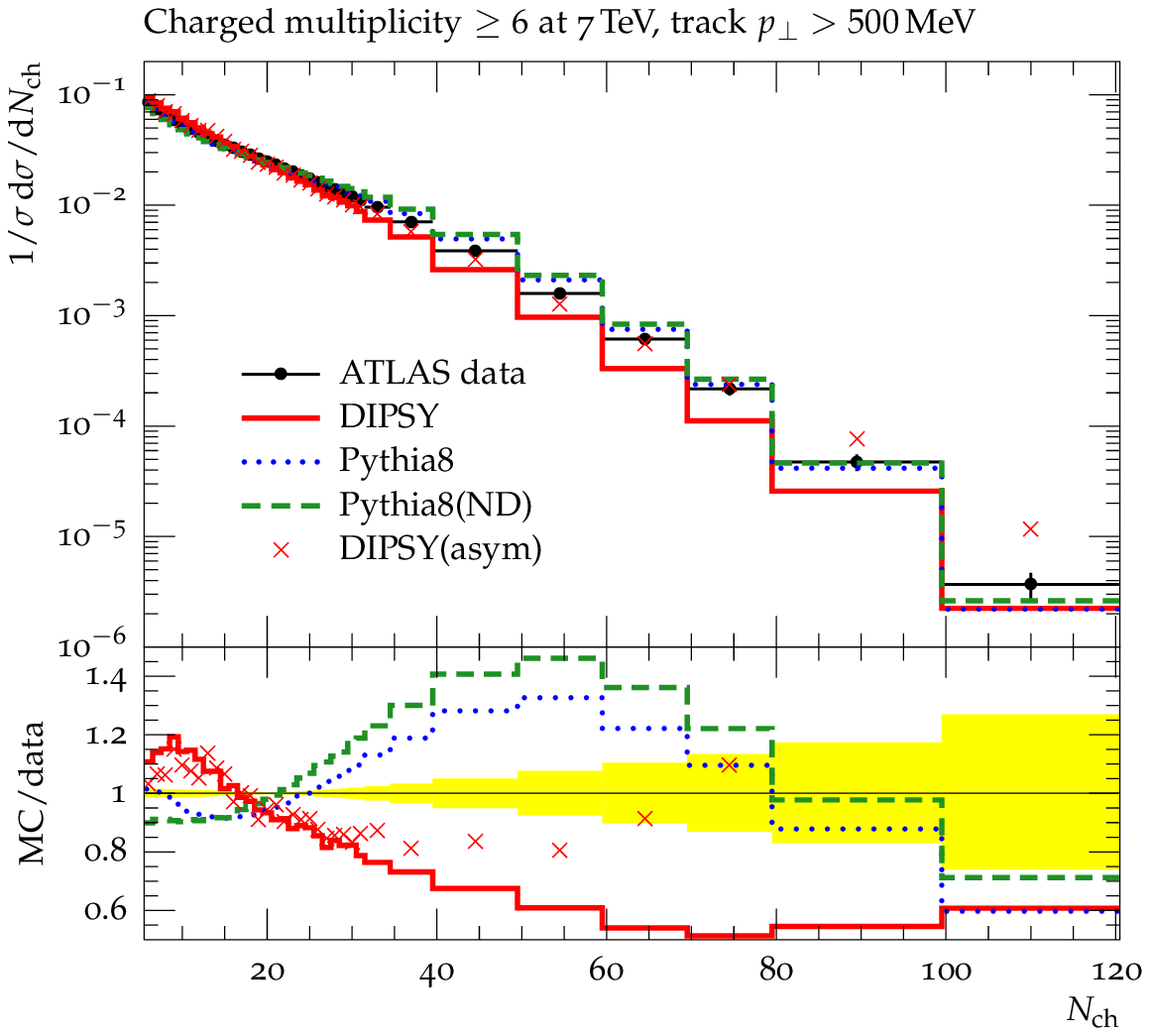,width=0.5\linewidth}
  \caption{\label{fig:MBNch} The charged multiplicity distribution at
    0.9 (left) and 7~TeV (right). The data points are the ones given
    in \rivet version 1.5.0 and are taken from \cite{atlas:2010ir} and
    include only tracks with $p_\perp>500$~MeV in events with more
    than 6 charged tracks. The full lines are the \dipsy results, the
    dotted lines are from \pythia with diffractive and non-diffractive
    events, the dashed lines are \pythia with only non-diffractive
    events, and the crosses are from a \dipsy simulation in an
    asymmetric frame.}

}

\FIGURE{
  \epsfig{file=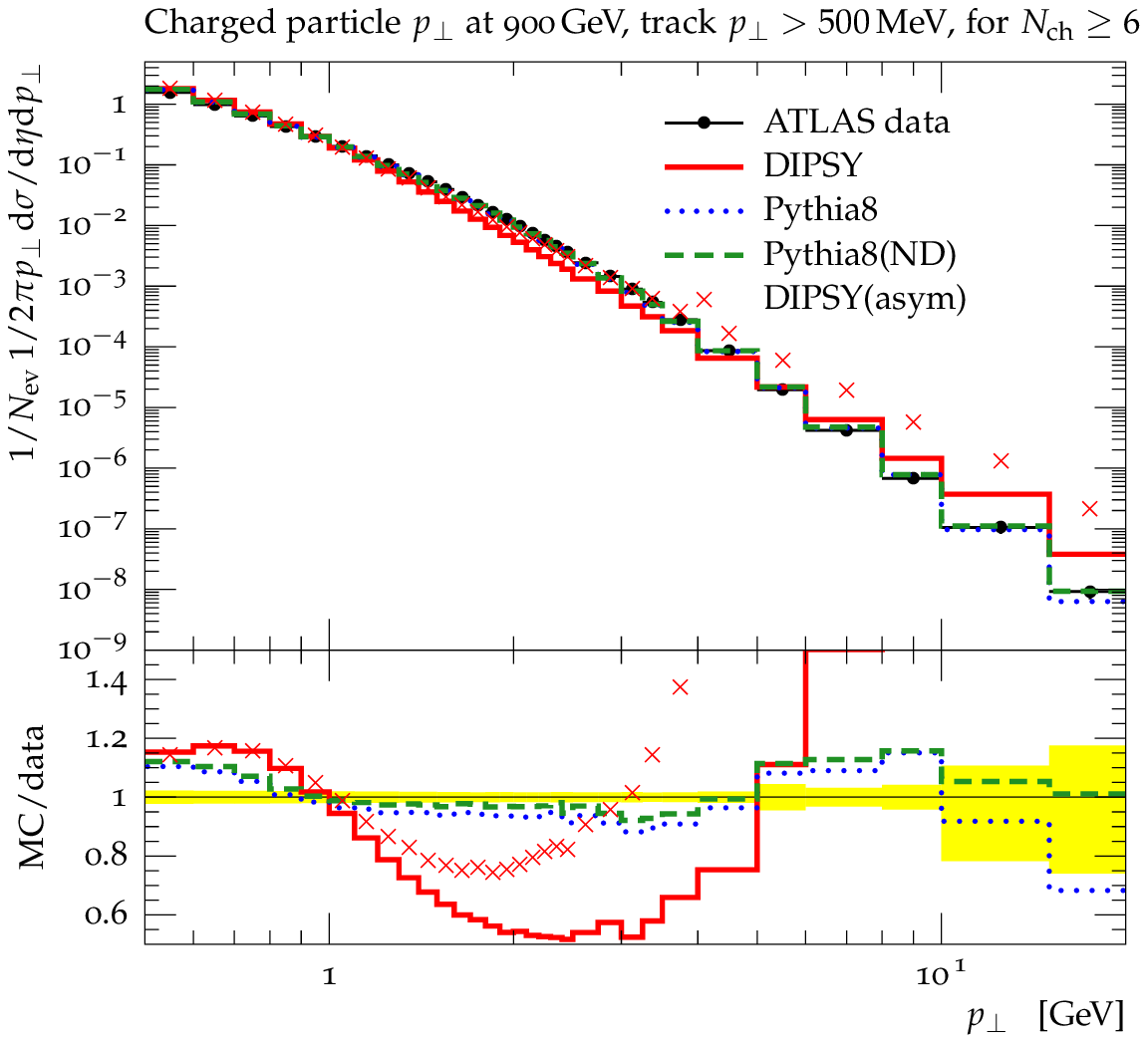,width=0.5\linewidth}%
  \epsfig{file=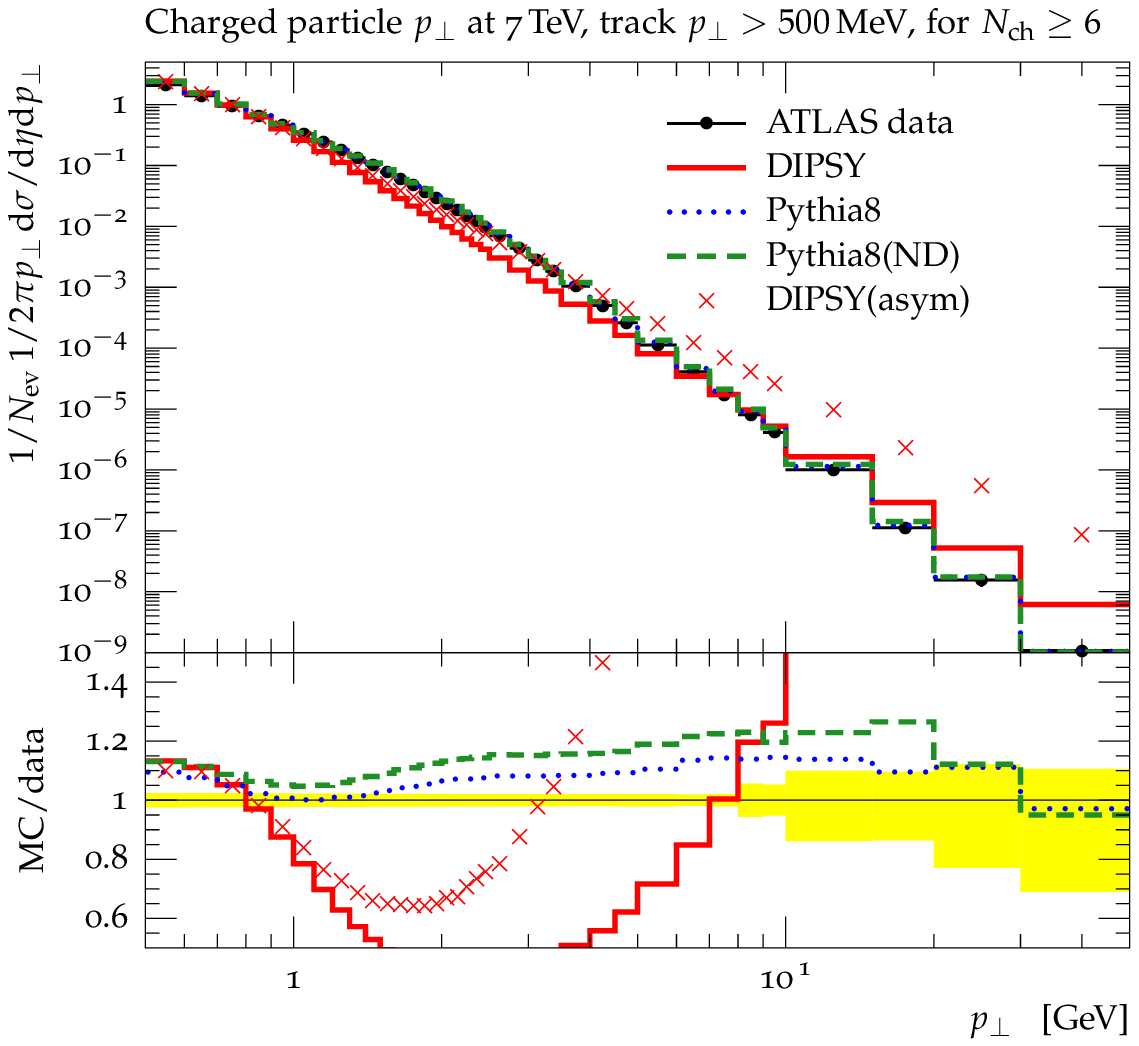,width=0.5\linewidth}
  \caption{\label{fig:MBptch} The transverse momentum distribution of
    charged particles at 0.9 (left) and 7~TeV (right). Data points and
    lines as in fig.\ \ref{fig:MBNch}.}
}

\FIGURE{
  \epsfig{file=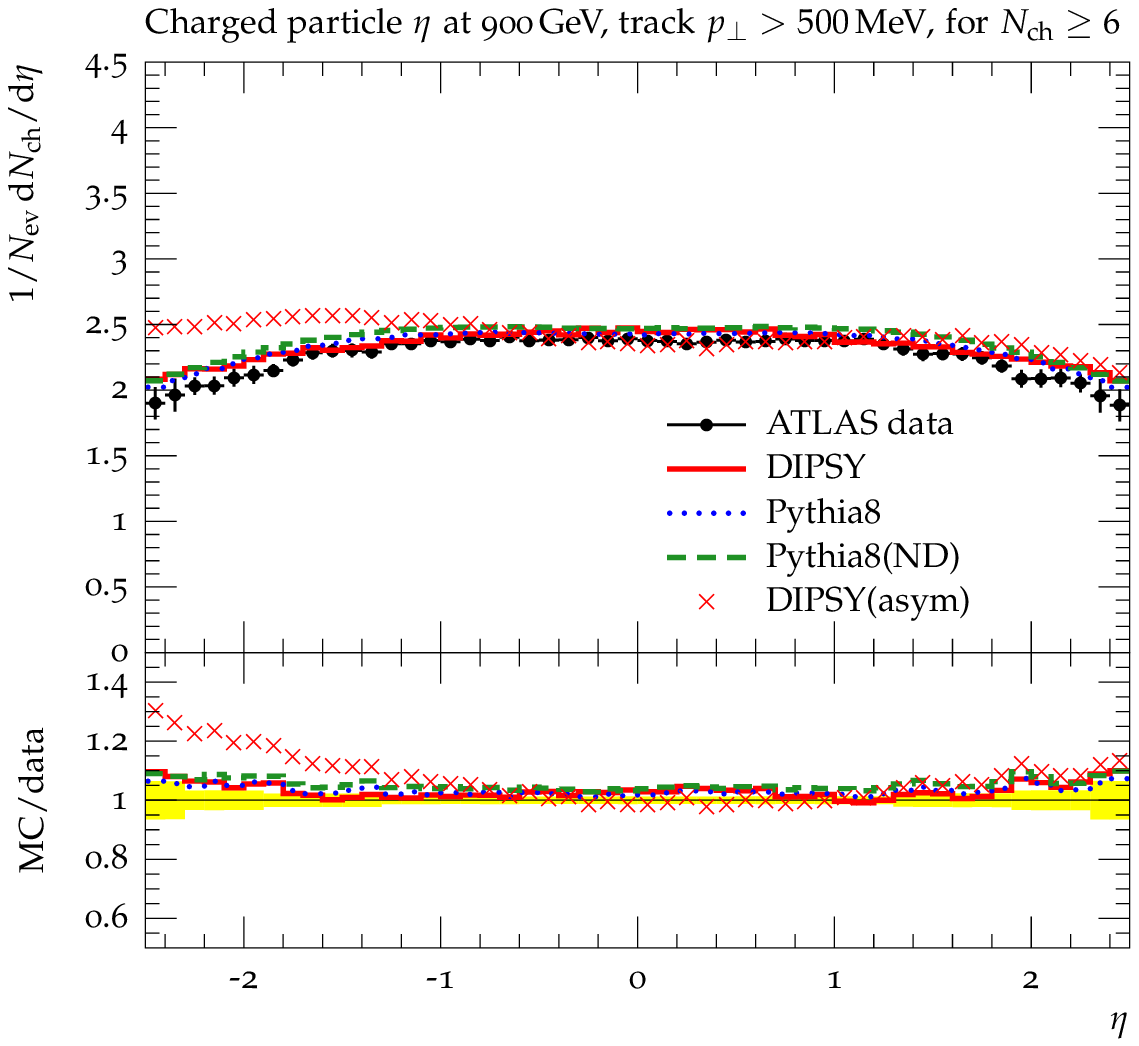,width=0.5\linewidth}%
  \epsfig{file=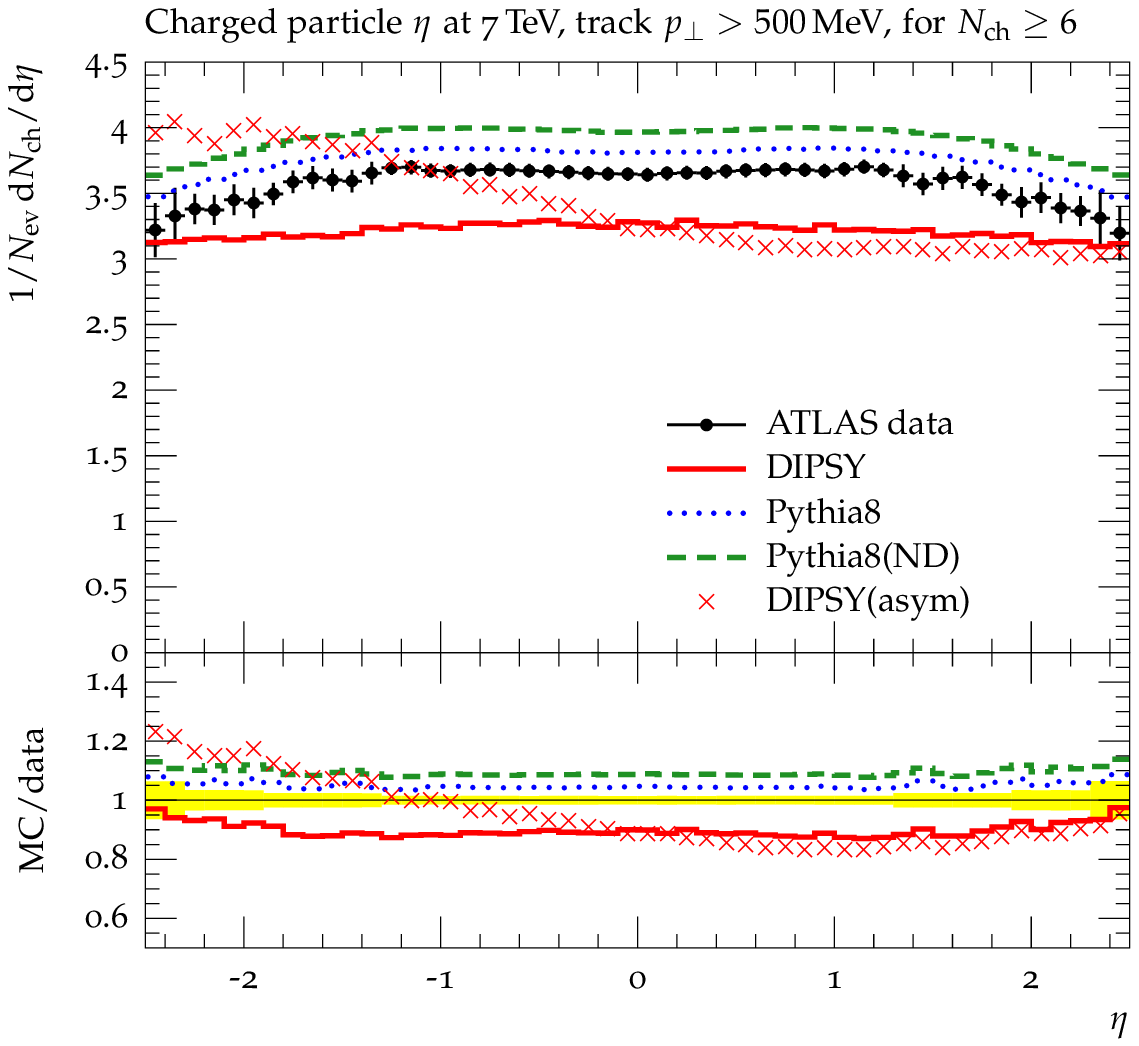,width=0.5\linewidth}
  \caption{\label{fig:MBeta} The pseudo-rapidity distribution of
    charged particles at 0.9 (left) and 7~TeV (right). Data points and
    lines as in fig.\ \ref{fig:MBNch}.}

}

\FIGURE{
  \epsfig{file=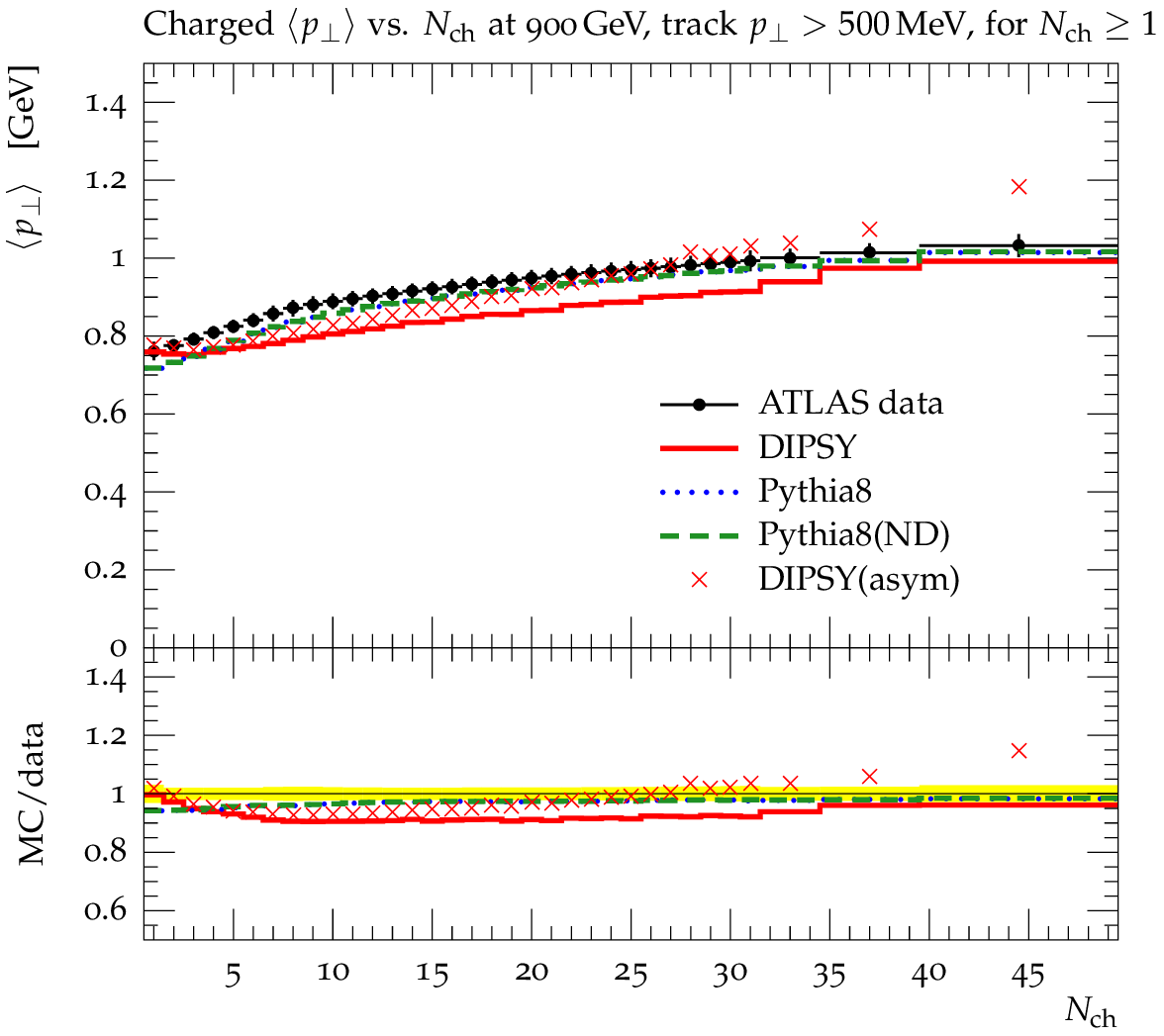,width=0.5\linewidth}%
  \epsfig{file=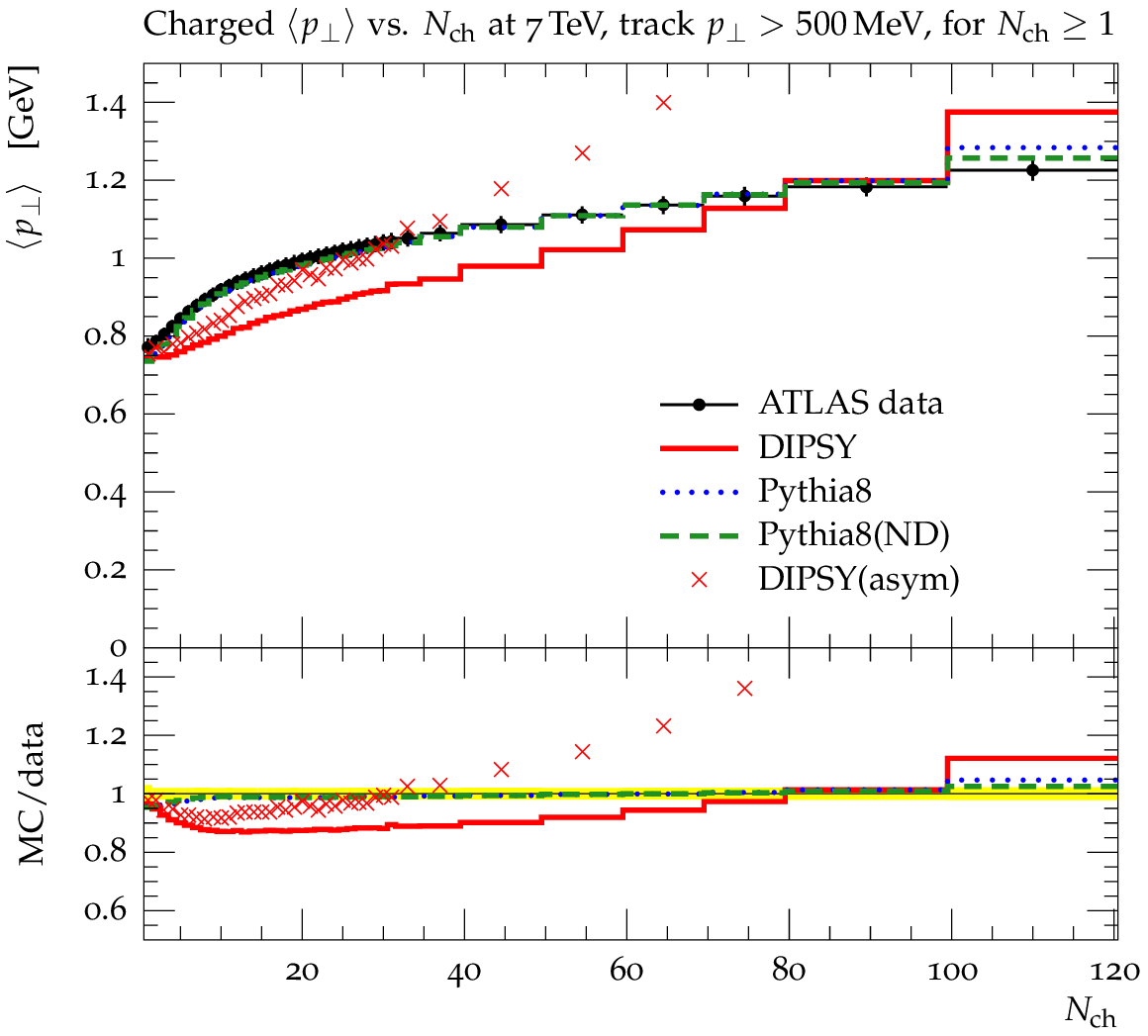,width=0.5\linewidth}
  \caption{\label{fig:MBavpt} The average transverse momentum as a
    function of charged particle multiplicity at 0.9 (left) and 7~TeV
    (right). Data points and lines as in fig.\ \ref{fig:MBNch}.}

}

In figures \ref{fig:MBNch}--\ref{fig:MBavpt} we show some standard
minimum-bias observables as measure by ATLAS at $900$~GeV and
$7$~TeV. A general observation is that \dipsy has a slightly too weak
energy dependence. Looking, \eg, in figure \ref{fig:MBNch} we see that
\dipsy tend to overestimate high multiplicities at low energies and
underestimate them at high energies. Note, however that the cross
section for \eg\ multiplicities between 40 and 50 increases by a
factor 40, and a large part of the increase is described by \dipsy (a
factor 25). The effect is more clearly seen in the pseudo-rapidity
dependence of the average multiplicity in figure \ref{fig:MBeta},
where again the energy dependence in \dipsy is a bit too weak.

For the transverse momentum distribution in figure \ref{fig:MBptch},
the energy dependence is a bit better, but \dipsy tend to overshoot
both at small and large transverse momenta, while undershooting at
medium values. For large transverse momenta, we believe that our
description may improve if we include $2\to2$ matrix element
corrections for local $q_\perp$ maxima, both in the evolution and in
the interactions. For the softer part of the spectrum, we believe that
the discrepancy is related to the point-like valence partons in the
proton wavefunction. Even when evolved to mid-rapidity, the valence
partons will still have a large $q_+$ and will allow emissions at
large $q_\perp$. Even though these emissions will be in mid-rapidity,
the balancing recoil will go with the valence partons down the
beam-pipe, making the spectrum in the main ATLAS detectors too soft.

In the transverse momentum distribution we also see that \dipsy in the
asymmetric frame gives a much harder spectrum, indicating that the
transverse momentum generated in the virtual cascade is harder than
the what is generated in the dipole--dipole interactions. The frame
dependence is also clearly seen in the pseudo-rapidity distribution in
figure \ref{fig:MBeta}, where we see an increased activity for large
rapidities in the asymmetric case, where the particles mainly
originate from the evolution rather than the interactions.

In figure \ref{fig:MBavpt} we see that \dipsy underestimate the
average transverse momentum at small and medium multiplicities, which
is related to the underestimate of the number of particles with medium
transverse momentum.

\paragraph{Underlying-event observables:}

\FIGURE{
  \epsfig{file=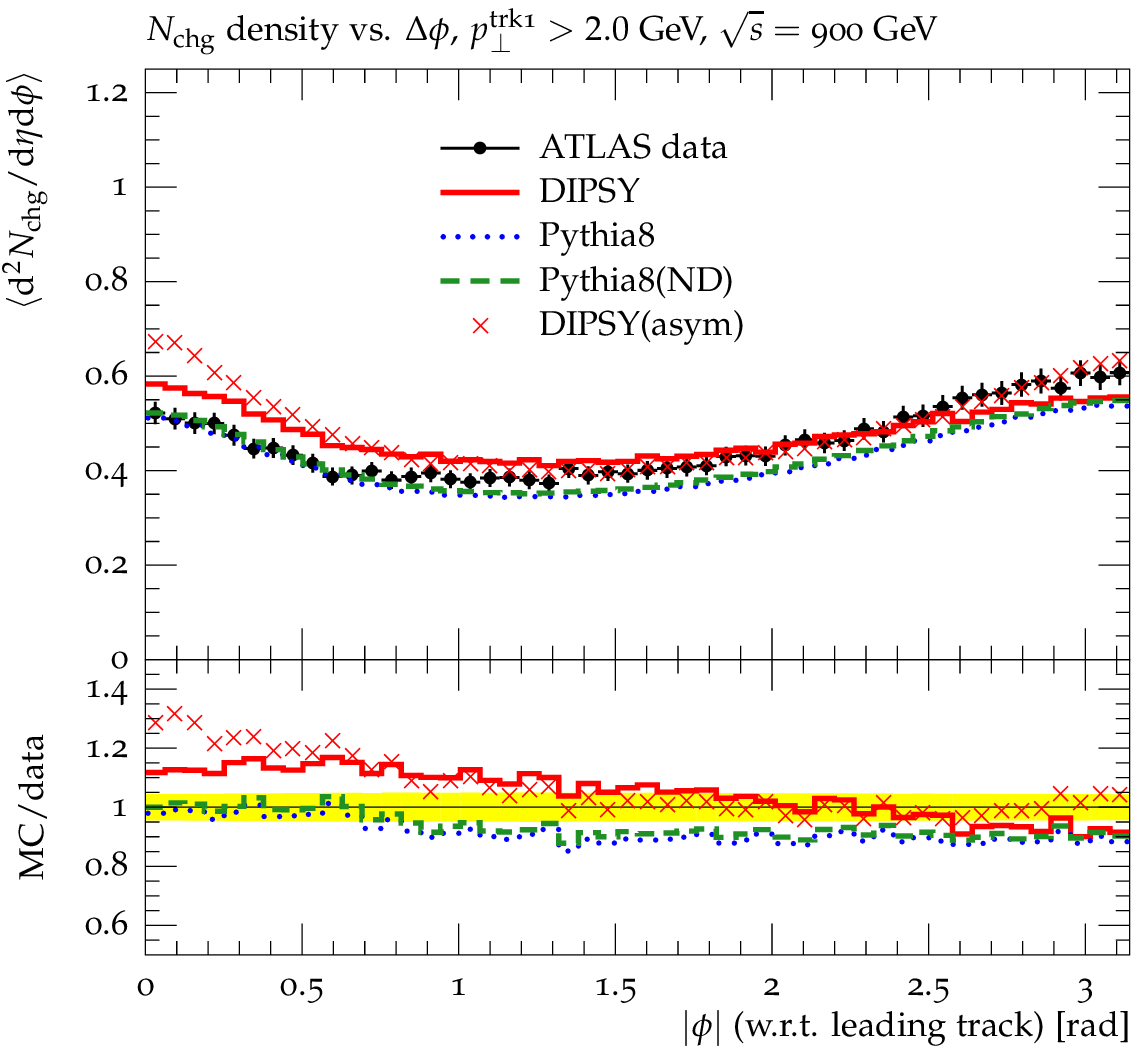,width=0.5\linewidth}%
  \epsfig{file=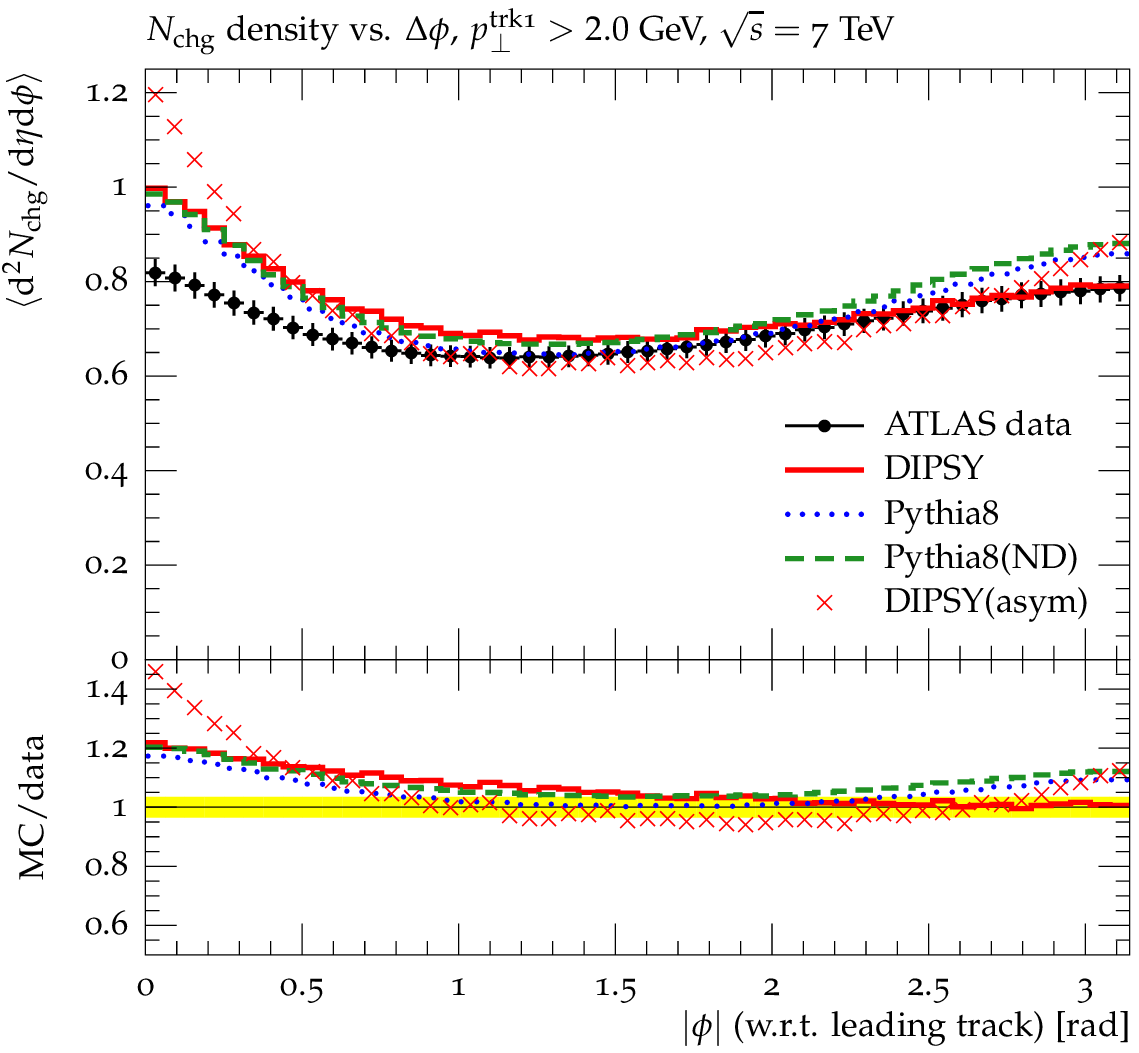,width=0.5\linewidth}\\[5mm]

  \epsfig{file=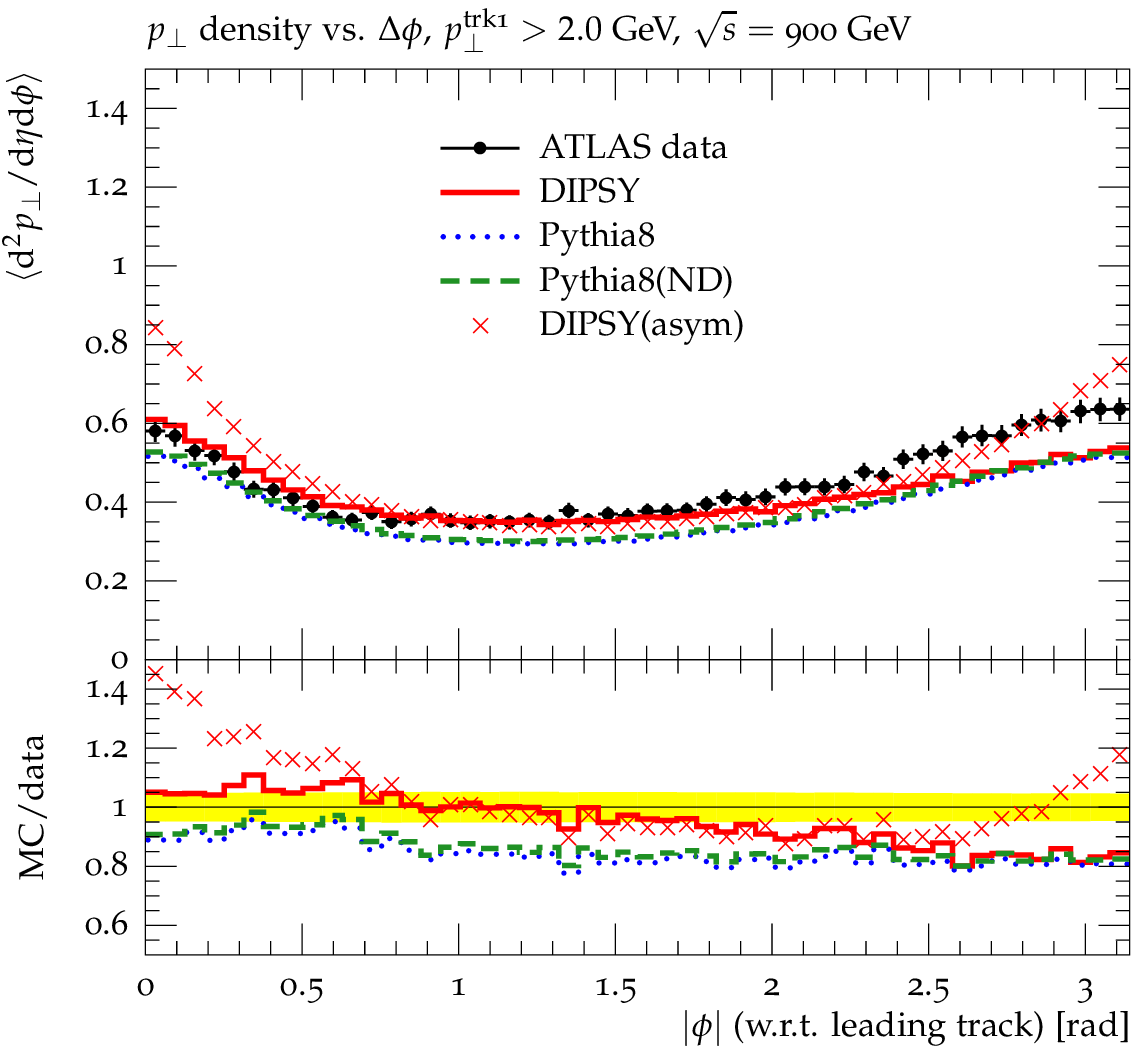,width=0.5\linewidth}%
  \epsfig{file=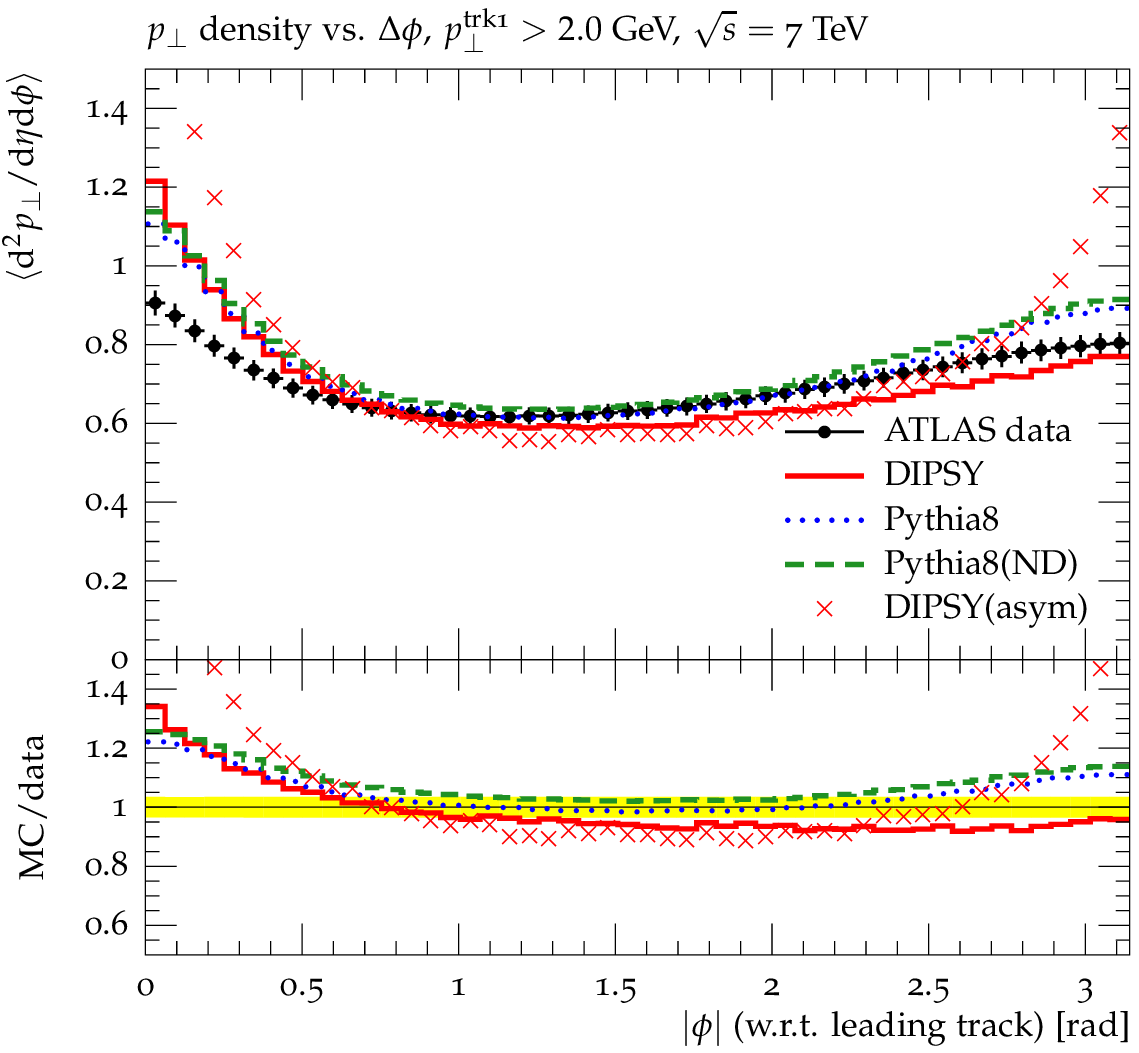,width=0.5\linewidth}

  \caption{\label{fig:UEphi} The multiplicity (top) and scalar sum of
    transverse momenta (bottom) of charge particles as a function of
    azimuth angle w.r.t.\ a leading charged particle of at least 2~GeV
    transverse momenta at 0.9~TeV (left), and 7~TeV (right). The data
    points are the ones given in \rivet version 1.5.0 and are taken
    from \cite{Aad:2010fh}. The generated data as in fig.\
    \ref{fig:MBNch}.}

}

\FIGURE{
  \epsfig{file=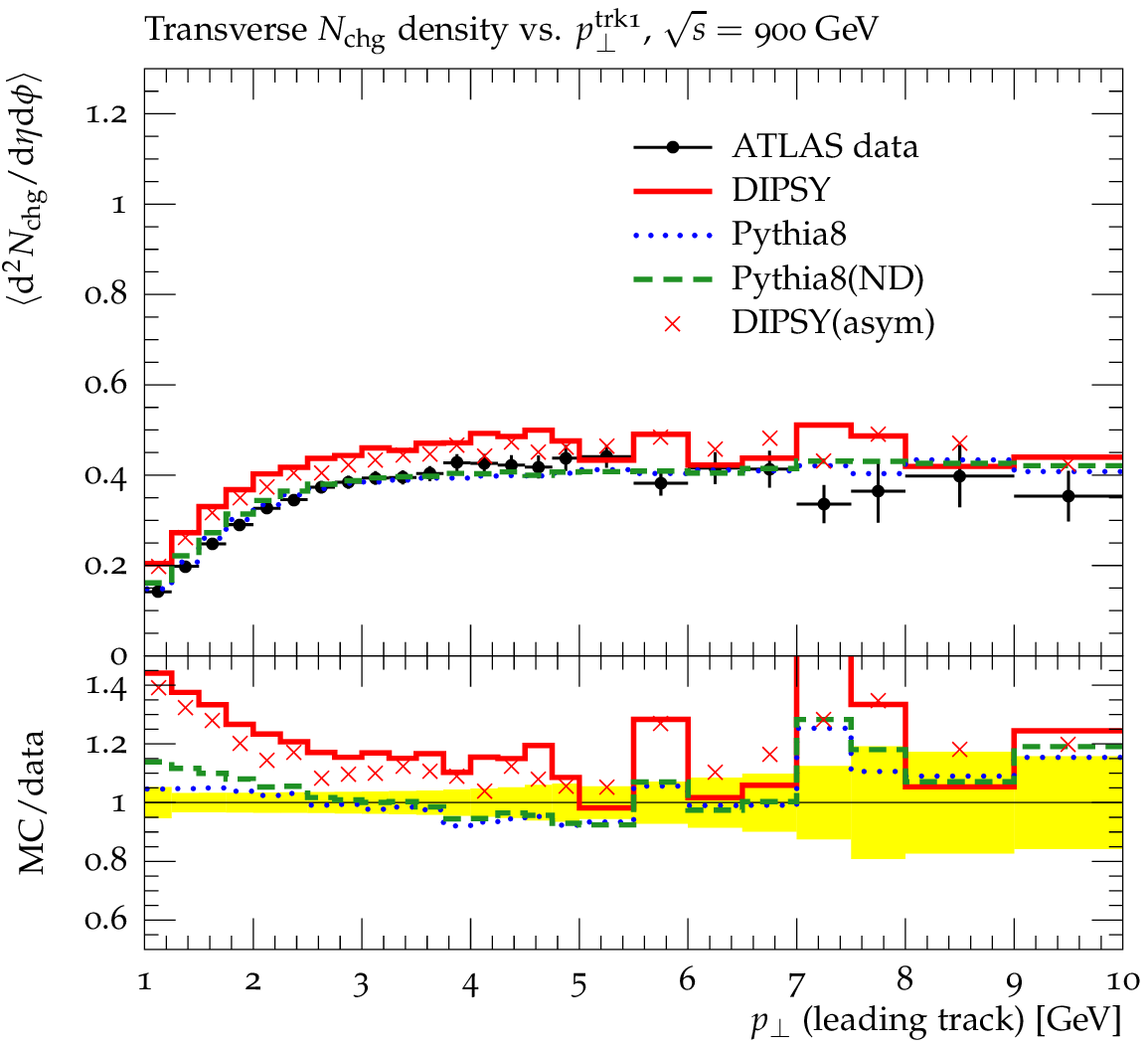,width=0.5\linewidth}%
  \epsfig{file=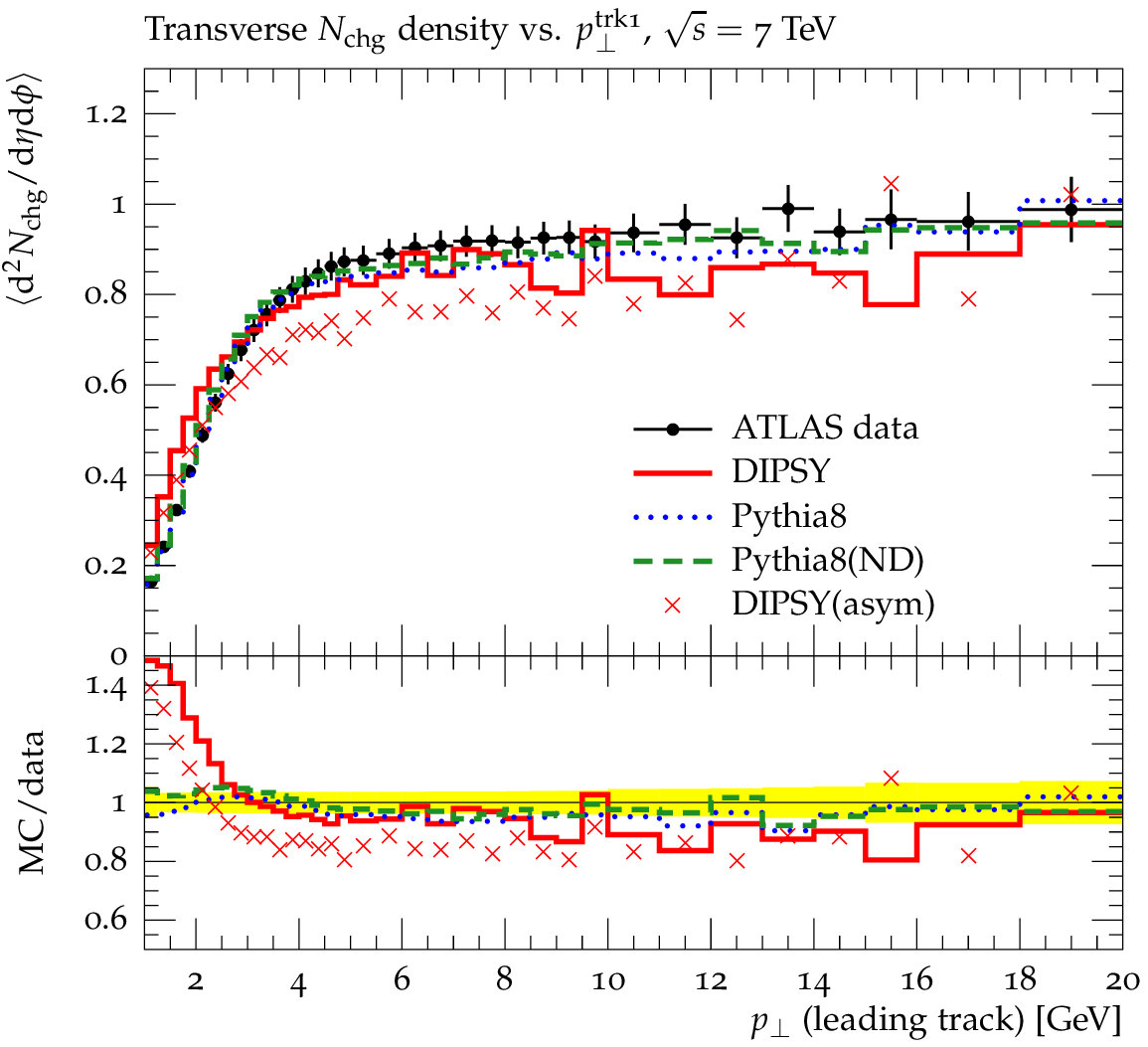,width=0.5\linewidth}\\[5mm]
  \epsfig{file=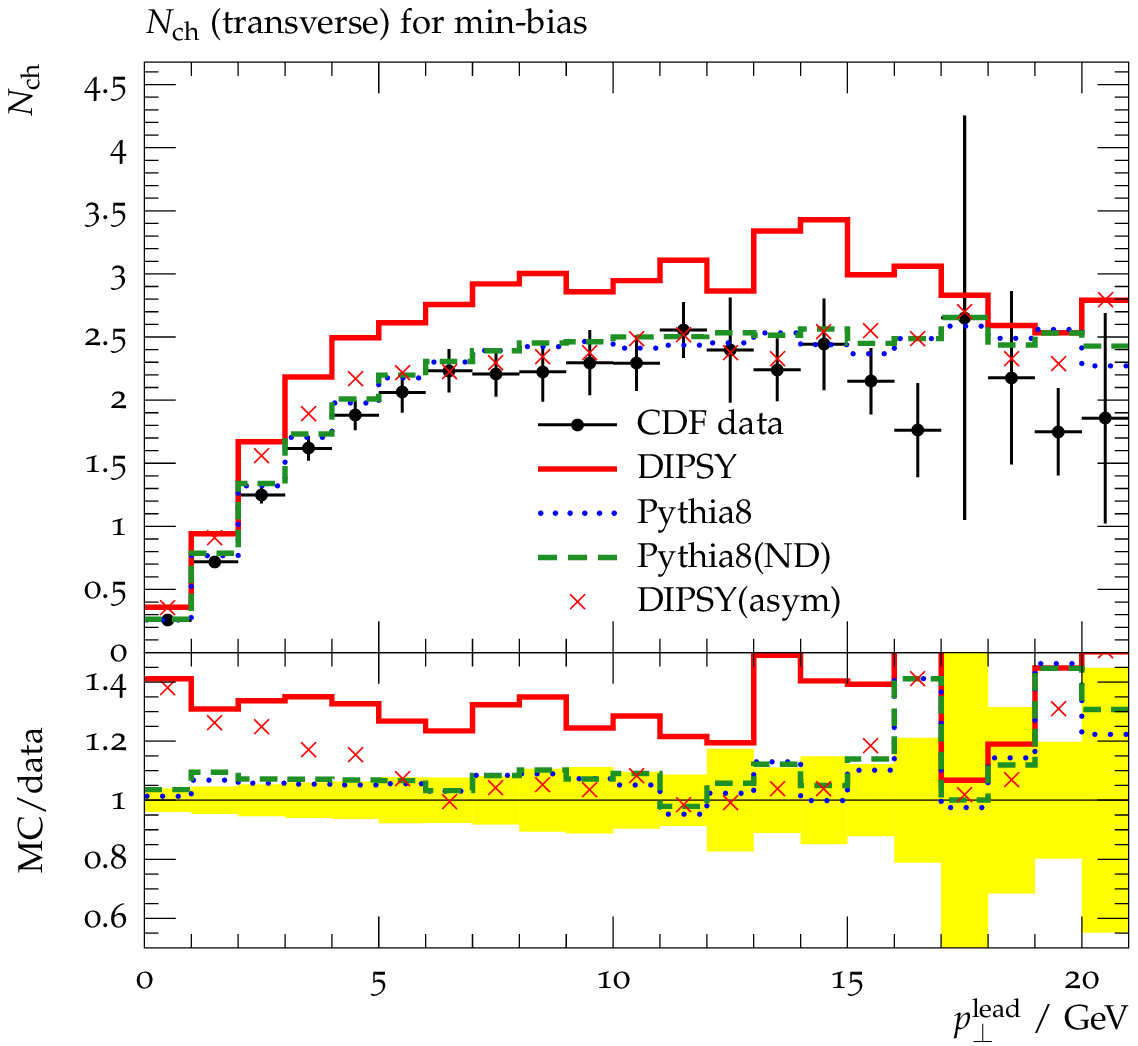,width=0.5\linewidth}

  \caption{\label{fig:UEtransNch} The multiplicity of charge particles
    in the \textit{transverse} region, as a function of the transverse
    momentum of the leading charged particle at 0.9 (top left), 1.8
    (bottom) and 7~TeV (top right). Data points and lines as in fig.\
    \ref{fig:UEphi}. We also show data from CDF
    \cite{Affolder:2001xt}, but note that this data is differently
    normalized.}

}

\FIGURE{
  \epsfig{file=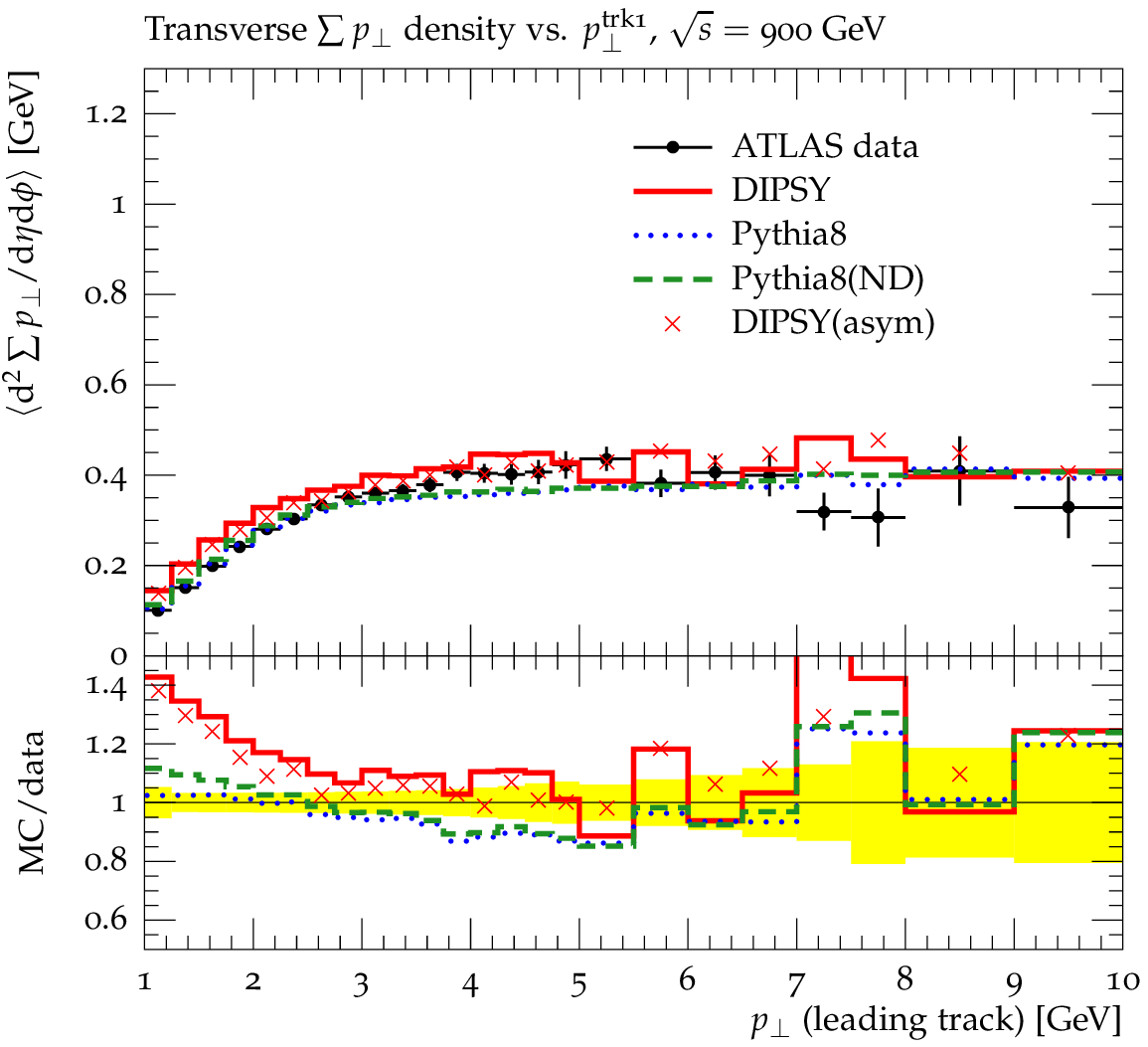,width=0.5\linewidth}%
  \epsfig{file=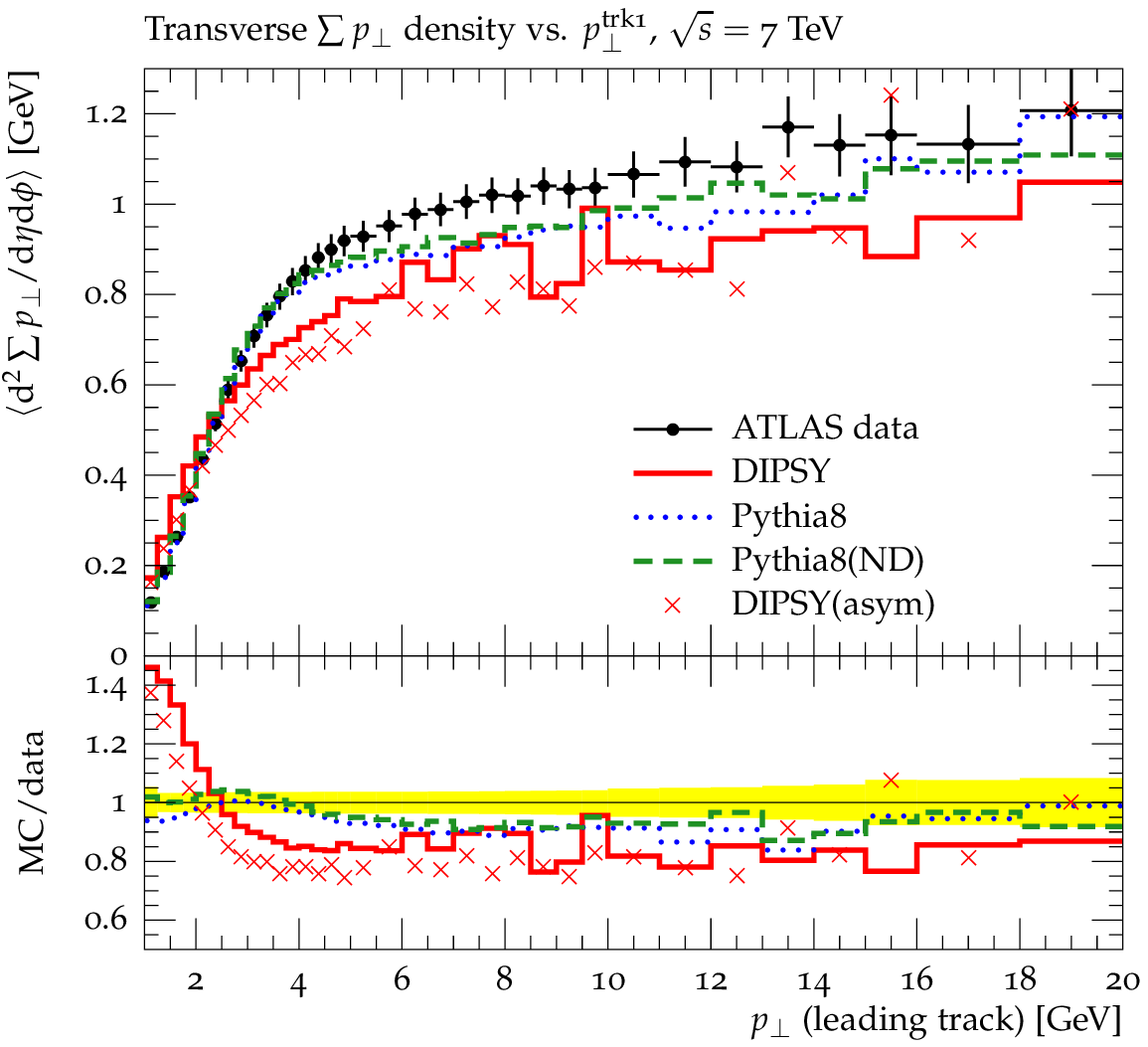,width=0.5\linewidth}\\[5mm]
  \epsfig{file=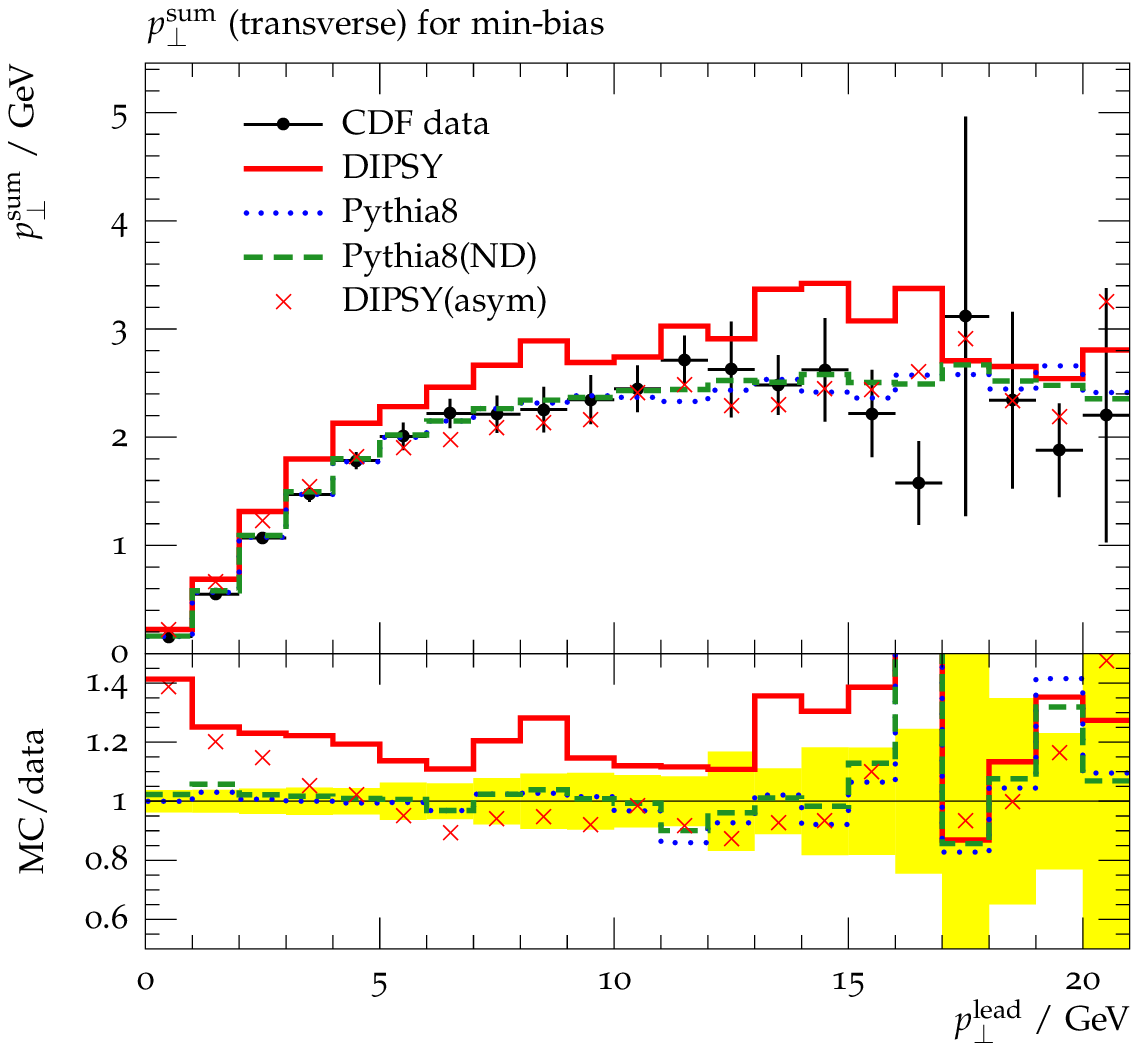,width=0.5\linewidth}

  \caption{\label{fig:UEtranspt} The scalar sum of transverse momenta
    of charge particles in the \textit{transverse} region, as a
    function of the transverse momentum of the leading charged
    particle at 0.9 (top left), 1.8 (bottom) and 7~TeV (top
    right). Data points and lines as in fig.\ \ref{fig:UEtransNch}.}

}

In figures \ref{fig:UEphi}--\ref{fig:UEtranspt} we show
observables related to the underlying event, many of which were
introduced by Field \cite{Field:2005qt}. The emphasis is to look at
particle flow and transverse momenta away from jets and/or high
transverse momentum charged particles to as much as possible isolate
effects of the underlying events.

In figure \ref{fig:UEphi} we show the multiplicity and summed
transverse momentum as a function of the azimuth angle from a leading
charged track with transverse momentum larger than $2$~GeV. We see
that all programs have difficulties reproducing the activity close to
the trigger particle (the \textit{towards} region) and in the
opposite (\textit{away}) region which should be dominated by the
recoil from the leading track. In fact, all programs have similar
problems, especially in the towards region, and while \dipsy in an
asymmetric frame is clearly the worst, giving too much back-to-back
activity, the default \dipsy actually gives a reasonable shape.

It is, however, the \textit{transverse} region, defined as
$60^\circ<\Delta\phi<120^\circ$, which is the most sensitive to the
underlying event. In figures \ref{fig:UEtransNch} and
\ref{fig:UEtranspt} we show the number and summed transverse momenta
of charged particles in this region as a function of the transverse
momentum of the leading track. We see that the \pythia here gives a
better description of data, while \dipsy as before seem to has a bit
too weak energy dependence and consistently seem to overestimate the
activity at small transverse momentum of the trigger particle.

\section{Conclusions}
\label{sec:conclusions}

We have here presented a completely new model for exclusive final
states in non-diffractive hadronic collisions, and implemented it in a
Monet Carlo program called \dipsy \footnote{The DIPSY program is
  dependent on the Ariadne program for final-state radiation. While
  this part of Ariadne is well tested and working, other parts of this
  program is still being worked on and Ariadne is therefore not
  officialy released. The official release of DIPSY will be done
  together with that of Ariadne. Meanwhile pre-releases of both DIPSY
  and Ariadne are available on request from the authors.}. 

As mentioned in the introduction our aim is not to produce the most
precise model for proton collisions, but rather to study the features of
small $x$ evolution and effects of saturation.
The model is based on BFKL evolution, including resummation of sub-leading
logs, and saturation effects which includes pomeron loops not only due to
multiple interactions but also within the cascade evolution. In earlier
publications we have demonstrated that a single perturbative pomeron can
describe not only total and and elastic scattering, including the scale
dependence of the effective power $\lambda_{\mathrm{eff}}(Q^2)$
\cite{Avsar:2006jy}, but also diffractive excitation in $pp$ collisions
and DIS \cite{Avsar:2007xg, Flensburg:2010kq}.  In this paper we have
extended the model to also describe exclusive final states in
non-diffractive hadronic collisions.

The model is dramatically different from conventional
multiple-interaction scenaria implemented in state-of-the-art general
purpose event generators such as \pythia and \herwig, which are based
on DGLAP evolution with structure functions adjusted to experimental
data. Also on the technical side, the differences are large. We use a
forward evolution in impact-parameter space and rapidity, while
conventional programs use backward evolution in momentum space. Our
procedure has the advantage that colour screening and saturation
effects are easily treated. In addition we have a detailed description
of fluctuations and correlations in the proton wave function, which
are typically averaged over in conventional approaches. The
fluctuations are essential in diffractive excitation, and correlations
are important e.g. in estimates of the effective cross section in
double-parton scattering \cite{Flensburg:2011kj}.

On the down-side, although the evolution of inclusive observables is
rather well constrained by the resummation of sub-leading logs, the
properties of exclusive states suffer from a number of non-leading
effects, which influence our results. Small dipoles have little influence
on inclusive cross sections, but a large effect on the final state. We
have here had to make several choices, which in many cases lack guidance
from perturbative QCD, and had to be determined by semi-classical and
phenomenological arguments.

Throughout the the development of our model the requirement that the
results should be independent of the Lorentz frame in which the
interactions are performed has been a very severe constraint. In
addition, our choices have been constrained by the requirement that
our description of inclusive and semi-inclusive observables, such as
the total and elastic cross sections remain satisfactory.

Although a big part of the framework from previous publications was
reevaluated to describe exclusive observables, the part of the model
that describes the inclusive observables remains basically
unchanged. The reweighting of the final state gluons does not affect
the interaction probability, and the ordering in the virtual cascade
and interaction, which were tuned for frame-independence, are very
similar to what we have presented in previous publications.

Thus, while the \dipsy program contains a fair amount of parameters
and switches, maybe even as many as in the multiple-interaction model
in \pythia, once the frame independence and inclusive cross sections
are accounted for, there is little left that will influence the
exclusive final states produced. As an example, the dependence on the collision
energy of final-state observables is locked to the energy dependence of the
total and elastic cross sections, and thus determined by the BFKL
evolution and saturation effects. In the
case of the multiple-interaction model in \pythia, in contrast, the
energy dependence is a free parameter, although indirectly related to
the increase of the total cross section with energy.

\subsection{Outlook}
\label{sec:outlook}

Although the basic cascade evolution appears to be fairly stable, the
generation of exclusive states introduces larger uncertainties, where we
have to be guided by semiclassical arguments. Also, as pointed out above,
the Lorentz-frame independence requirement is not quite satisfied. With
this in mind, we are very pleased that although not perfect, our model
gives quite reasonable results for minimum-bias and underlying-event
observables at the Tevatron and the LHC. This implies that our model can be used to study if
tuned parameters in conventional event generators, like \emph{e.g.} the
low-$p_\perp$ cutoff for parton subcollisions, agree with expectations for
low-$x$ evolution and saturation, and if large effects of fluctuations are
expected to be important.

For future improvements we believe the
most important are the matching to matrix elements for the hard
subcollisions, and the inclusion of quarks. The latter is particularly
important for EW processes, like $Z$, $W$, and Higgs production, and for
forward production where our simple proton model is not expected to be
sufficient.

We also plan to apply our model to collisions involving nuclei, where
saturation effects are particularly important. We are here able to produce
complete partonic final states in $eA$, $pA$, and $AA$ collision,
including finite size effects and non-trivial correlations between partons
in different nucleons in the nuclei. Our model does not include any
hydro-dynamical evolution, but in case a quark-gluon plasma is formed, the
parton state can be used as initial condition for a hydro expansion. In
the absence of plasma, \emph{e.g.} in $eA$, $pA$, or collisions between
light nuclei,  we are also able to generate full hadronic states. Also in
collisions between heavier nuclei it can be interesting to study the
collective effects by comparing the results with and without hydro.

\section*{Acknowledgments}

Work supported in part by the EU Marie Curie RTN MCnet
(MRTN-CT-2006-035606), and the Swedish research council (contracts
621-2008-4252 and 621-2009-4076).

L.L.~gratefully acknowledges the hospitality of the CERN theory unit.


\section*{Appendices}
\appendix
\section{$q_\perp$ max reweighting in \dipsy}
\label{sec:reweighting}
As was described in sec.~\ref{sec:RutherfordAlgorithm}, the maxima in in $q_\perp$ corresponding to an ``outer'' dipole in the cascade, that is a dipole that have emitted no on-shell emissions, come in with the wrong weight, and need to be reweighted.

This is in \dipsy obtained by finding and reabsorbing
some of the outer maxima to restore the correct weight. The inner and
outer dipoles are only known after the backbone gluons are identified, and thus
the reweighting can be performed only after selecting the interactions and the
identification of the backbone gluons.

To find the emissions that need to be reweighted, we note that an outer dipole corresponds to a parton that has not emitted on one side in colour flow, as parton 3 in fig.~\ref{fig:outsidemaxkt}. The maximum in $q_\perp$ is associated with the the backwards dipole being smaller than the forward dipole, that is, dipole $c$ being smaller than dipole $d$.

Starting from the interaction frame, the backbone gluons are checked for partons fitting these criteria, and the ones found are reweighted by an extra factor
\begin{equation}
R = \frac{r_<^2\al_s(r_<)}{r^2_>\al_s(r_>)}, \label{eq:rutherford}
\end{equation}
where the running couplings have been included for better accuracy. This changes the weight of the two dipoles from $W_{\text{before}}$ to $W_{\text{after}}$, with
\begin{equation}
W_{\text{before}} = d^2r_<d^2r_>\frac{\al_s(r_<)}{r_<^2}\frac{\al_s(r_>)}{r_>^2}, \qquad
W_{\text{after}} = d^2r_<d^2r_>\frac{\al_s^2(r_<)}{1}\frac{1}{r_>^4}.
\end{equation}
In momentum space, this corresponds to
\begin{equation}
W_{\text{before}} = d^2q_>d^2q_<\frac{\al_s(q_>)}{q_>^2}\frac{\al_s(q_<)}{q_<^2}, \qquad
W_{\text{after}} = d^2q_>d^2q_<\frac{\al_s^2(q_>)}{q_>^4}.
\end{equation}
Notice that the small dipole size $r_<$ is associated with the large recoil $q_>$, and conversely for the large dipole. The reweighting is done by reabsorbing partons of this type with a probability $1-R$, as is illustrated in fig.~\ref{fig:maxmerged}. This reabsorption is done to the outer parent, so that their recoil will be canceled out, and the large $q_\perp$ vanish.

\FIGURE{
\includegraphics[scale=1]{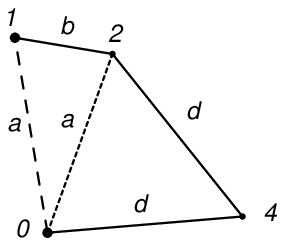}
\hspace{3cm}
\includegraphics[scale=1]{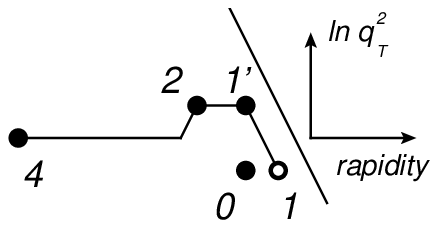}
\caption{\label{fig:maxmerged}The backbone gluons after the outer $q_\perp$ maximum in fig.~\ref{fig:outsidemaxkt} has been merged. The left figure is in impact-parameter space. }
}

\FIGURE{
\includegraphics[scale=1]{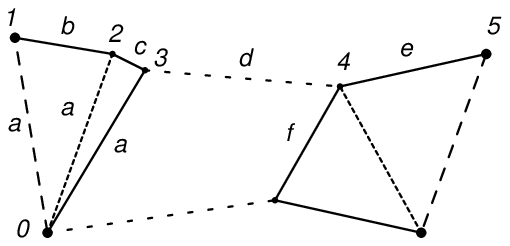}
\caption{\label{fig:outsideint} Partons 1 to 3 incoming from the left side collides with a cascade from the right, with partons 4 and 5. Dipole $a$ interacts with dipole $f$, which makes dipole $c$ an outer $q_\perp$ maximum. The figure is in impact-parameter space. }
}

Some care has to be taken with the definition of an outer $q_\perp$ maximum close to the interacting frame, where it is not enough to look at the cascade from just one side. A dipole interacts with a probability roughly proportional to $r^2$, canceling out the $1/r^2$ weight associated with its emission, and the colour flow will be recoupled over to the other state. Looking at fig.~\ref{fig:outsideint}, parton 3 will not fit the above definition for outer $q_\perp$, as it has no emissions on either side. However, since the dipole on one side is interacting, and the dipole on the other side is not, dipole $c$ will still be a $q_\perp$ maximum with incorrect weight. The Monte Carlo uses a generalized algorithm that covers also these cases, and uses the dipole $d$ in fig.~\ref{fig:outsideint} as the larger dipole to reweight and possible reabsorb parton 3.

Outer $q_\perp$ maxima from the interaction, for example if $d$ would be smallest dipole, does not have to be corrected as the interaction amplitude $f_{ij}$ already has the correct weight for large $q_\perp$.

This procedure ensures that the spectrum of large $q_\perp$ from the
cascade and interaction is indeed going as
$\frac{d^2q_>\al_s^2(q_>)}{q_>^4}$. The proportionality constant is set from the known amplitude for a strictly increasing chain in $q_\perp$, and the known ``penalty'' for being unordered.

This extra reweighting was not
needed before, as only inclusive observables were considered which are
not affected by this procedure. An interacting dipole is always
counted as an inner dipole in the final state, which means that no
interacting dipole will be removed in this way, and the inclusive
interaction probability for the two cascades is unchanged.

\section{Absorbed partons and ordering}
\label{sec:phasespace}
\FIGURE{
\includegraphics[scale=0.8]{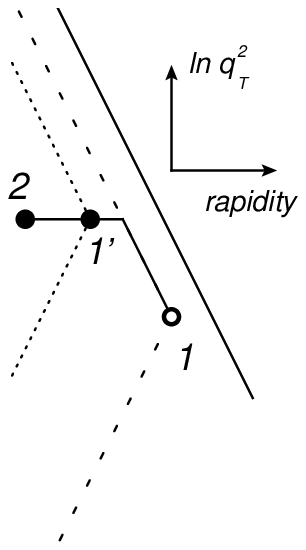}
\caption{\label{fig:mergedPS} An emission limits the ordered phase space of the parents.}

}

In the previous sections, we describe how the backbone gluons are
controlled for ordering and reweighted, and some partons are
absorbed. An absorbed parton returns the $q_+$ to its parents, and the
transverse recoils are undone. This will in general move the parents
back in rapidity and down in $q_\perp$, which opens up a larger
ordered phase space for the parents as in
fig.~\ref{fig:mergedPS}. After parton 2 is emitted, taking part of the
$q_+$ from the parent and giving a $q_\perp$ recoil, 1' is still
allowed to continue emitting. For these emissions to be ordered in
$q_+$ and $q_-$, they have to be between the dotted lines in
fig.~\ref{fig:mergedPS}. If parton 2 was not emitted though, the
larger order phase space between the dashed lines would have been
ordered with respect to 1.

If the virtual cascade would only allow emissions ordered in $q_+$ and $q_-$, every absorbed gluon would open a part of phase space for emissions from the parents that has not been considered in the virtual cascade. This is solved by partially relaxing the ordering in the virtual cascade, and thus overestimating the ordered phase space to cover the case of reabsorbed partons. Care has to be taken though, as the virtual cascade, before any reabsorbtions, will be used to calculate inclusive observables. As the backbone gluons are not yet decided when the cascade is made, it is from this approach not possible to use the correct ordered phase space already for inclusive observables, and at the same time cover all necessary phase space for all possible sets of backbone gluons.

However, by estimating what the backbone gluons may look like already in the virtual cascade one can find a sufficiently good middle way: not too strict ordering so that an important part of the possible final states get cut away, but not too open overestimate so that the inclusive cross sections inflates.

\subsection{Coherence}
\label{sec:coherence}
Three observation at this point will all point towards the same solution:
\begin{itemize}
\item Only large $q_\perp$ emissions, like the one in fig.~\ref{fig:mergedPS}, significantly change the parents momenta. Thus it is only absorption of a step up in $q_\perp$ that risks shadowing emissions from the parent.
\item The phase space that get shadowed mainly concerns small $q_\perp$ emissions. That is, the issue is when a large $q_\perp$ parton is emitted and later absorbed, where the parent has not been allowed to emit with small $q_\perp$.
\item Many of the absorbed backbone gluons are absorbed due to an outer $q_\perp$ maximum. This is exactly the scenario described in the previous two points, that is an absorbed step up in $q_\perp$ that shadows a step down in $q_\perp$.
\end{itemize}
In \dipsy, when a dipole is emitted in the virtual cascade, it is not only emitted by the parton at the end of the dipole, but coherently by all partons being close to the end of the dipole. This means that recoil and ordering is using an effective parton with the summed momenta of all the partons within range, where the range is set by the distance from the parent to the emitted gluon.

\FIGURE{
  \begin{minipage}{0.3\linewidth}
    \begin{center}
      \includegraphics[scale=1]{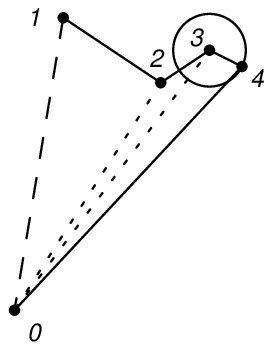}

      \includegraphics[scale=1]{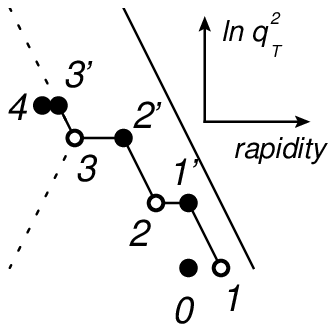}

      \vspace{6mm}
      $(a)$
    \end{center}
  \end{minipage}
  \begin{minipage}{0.3\linewidth}
    \begin{center}
      \includegraphics[scale=1]{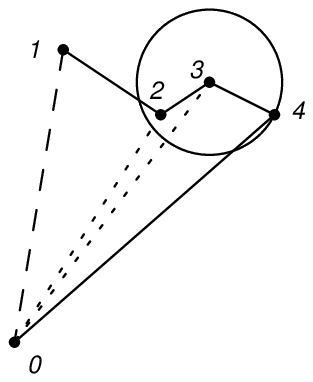}

      \includegraphics[scale=1]{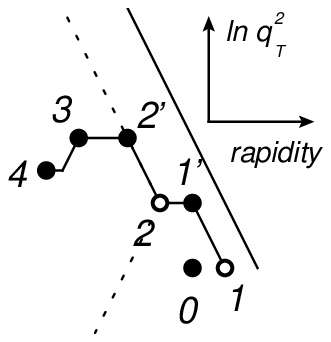}

      \vspace{3mm}
      $(b)$
    \end{center}
  \end{minipage}
  \begin{minipage}{0.3\linewidth}
    \begin{center}
      \includegraphics[scale=1]{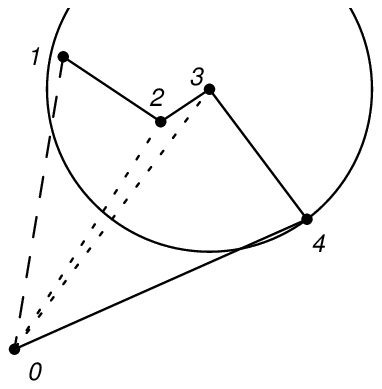}

      \includegraphics[scale=1]{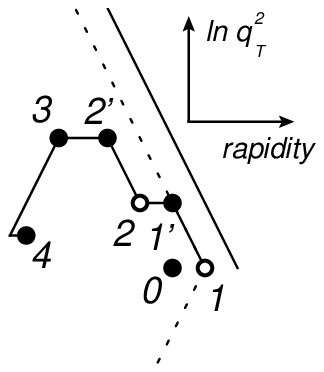}

      $(c)$
    \end{center}
  \end{minipage}

    \caption{\label{fig:coherence} A parton emission (4), being emitted at different distances. The upper figures are in impact-parameter space, and the lower figures are in $y$-$q_\perp$ representation, where the dotted lines show the limits of the $q_+$ and $q_-$ ordering. The circles in the upper figures show the coherence range of the emission.}
}

This procedure is shown in fig.~\ref{fig:coherence}, where parton 4 is emitted at different distances from its parent. In $(a)$, the emission is a smaller dipole than the previous one, and the coherence range is smaller than the distance to the previous parton 2, so the recoil and ordering will be done by parton 3 only. If the emission is at a larger distance than the previous emission, as in $(b)$, then the emission will be done by an effective parton with the merged momenta of parton 2 and 3. Notice that the recoil between 2 and 3 will cancel when their momenta are added, and the effective parton will have the momentum of the original parton 2, as shown in the lower figure $(b)$. Emissions at larger distance will continue to add parton to the effective parton, and can allow emissions that are not ordered with the parent itself, as in $(c)$.

An outer $q_\perp$ maximum is always a small dipole followed by a
larger dipole, and reweighting will sometimes merge the two close by
partons to one. The extra ordered phase space opened by this merging
is exactly the part of phase space used by the coherent emissions at
long distance, which can be seen by comparing
figures~\ref{fig:outsidemaxkt}, \ref{fig:maxmerged} and
\ref{fig:coherence}$(b)$. This algorithm thus ensures that an absorbed
outer $q_\perp$ maximum can not open up a part of phase space that
was not considered in the virtual cascade. By solving this issue, we
have been forced to introduce coherence, where an emitted gluon cannot
resolve partons that are closer together than its wavelength.

While this increased ordered phase space will cover all situations in the final state, some care has to be taken. First, some of the small dipoles that emit coherently will not be absorbed, and in that case the virtual cascade has emitted large dipoles in unordered regions. These large dipoles have a significant effect on inclusive observables, and the total cross sections may get inflated.

Second, the large unordered dipoles are emitted with the motivation that the previous small dipole may get emitted. However, the situation can be that the small dipole will be absorbed \emph{because} the large dipole was emitted, as the large emission makes the previous small dipole a $q_\perp$ maximum. Thus coherence is an estimate of how the final state will look which is to some extent self-fulfilling.

With this in mind, full coherence as described above may need to be
scaled back to give a more accurate description. This was addressed in
sec.~\ref{sec:PStuning} where frame independence was used in deciding
exactly what phase space should should be considered as ordered in the virtual cascade.

\subsection{Ordering in the interaction}
\label{sec:intorder}
Like with the virtual cascade it is impossible to use an ordering that allows the correct part of phase space for both inclusive and exclusive observables, but again some observations can be made.
\begin{enumerate}
\item \label{item:same}The final result should not depend on whether a recoil comes from the cascade or the interaction. Thus the ordering in the interaction should allow the same backbone gluons as the ordering in the cascade.

\item \label{item:different}It is not possible to use the exact same ordering as in the virtual cascade, as the situation is fundamentally different. At a certain rapidity in the virtual cascade, the partons have not yet gotten recoils from emissions at later rapidities. Thus, ordering in the virtual cascade is checked when only recoils with partons in one direction have been made. The same is not true in the interaction, where all recoils in both cascades are already made.
\item \label{item:coherence}Events where an outer $q_\perp$ maximum next to an interaction, as in fig.~\ref{fig:outsideint}, is absorbed will allow more phase space for the interaction in exactly the same way as in the cascade. In the same way, coherence can be used to make sure all final states are still being considered.

\item \label{item:onlyone}In the virtual cascade, $q_+$ and $q_-$ held two fundamentally different roles where $q_+$ was the lightcone momentum the particle came in with, and $q_-$ measured the increasing virtuality. In the interaction the situation is completely symmetric in $q_+$ and $q_-$, so the orderings must be the same.
\item \label{item:limit}In the cascade, there is no absolute limit on how small $q_-$ can be allowed at a certain rapidity, as no momentum, $q_+$ or $q_-$, will be emitted in the limit of a soft gluon. However, energy conservation sets a hard limit on how much $q_+$ can be allowed, namely the $q_+$ of the parent.
\end{enumerate}
Putting these observations together, we see that there is a limit for how much phase space can be allowed, as there always have to be enough energy for the interaction recoil to come on shell. Notice that this corresponds to both $q_+$ and $q_-$ ordering, as the particles from both states must bring sufficient energy. As in the previous section, the possibility of merged partons forces coherent ordering when the interaction distance is larger than the previous dipoles in the cascade. The allowed phase space is illustrated for two different impact-parameters in fig.~\ref{fig:intorder}.

\FIGURE{
  \begin{minipage}{0.45\linewidth}
    \begin{center}
      \includegraphics[scale=1]{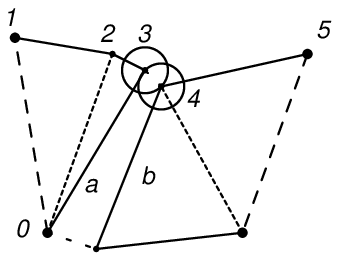}

      \vspace{5mm}

      \includegraphics[scale=1]{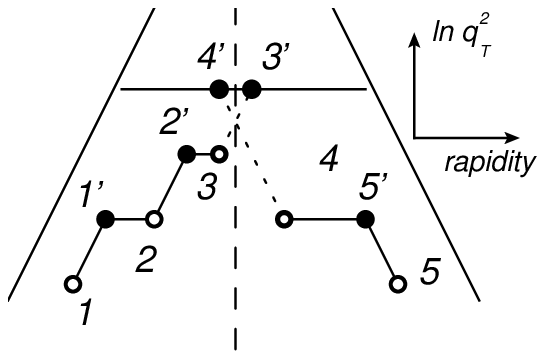}

      $(a)$
    \end{center}
  \end{minipage}
  \begin{minipage}{0.45\linewidth}
    \begin{center}
      \includegraphics[scale=1]{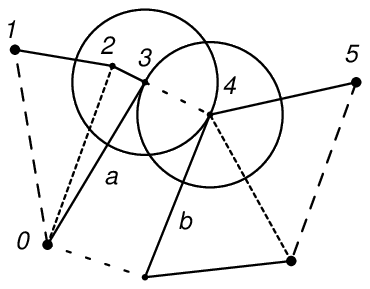}

      \vspace{5mm}

      \includegraphics[scale=1]{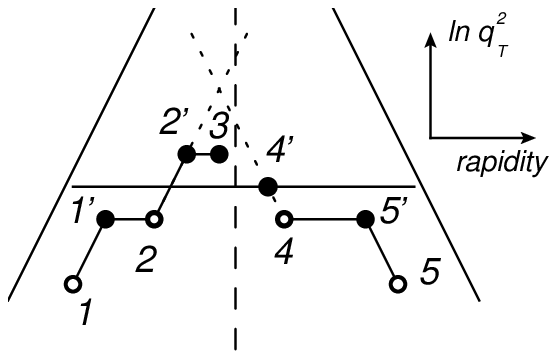}

      $(b)$
    \end{center}
  \end{minipage}

  \caption{\label{fig:intorder} Ordering in an interaction for two
    different impact-parameters, where parton 1-3 are belonging to one
    cascade, and 4 and 5 to the other. The upper figures are in
    impact-parameter space, and the circles shows the coherence range in the
    interaction for parton 3 and 4. The lower figures are in
    $y$-$q_\perp$ representation, where vertical dashed line is the
    interaction frame $y_0$, the horizontal full line marks the
    interaction $q_\perp$, and the diagonal dotted lines marks the
    $q_-$ and $q_+$ parton 3 and 4 respectively brings to the
    interaction.}
}

In $(a)$, parton 3 and 4 are close to each other, corresponding to a large interaction~$q_\perp$ that will recoil the partons past each other as can be seen in the lower figure. As 4' needs more $q_-$ than 3 can supply, and conversely parton 3' needs more $q_+$ than 4 supplies, the interaction~$q_\perp$ can not be set on shell, and $f_{ij}$ will be set to 0. In this case, there is no outer $q_\perp$ maximum in either cascade as the maximum is in the interaction, so no parton risk being absorbed and coherence will not include any extra partons.

If the impact-parameter between the two cascades would be such as in $(b)$ where the partons 3 and 4 are farther away from each other, the interaction~$q_\perp$ is smaller and the coherence range larger. As can be seen in the lower figure, the smaller virtuality and the coherent interaction of partons 2 and 3 means that there is enough $q_+$ and $q_-$ to set the interaction on shell, and $f_{ij}$ is calculated as normal.

Here it should also be taken into account that by transferring a fraction $1-z$ of the supplied $q_+$ to the other state at fix $q_\perp$ will increase the virtuality by a factor $1/z$, and conversely for the colliding parton, which gives the limit for allowed interactions:
\begin{equation}
q_+q_- > 16 q_{\perp \text{int}}^2, \label{eq:intPS}
\end{equation}
where $q_+$ and $q_-$ are the supplied lightcone momenta from the interacting effective partons, and $q_{\perp \text{int}}$ is the transverse recoil from the interaction.

This is a minimum for what has to be cut away in the ordering, and
from frame independence it turns out that it can not be stricter (see
section \ref{sec:findep}), thus motivating us to set the allowed phase
space in $f_{ij}$ to \eqRef{eq:intPS}.

\section{Saturation effects}
\label{sec:finalsat}

All motivations and examples discussed above have been based on a single backbone gluon chain, but in LHC events there will be several chains branching of and merging due to multiple interactions and the swing. This complicates the situation for most of the arguments presented previously in this section. However, the algorithms have been designed with saturation effects in mind, to give realistic results in all situations. Here follows some of the larger effects from saturation, and how the above algorithms handle them.

\subsection{Multiple interaction}
\FIGURE{
  \includegraphics[scale=1]{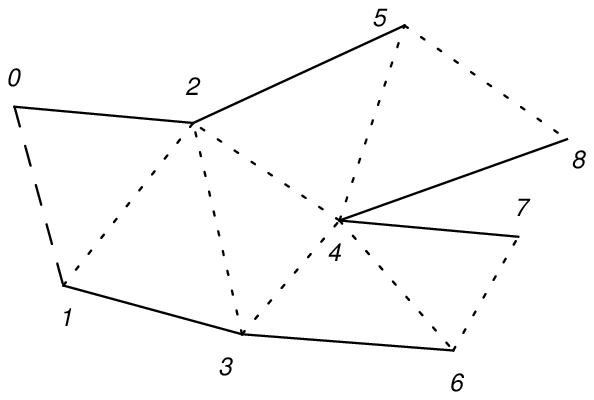}
  \caption{\label{fig:outsidemultint} Two interactions makes a chain
    split in two. Dotted lines show parent structure, full lines show
    colour flow. The picture is in impact-parameter space.}
}

In the interaction, multiple scatterings are allowed, which means that
there will be several chains cutting the interaction frame. These
chains will merge at some point when following them back, if not
before, they will always meet at the valence partons. This does not
introduce any complications for the ordering. Each emission is
controlled at the time of the emission, and if another branch has
taken energy from the parents, then the ordered phase space for
emission will be smaller, which is expected.

One has to be a bit careful with the
reweighting however, but the algorithm described in
sec.~\ref{sec:RutherfordAlgorithm} is made to handle both splitting and
merging chains. The dipoles created with a weight of
$d^2\pmb{r}/\pmb{r}^2$ are still the partons that have on-shell children
on one side, but not on the other, and therefore the definition of an outer
$q_\perp$ maximum can remain.

This is illustrated in fig.~\ref{fig:outsidemultint}, where a backbone
chain splits up in two due to multiple interaction. Dipoles (24) and
(34) both get the weight $d^2r$ corresponding to
$d^2q_\perp/q_\perp^4$, and they will get two factors of $\al_s(r)$ in
case of $q_\perp$ maximum. Thus they do not need any extra
suppression, and they do indeed not fit the definition of outer
dipole, as parton 4 has children on both sides. For all other partons
and dipoles, the arguments for the single chain can be applied as
normal, so the reweighting algorithm works also with multiple
interactions.

\subsection{The Swing}
\label{sec:satswing}
\FIGURE{
  \includegraphics[scale=1]{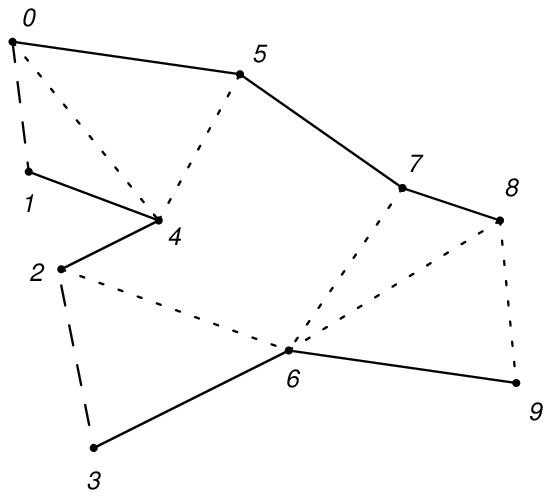}
  \caption{\label{fig:outsideswing} A swing between (45) and (26) causing two chains of backbone gluons to merge. Dotted lines show parent structure, full lines show colour flow. The picture is in impact-parameter space.}
}
The dipole swing in the virtual cascade does not cause any recoils, but it
does affect the parent structure of future emissions. In a non-swinged, single
interaction chain, one of the two parents of each parton will also be parent to the other parent, making for the typical ladder-like structure seen in previous figures. However, with swinged emissions this is no longer true, and the two parents can have 4 different parents.

This is seen in fig.~\ref{fig:outsideswing} where a swing has recoupled dipole (45) and (26) into (24) and (56), (56) then evolves further and interacts, while (24) does not interact. Here the parents of parton 8 and 9 cover all the partons (notice that the parents of 7 are parton 5 and 6), so all partons are set on-shell. Notice in the figure that even though parton 4 is colour connected to 2 from the swing, parton 4 was not emitted by 2 (it was emitted by 0 and 1); thus no recoil has been made between the two partons, and no reweighting has to be made.

Dipole (14) has the weight $1/r^2$ from the cascade and is an outer dipole as parton 4 has no children on that side. Thus (14) will, if smaller than (45), be reweighted by
\begin{equation}
\frac{r_{14}^2\al_s(r_{14})}{r^2_{45}\al_s(r_{45})}
\end{equation}
from \eqRef{eq:rutherford}, and reweighting again produces the correct
weights. Correspondingly for parton 2, which is in the same situation as 4, but since 2 is a valence parton with no parents, there is no reweighting to be done.
Dipole (45) is not an outer dipole because parton 7 is a child of 5 on the same side as (45), so no reweighting will be done. If (45) is a $q_\perp$ maximum, it will be emitted with a weight $\al(r_{45})/r_{45}^2$, and the swing is made with a probability $\frac{r_{45}^2r_{26}^2}{r_{24}^2r_{56}^2}$, giving the weight
\begin{equation}
r_{45}^0 \al(r_{45}) \qquad \leftrightarrow \qquad \frac{\al(r_{45})}{q_{45\perp}^4} .
\end{equation}
Notice that while the power of $q_\perp$ is the correct one, a power in the running coupling is missing. This power of $\al_s$ will instead come from the smallest dipole in the emission of parton 7 from 5 and 6. Thus, a saturated cascade can give rise to $q_\perp$ maxima where one of the powers in $\al_s$ is at the wrong scale.

This flaw can possibly be mended by introducing a running $\al_s$ in the swing amplitude, but in the current version of \dipsy it is not implemented. As it is a next to leading order factor in the weight occurring only in special configurations of a saturated cascade, it is estimated to be a small effect.

In conclusion, the algorithm in section \ref{sec:RutherfordAlgorithm} will still provide a satisfactory treatment of reweighting outer $q_\perp$ maxima.

\subsection{colour flow in saturated cascades}
For a saturated cascade, the colour flow does not necessarily return to the original structure when the virtual emissions are removed according to \ref{sec:colourflow}. The swings are changing the colour structure, and after all the virtual emissions are reabsorbed, the colour flow does not necessarily reproduce the structure without the virtual emissions. An example of such a virtual cascade and backbone gluons is shown in fig.~\ref{fig:virtualswing}.

\FIGURE{
  \includegraphics[angle=0, scale=0.65]{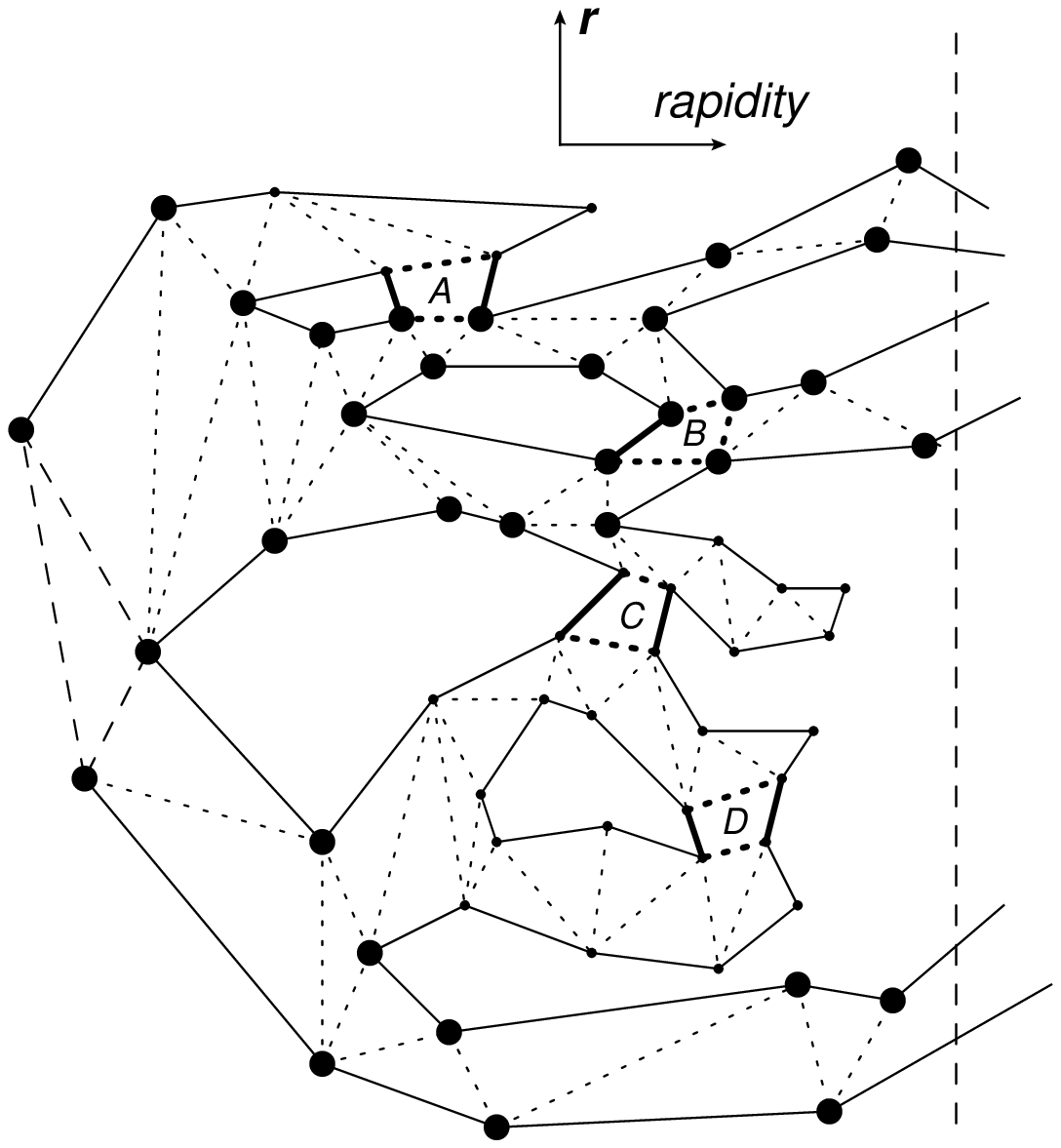}
  \includegraphics[angle=0, scale=0.65]{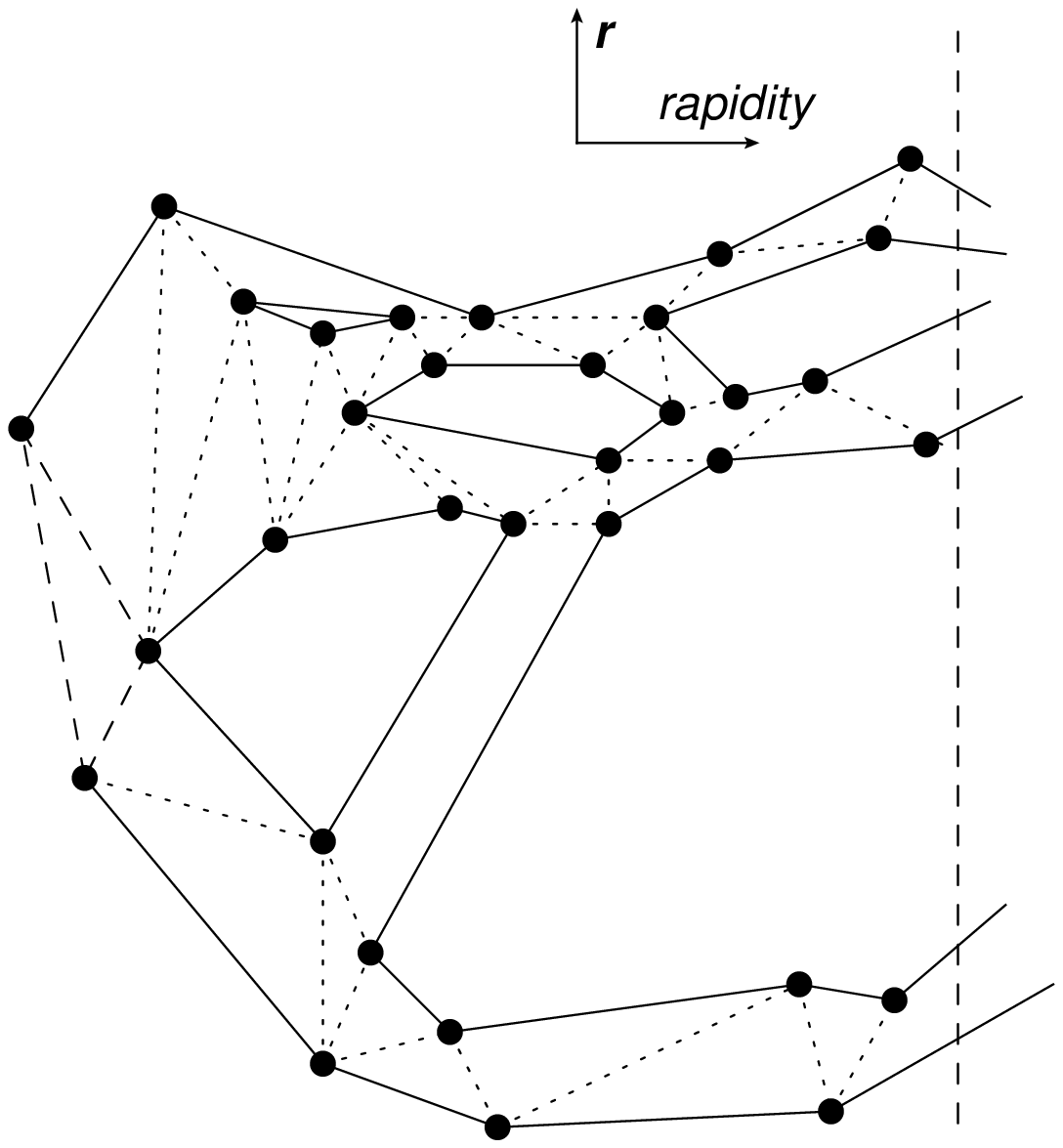}
  \caption{ \label{fig:virtualswing} A cascade with swings and
    multiple interactions. The long dashed lines on the left side are
    the three valence dipoles of a proton. Full lines mark colour flow
    and dotted lines mark emissions. There are three interactions,
    which are the colour lines passing over the interaction rapidity
    marked by a vertical dashed line to the right. The big dots are
    the backbone gluons, selected through tracing each interaction back
    towards the valence partons, and the rest of the partons are
    virtual. The left figure is the virtual cascade, and four swings A
    to D are highlighted with thick lines. The right figure shows the
    backbone gluons after the virtual partons have been absorbed.}
}

Swings where one of the swinged dipoles go on to interact will merge
two backbone chains, as was seen in sec.~\ref{sec:satswing}, and are very
important for a frame-independent description. Swings by an
outer dipole, such as A, will not influence the choice of backbone
gluons, but can reconnect the colour flow between the backbone gluons. A
colour reconnection between backbone gluons can also happen by a swing
between virtual dipoles, if the dipoles originate from different parts
of the backbone chain, as is the case for swing C. If the two swinging
virtual dipoles originate from the same part on the backbone chain, as
swing D, no difference will be seen in the colour flow between the
backbone gluons.

In this way, not only exchange of individual soft gluons between the chains are simulated, but virtual chains of soft and hard gluons can carry colour flow between the chains, or between different parts of the same chain. The virtual cascade can thus give a very dynamic and detailed description on how the colour flow changes in saturated environments, which will be increasingly important at high energies, and can be an essential ingredient in heavy ion collisions.

\bibliographystyle{utcaps}
\bibliography{/home/william/people/leif/personal/lib/tex/bib/references,refs}

\end{document}